%% file: bolu.tex
\documentstyle[revtex]{aps}
\widetext
\begin{document}
\draft
\begin{title}
Single Squark Production at TeV Scale $\gamma p$ and $\gamma e$ Colliders
\end{title}
\author{Z.Z. Aydin, N. Karag\"{o}z \thanks{Abant Izzet Baysal University,
Physics Department, Bolu - Turkey} and A.U. Y\i lmazer}
\begin{instit}
Ankara University, Faculty of Engineering \\
Department of Engineering Physics \\
06100 Tando\u{g}an, Ankara - Turkey
\end{instit}
\begin{abstract}
We calculate the total cross sections for squark productions
at future $\gamma p$ and $\gamma e$
colliders in $R_p$ violating MSSM. The $\lambda'_{ijk}$ couplings
offer several ways of producing single squarks in photon
colliders. The results are discussed  and compared with
those of the existing colliders.
\end{abstract}

\section{Introduction}
\label{sec:intro}

   Supersymmetric particles can be produced singly in the R-parity
violating version of the minimal supersymmetric standard model
(${R\!\!\!/}_p - MSSM$).
Although the production cross section is suppressed
by ${R\!\!\!/}_p$ coupling constants the relevant processes have higher
kinematical reaches compared with the case of pair (or associated)
productions of superparticles through the $R_p$-conserving mechanisms. Also
$R_p$ violation leads to a totally different phenomenology and the detailed
studies for the specific colliders (LEP2, HERA, TEVATRON)
have already appeared in the literature [1].

On the other hand in recent years in addition to
the existing colliders the possibilities
of the realization of $\gamma e$, $\gamma \gamma$ and $\gamma p$
colliders have been proposed and discussed in detail [2].
Colliding the
beam of high energy photons produced by Compton backscattering of
laser photons off linac electrons with
the beam of a proton ring is the idea leading
to TeV scale high luminosity
$\gamma p$ colliders. Also $\gamma e$ option has been
included in the conceptual
design of the linear colliders such as TESLA. Physics program of the photon
colliders are studied in [3].
Also search for SUSY in polarized  $\gamma p$
collisions have been investigated in [4].

In searching the superpartners
squarks (except the third generation ones)
might be too heavy to be produced at HERA, LEP or
TEVATRON, but sleptons are generally expected to be lighter than squarks
so single slepton production would be interesting. Also pair production of
sleptons via $R_p$ conserving mechanisms might be closed kinematically.
The s-channel slepton resonance production
via ${R\!\!\!/}_p$ interactions in $e^+ e^-$
collisions through
$e^+e^-\rightarrow \tilde{\nu}\rightarrow l^+l^-$, and
in $p\bar p$ collisions through
$p\bar p \rightarrow \tilde{\nu}\rightarrow l^+l^-$,
$p\bar p \rightarrow \tilde{l}^+ \rightarrow l^+ \nu$ have been examined
in the literature [5]. In $e^+p$ collisions at HERA squark resonance
productions via
$e^+d^k_R \rightarrow \tilde{u}^j_L$  ($\tilde{u}^j=
\tilde{u}, \tilde{c}, \tilde{t}$),
$e^+ \bar u^j_L \rightarrow \bar {\tilde{d}^k}_R$, ($\tilde{d}^k=
\tilde{d}, \tilde{s}, \tilde{b}$) have been investigated in [6].
Although HERA, LEP, FERMILAB and LHC should be sufficient to check
the low energy SUSY however
experiments at all possible types of colliding beams would be useful
to explore the new physics around the TeV scale, and hence
with their additional polarization facilities future
$\gamma e$ and $\gamma p$ colliders might play
complementary roles to the existing accelerators.
                     
In  this letter we mainly focus on the single squark
productions via
${R\!\!\!/}_p - MSSM$ interactions
at photon-proton and photon-electron collisions
as an alternative to the above-mentioned s-channel
resonance production.
                     
\section{Single squark production with an escaping neutrino
at gamma-proton colliders}
\label{sec:pro}

$R_p$ violating superpotential terms can be written in four-component
Dirac notation as

\begin{eqnarray}
{\cal L}_{{R\!\!\!/}_p} &=& 
\lambda_{[ij]k} \Big[\tilde\nu_{iL} \bar
e_{kR} e_{jL} + \tilde e_{jL} \bar e_{kR} \nu_{iL} + \tilde
e^\star_{kR} \overline{(\nu_{iL})^C} e_{jL} - \tilde\nu_{jL} \bar
e_{kR} e_{iL} \nonumber \\[2mm] && - \tilde e_{iL}
\bar e_{kR} \nu_{jL} + \tilde e^\star_{kR} \overline{(\nu_{jL})^C}
e_{iL}\Big] + \lambda'_{ijk} \Big[\tilde\nu_{iL} \bar d_{kR}
d_{jL} + \tilde d_{jL} \bar d_{kR} \nu_{iL} \nonumber \\[2mm]
&& + \tilde d^\star_{kR} \overline{(\nu_{iL})^C} d_{jL} - \tilde e_{iL} \bar
d_{kR} u_{jL} - \tilde u_{jL} \bar d_{kR} e_{iL} - \tilde
d^\star_{kR} \overline{(e_{iL})^C} u_{jL}\Big] 
+ \lambda^{\prime\prime}_{i[jk]} \epsilon_{\alpha\beta\gamma} 
\nonumber \\[2mm] 
&& \Big[\tilde u^\star_{iR\alpha} \bar
d_{kR\beta} d^C_{jR\gamma} + \tilde d_{jR\beta} \bar
e_{kR\gamma} u^C_{iR\alpha} + \tilde d^\star_{kR\gamma}
\overline{(u_{iR\alpha})^C} d_{jR\beta}\Big] + h.c.
\end{eqnarray}
where $i, j, k$  are the generation indices. In photon-proton
collisions a single slepton or a single squark production
is possible. Single productions of charged sleptons (stau)
and sneutrinos have been considered in a previous letter [7]. Here let
us examine the single squark production associated with an escaping neutrino
$\gamma p \rightarrow \tilde{d}^j_R \nu_LX$. We note that
$\gamma p \rightarrow \tilde{d}^j_L \nu_LX$ is also possible from the
second $\lambda'$ term in Eq.(1).
One of the relevant subprocesses, $\gamma d \rightarrow \tilde{d}^j_R \nu^k_L$,
proceeds via the s-channel d-quark, t-channel squark
exchanges.
The invariant amplitude in four-component Dirac notation
(which could be written equally in two-component Weyl language) is

\begin{eqnarray}
M=e_d g_e \lambda' \epsilon_{\mu}(k)\bar{u}(p')Q^{\mu} u(p)\\[4mm]
Q^{\mu}={1\over2} (1-\gamma_5)\biggr [\frac{\not\! k+\not\! p}
{\hat s-m^2_d}\gamma ^{\mu}+
\frac {(p-p'+k')^{\mu}}{\hat t -{m_{\tilde d}}^2}
\biggr ]
\end{eqnarray}
where $g_e=\sqrt {4\pi \alpha}$, $\epsilon_{\mu}(k)$ is the
photon polarisation, $q_d$ is the charge factor
and $k, p, k'$ and $p'$ are the four momenta
of the photon, d-quark in the proton, squark and escaping neutrino
respectively. The differential cross-section for the subprocess is
given by
\begin{eqnarray}
\frac{d\hat{\sigma}}{d\hat{t}}=\frac{1}{16\pi\hat{s}^2}M^2
\end{eqnarray}
and to obtain the total cross section for the subprocess
$\gamma d \rightarrow {\tilde d}^j_R \nu^k_L$ one should
perform integration over $\hat t$. The total cross-section
for the main process
$\gamma p \rightarrow {\tilde d}^i_R \nu^k_L X$ is obtained
after the integration of $\hat{\sigma}$
over the quark and photon distributions.
For this
purpose we make the following change of variables: first
expressing $\hat{s}$ as $\hat{s}=x_1x_2s$ where
$\hat{s}=s_{{\gamma}q}$, $s=s_{ep}$, $ x_1=E_{\gamma}/E_e$,
$x_2=E_q/E_p$ and furthermore calling $\tau =x_1x_2$, $x_2=x$
then one obtains $dx_1dx_2 = dx d\tau/x$. The limiting values
are $x_{1,max}=0.83$ in order to get rid of the background effects
in the Compton backscattering, particularly $e^+e^-$ pair
production in the collision of the laser with the high energy
photon in the conversion region,
$x_{1,min}=0$, $x_{2,max}=1$,
$x_{2,min}=\frac{\tau}{0.83}$,
$\hat{s}_{min}=m^2_{\tilde{d}}$.
Then we can write the total cross-section as :

\begin{eqnarray}
\sigma=\int^{0.83}_{ {m^2_{\tilde{d}}}/s }
d\tau\int^{1}_{\tau/0.83}
dx\frac{1}{x}f_{\gamma}(\frac{\tau}{x})f_q(x)\hat{\sigma}
(\tau s,m_{\tilde{e}})
\end{eqnarray}
where $f_q(x)$ is the distribution of down-quarks inside
the proton [8]
\begin{eqnarray}
f_q(x)=0.67 x^{-0.6}(1-x^{1.5})^{4.5}
\end{eqnarray}
and $f_{\gamma}(y)$ is the energy spectrum of the high energy
real photons (Ginzburg et.al. in ref.[2]
\begin{eqnarray}
f_{\gamma}(y)=\frac{1}{D(\kappa)}\biggr[1-y-\frac{1}{1-y}-\frac{4y}
{\kappa(1-y)}+\frac{4y^2}{{\kappa}^2(1-y)^2}\biggr]
\end{eqnarray}
with $y=\frac{E_{\gamma}}{E_e}, \kappa\cong 4.8, D(\kappa)
\cong 1.84$,and $y_{max}\cong 0.83$. $Q^2$ independent proton structure
function used above is satisfactory for the present analysis;
use of the $Q^2$ dependent quark distributions $f(x,Q^2)$
does not alter the results significantly.

The current upper bounds on
the relavant ${R\!\!\!/}_p$ coupling constants are
$\lambda'_{311}=\lambda'_{312}=\lambda'_{313}
= 0.11 \times \frac {m_{\tilde d}}{100 GeV}$ [12],
consequently
we may look for the productions of
$\tilde d_R$, $\tilde s_R$ or $\tilde b_R$.
The results of the numerical integration
for $\gamma p \rightarrow \tilde q_R \nu_{\tau L} X$, where
$\tilde q_R$ being either one of the above three down-type squarks
is plotted in Fig.1.

As can be seen from the figure the total cross-section for
a $\tilde d$ mass of 400 GeV is about 0.1 $pb$. Hence around
one thousand
events per running year can be seen at HERA+LC up to
$\tilde d$ masses of  400 GeV. For comparison we note that
for the resonant production of
squarks of masses up to 200 GeV at  the HERA the total cross-section is
0.1 - 1 $pb$ (see E.Perez et.al. in ref.[6]).
On the other hand, if one uses the Weizsacker-Williams
approximations for the quasi-real photon distribution at the HERA machine
a similar
$\gamma^* p \rightarrow {\tilde d}^i_R \nu^k_L X$
process is possible but with
an almost hundred times smaller cross-section since WW-spectrum
is much softer than the real $\gamma$-spectrum. Clearly for the HERA machine
resonant production of the sparticles in R-parity violating
MSSM is the dominant process.\\

{\bf {Signature:}}
In models with R-parity violation the LSP is unstable, which leads
to signatures which differ strongly from the characteristic
missing energy signals in usual MSSM. In our case
the produced squark ($\tilde d$,$\tilde s$ or $\tilde b$)
will decay either by direct
${R\!\!\!/}_p$ couplings into matter fermions
$\tilde d_R \rightarrow \tau+u$ (or $\nu_{\tau}+d$)
through the same coupling
$\lambda'_{ijk}$) leading to the signals
1 $lepton$ + 1 $jet$ + ${E\!\!\!\!/}_T$
or 1 $jet$ + ${E\!\!\!\!/}_T$ (however there are large SM backgrounds);
or by cascading
through MSSM to the LSP which in turn decays via  ${R\!\!\!/}_p$:
$\tilde d^i_R \rightarrow q_R + \tilde{\chi}^o_1$,
$\tilde{\chi}^o_1 \rightarrow \tau^+ d \bar u
 \  {\rm or} \  d \bar d \nu_{\tau}$
leading to the signals 2 $leptons$ + 1 $jet$ + ${E\!\!\!\!/}_T$ or
1 $lepton$ + 3 $jets$ + ${E\!\!\!\!/}_T$.
Within MSSM it is forbidden for $\tilde q_R$ to decay into charginos.
These LSP decays depend on couplings $\lambda '$ but also on the
supersymmetry parameters $M_2$, $\mu$ and $\tan {\beta}$
(see E.Perez et.al. ibid). In  the special cases involving only
the operators $L_iQ_j\bar D_j$ the LSP can also dominantly decay
via the radiative process
$\tilde {\chi}^o_1 \rightarrow \gamma + \nu $ \cite{dreiner91}.
On the other hand the above ${R\!\!\!/}_p$ decays of the produced
squark have the following partial decay widths :
\begin{eqnarray}
\Gamma_{\tilde q_R \rightarrow \nu_{\tau L} + q_d}=\frac
{m_{\tilde q}(\lambda'_{313})^2}{16\pi}\\
\Gamma_{\tilde q_R \rightarrow \tau + q_u}=\frac
{m_{\tilde q}(\lambda'_{313})^2}{16\pi}
\end{eqnarray}
thus they are equal to each other.

\section{Single squark production associating tau lepton
at gamma-proton colliders}
\label{sec:sneutr}
${R\!\!\!/}_p$ Yukawa couplings $\lambda'_{ijk}$ offer also the
opportunity to produce a single squark with an associating
lepton in gamma-proton collisions.
The relevant subprocess
$\gamma u \rightarrow \tilde d^i_R e^{+k}_L$,
proceeds via the s-channel u-quark, t-cannel squark and
u-channel lepton exchanges. The invariant amplitude is given as in Equ.(2),
with $Q^{\mu}$ as,

\begin{eqnarray}
Q^{\mu}={1\over2} (1-\gamma_5)\biggr [e_u\frac{\not\! k+\not\! p}
{\hat s-m^2_u}\gamma ^{\mu}+e_d
\frac {(p-p'+k')^{\mu}}{\hat t -{m_{\tilde d}}^2}
+e_{\tau}\gamma ^{\mu}\frac{\not\! p'-\not\! k}
{\hat u-m^2_{\tau}}\biggr ]
\end{eqnarray}

${R\!\!\!/}_p$ coupling constants
$\lambda'_{311}=\lambda'_{312}=\lambda'_{313}
= 0.11 \times \frac {m_{\tilde d}}{100 GeV}$ favor the productions
of $\tilde d_R$,$\tilde s_R$ or $\tilde b_R$ respectively, associating
the $\tau$-lepton. Since instead of a heavy sparticle, a $\tau$-lepton is
exchanged in the u-channel the cross section for squark production
becomes considerably bigger than the case analysed in the previous
section. The details
of the calculation is similar and  taking again
the simplistic $Q^2$ independent
distribution of up-quarks inside the proton as [8]
\begin{eqnarray}
f_q(x)=2.751 x^{-0.412}(1-x)^{2.69}
\end{eqnarray}
the results of the numerical integration for
$\gamma p \rightarrow \tilde d^i_R \tau_L X$ is plotted
in Fig.2. Similarly d-quark in the proton gives rise to the production
$\gamma d \rightarrow \tilde u_R \tilde{\tau^c_L}X$, which contributes
roughly half of the u-quark distribution, and leads to the same signature.
Hence around $10^3$ events per running year can be seen at
HERA+LC up to $\tilde d_R$ masses of 300 GeV. For single
squark production at LEP and hadron colliders see [1],[5] and [9-10].

{\bf {Signature:}} Considering leptonic $\tau$-decay via SM- W boson to
$\tau^- \rightarrow \nu_{\tau}W \rightarrow \nu_{\tau} \nu_{l}l $ and also
taking into account the squark decays listed above we hence point out
that the signature might be
2 $leptons$ + 1 $jet$ + ${E\!\!\!\!/}_T$ or
1 $leptons$ + 1 $jet$ + ${E\!\!\!\!/}_T$ or
3 $leptons$ + 2 $jets$ + ${E\!\!\!\!/}_T$ or
3 $leptons$ + ${E\!\!\!\!/}_T$.

\section{Single squark production at electron-photon colliders}

The same ${R\!\!\!/}_p$ Yukawa couplings $\lambda'_{ijk}$
(i.e. eleventh term in Equ.(1) ) offer the opportunity
to produce squarks in photon-electron collisions.
One of the relevant subprocess
$\gamma e \rightarrow \tilde u_{jL} d_{kR}$,
proceeds via electron (left-handed) exchange
in s-channel,squark exchange in the t-channel and quark (right-handed)
exchange in u-channel.
 The invariant amplitude in two-component MSSM language is given as

\begin{eqnarray}
M=g_e \lambda' \epsilon_{\mu}(k)\psi_+(p')Q^{\mu} \psi_-(p)\\[4mm]
Q^{\mu}=\biggr [e_d \frac{(k+p)_\nu}
{\hat s-m^2_e}\sigma ^{\nu} \bar{\sigma} ^\mu+e_{\tilde t}
\frac {(p-p'+k')^{\mu}}{\hat t-{m_{\tilde u}^2}}+e_s
\frac {(p'-k)_{\nu}}{\hat u -{m_{q}}^2}\sigma ^{\mu} \bar{\sigma}^\nu
\biggr ]
\end{eqnarray}
where $\psi_+(p')$ ($\psi_-(p)$) is the Weyl spinor for the right-handed
quark (left-handed electron). Taking $\lambda'_{132}=0.33$
(present upper bound) the total cross section for the particular process
$\gamma e \rightarrow s \tilde t $  may be
written as

\begin{eqnarray}
\hat{\sigma} (\hat{s},\gamma e)= \int^{t_{max}} _{t_{min}}
\frac{1}{16\pi\hat{s}^2}M^2 d\hat t
\end{eqnarray}
where
\begin{eqnarray}
t_{max/min}=(\frac{m^2_{\tilde {t}}-m^2_{q}}{2 \sqrt{s}})^2
-\biggr \{ \frac{\sqrt s}{2} \mp
\biggr [ \frac {(s+m^2_{\tilde t}-m^2_{q})^2}{4s}
- m^2_{\tilde t} \biggr ]^{1/2}
\biggr \}^2
\end{eqnarray}
The total cross-section for $e^+ e^- \rightarrow e s \tilde {t}$
is given by

\begin{eqnarray}
\sigma (s, e^+ e^-)=2 \int^{0.83}_{ {m^2_{\tilde{\tau}}}/s }
f_{\gamma /e}(y)\hat{\sigma}
(ys,m_{\tilde{\nu}})dy
\end{eqnarray}
where $f_{\gamma /e}$ is the distribution of high energy real photons
at a given fraction $y=\frac{E_{\gamma}}{E_e}$.
The results
of numerical integration for $e^+ e^-$ $\rightarrow$ $e s \tilde{t}$
is plotted in Fig.3 for a linear collider with center-of-mass energy of 500 GeV.
As can be seen from the figure around $ten$ $thousands$ events per running year
can be observed up to $\tilde {t}$ masses of 275 GeV.
On the other hand using the Weizsacker-Williams approximation
for the photon flux the same process can also be studied at the ordinary
modes of LEP2 and NLC.
The relevant calculations are completely similar to the above analysis, the
only difference is in taking the upper limit of the integral
as unity in the Equ.(16)
and the insertion of the WW distribution, which is given by
\begin{eqnarray}
f_{\gamma /e}(y)=\frac{\alpha}{2\pi}\biggr[\frac{1+(1-y)^2}{y}
\ln \frac{Q_{max}^2}{Q_{min}^2}-2m_e^2
(\frac {1}{Q_{min}^2}-\frac{1}{Q_{max}^2})\biggr]
\end{eqnarray}
The results of the numerical integration for the total cross section
at a center of mass energy of 500 GeV
is shown in Fig.(4) as a function of the
sneutrino mass. The total cross section is approximately a
hundred times smaller than the value obtained at
the  $\gamma e$ collider mode
of NLC, and consequently the discovery mass reach for $\tilde {t}$
is less than 150 GeV.

Here we should note that the production of a single scalar leptoquark
in $ \gamma e$ collisions is a similar process and
has been studied in some detail in a series
of papers by Doncheski and Godfrey [11]. The total cross section is the
same up to a trivial relabeling of the couplings. They have also taken
into account the resolved photon contribution, which can
surprisingly give values
higher than the direct photon
interactions. On the other hand the best place to
produce squarks singly is an ep-collider (namely HERA)  [1].

In this paper we have first investigated three different ways of
producing single squarks (down-type)
at TeV scale $\gamma p$ colliders considering
R-parity violating $LQ\bar D$ interactions. In one of them  the particle
associating the squark is a neutrino and in the latter case
it is a tau-lepton.Then single production of squarks
through again $LQ\bar D$ interactions at $\gamma e$ mode of the
NLC colliders has been discussed.
For comparison,
it is shown that for $\tilde {t}$ production
the $\gamma e$ collider mode of a future NLC gives much higher values
of the total cross section than
the normal operation
mode through soft WW photons.
The production cross sections are
functions of only sparticle mass and ${R\!\!\!/}_p$
coupling constants, and
lead to detectable signals. Polarizations of the initial photon
beam [8-9], which can be accomplished relatively easily,
constitute additional advantages. Our results
show that $\gamma e$ and $\gamma p$ colliders can play complementary role
in searching for supersymmetry in the future.


\newpage

\widetext

\smallskip

\figure{ Production cross section of
a down-squark ( $\tilde d_R$, $\tilde s_R$ or $\tilde b_R$ )
as a function of its mass for HERA+LC
$\gamma p$ collider via the process
$\gamma p \rightarrow {\tilde d}^i_R \nu^k_L X$.\label{fig1}}

\figure{ Production cross section of
a down-squark ( $\tilde d_R$, $\tilde s_R$ or $\tilde b_R$ )
as a function of its mass for HERA+LC
$\gamma p$ collider via the process
$\gamma p \rightarrow {\tilde d}^i_R + \tau_L + X$.\label{fig2}}

\figure{ Production cross section of
a stop ( $\tilde t_R$ )
as a function of its mass for $\gamma e$ mode of LC collider
via the process
$\gamma e \rightarrow s \tilde t$.\label{fig3}}

\figure{ Production cross section of
a stop ( $\tilde t_R$ )
as a function of its mass for NLC ordinary mode
with WW-photons via the process
$\gamma^* e \rightarrow s \tilde t$.\label{fig4}}

\begin{center}
\input{fig1}
\smallskip
\smallskip

Fig.1
\end{center}
\bigskip

\begin{center}
\input{fig2}
\smallskip
\smallskip

Fig.2
\end{center}
\bigskip

\begin{center}
\input{fig3}
\smallskip
\smallskip

Fig.3
\end{center}
\bigskip

\begin{center}
\input{fig4}
\smallskip
\smallskip

Fig.4
\end{center}

\smallskip

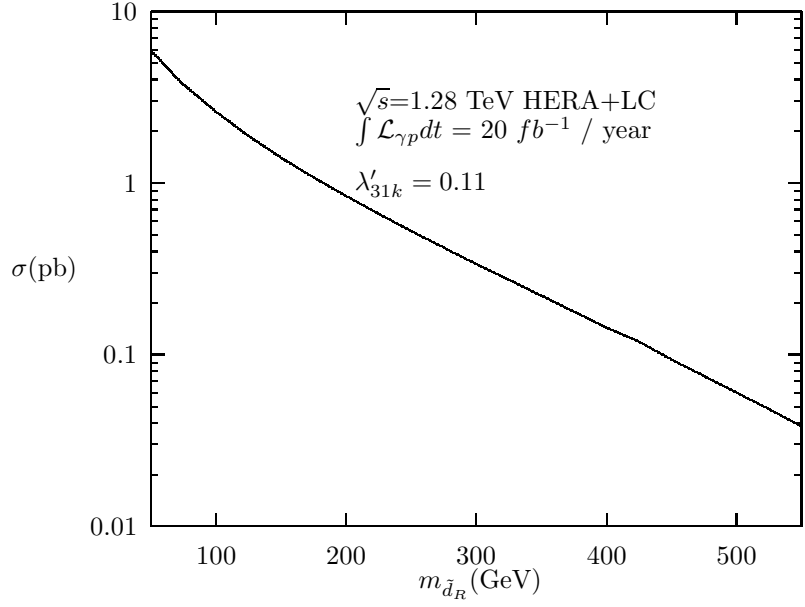
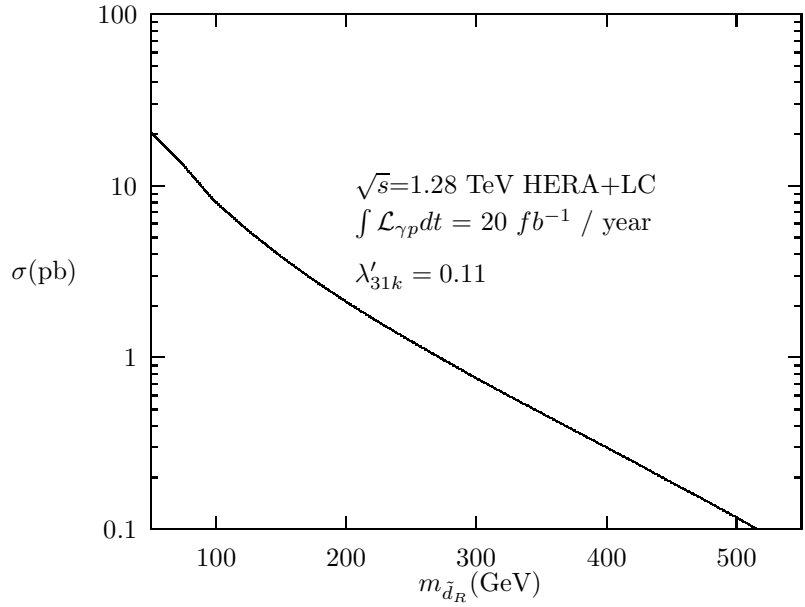
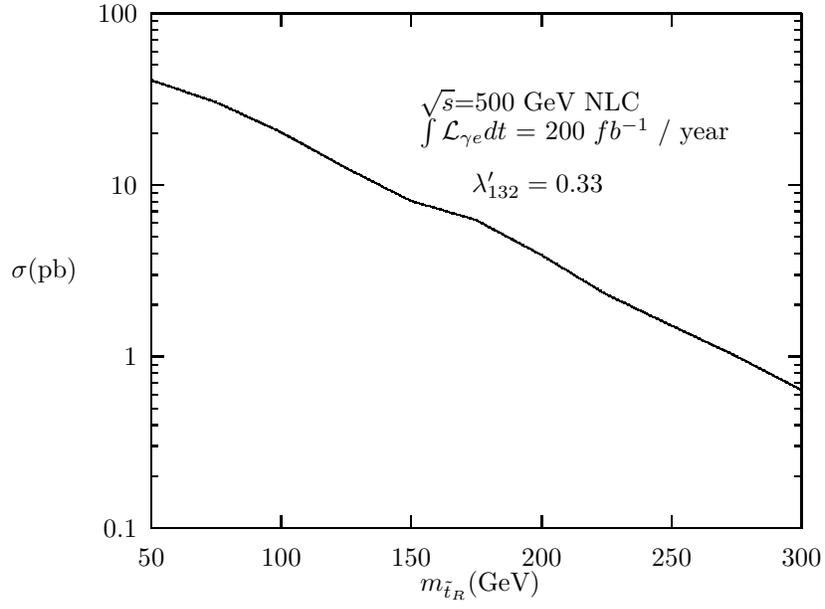
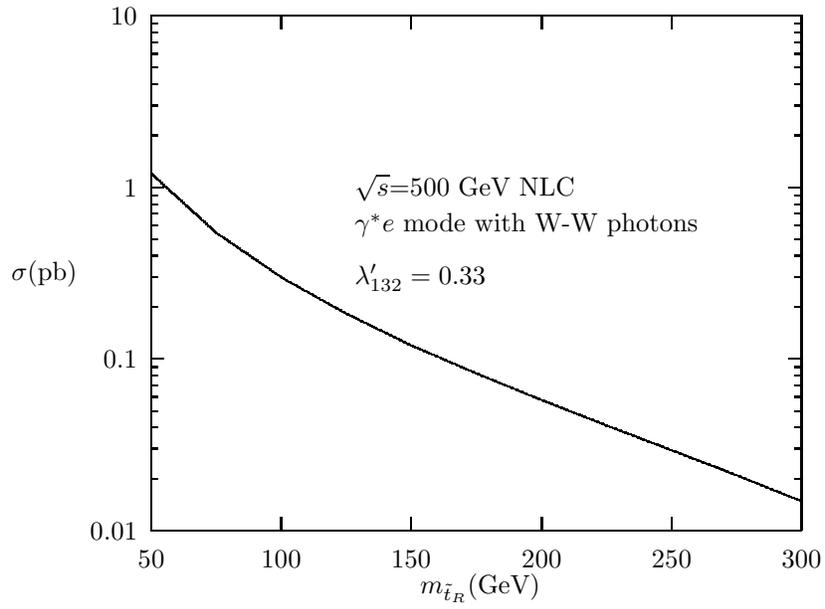
\end{document}

%% file: fig1.tex
\setlength{\unitlength}{0.240900pt}
\begin{picture}(1349,990)(0,0)
\tenrm
\ifx\plotpoint\undefined\newsavebox{\plotpoint}\fi
\put(264,113){\line(1,0){20}}
\put(1285,113){\line(-1,0){20}}
\put(242,113){\makebox(0,0)[r]{0.01}}
\put(264,194){\line(1,0){10}}
\put(1285,194){\line(-1,0){10}}
\put(264,242){\line(1,0){10}}
\put(1285,242){\line(-1,0){10}}
\put(264,275){\line(1,0){10}}
\put(1285,275){\line(-1,0){10}}
\put(264,301){\line(1,0){10}}
\put(1285,301){\line(-1,0){10}}
\put(264,323){\line(1,0){10}}
\put(1285,323){\line(-1,0){10}}
\put(264,341){\line(1,0){10}}
\put(1285,341){\line(-1,0){10}}
\put(264,357){\line(1,0){10}}
\put(1285,357){\line(-1,0){10}}
\put(264,370){\line(1,0){10}}
\put(1285,370){\line(-1,0){10}}
\put(264,383){\line(1,0){20}}
\put(1285,383){\line(-1,0){20}}
\put(242,383){\makebox(0,0)[r]{0.1}}
\put(264,464){\line(1,0){10}}
\put(1285,464){\line(-1,0){10}}
\put(264,511){\line(1,0){10}}
\put(1285,511){\line(-1,0){10}}
\put(264,545){\line(1,0){10}}
\put(1285,545){\line(-1,0){10}}
\put(264,571){\line(1,0){10}}
\put(1285,571){\line(-1,0){10}}
\put(264,593){\line(1,0){10}}
\put(1285,593){\line(-1,0){10}}
\put(264,611){\line(1,0){10}}
\put(1285,611){\line(-1,0){10}}
\put(264,626){\line(1,0){10}}
\put(1285,626){\line(-1,0){10}}
\put(264,640){\line(1,0){10}}
\put(1285,640){\line(-1,0){10}}
\put(264,652){\line(1,0){20}}
\put(1285,652){\line(-1,0){20}}
\put(242,652){\makebox(0,0)[r]{1}}
\put(264,734){\line(1,0){10}}
\put(1285,734){\line(-1,0){10}}
\put(264,781){\line(1,0){10}}
\put(1285,781){\line(-1,0){10}}
\put(264,815){\line(1,0){10}}
\put(1285,815){\line(-1,0){10}}
\put(264,841){\line(1,0){10}}
\put(1285,841){\line(-1,0){10}}
\put(264,862){\line(1,0){10}}
\put(1285,862){\line(-1,0){10}}
\put(264,880){\line(1,0){10}}
\put(1285,880){\line(-1,0){10}}
\put(264,896){\line(1,0){10}}
\put(1285,896){\line(-1,0){10}}
\put(264,910){\line(1,0){10}}
\put(1285,910){\line(-1,0){10}}
\put(264,922){\line(1,0){20}}
\put(1285,922){\line(-1,0){20}}
\put(242,922){\makebox(0,0)[r]{10}}
\put(366,113){\line(0,1){20}}
\put(366,922){\line(0,-1){20}}
\put(366,68){\makebox(0,0){100}}
\put(570,113){\line(0,1){20}}
\put(570,922){\line(0,-1){20}}
\put(570,68){\makebox(0,0){200}}
\put(774,113){\line(0,1){20}}
\put(774,922){\line(0,-1){20}}
\put(774,68){\makebox(0,0){300}}
\put(979,113){\line(0,1){20}}
\put(979,922){\line(0,-1){20}}
\put(979,68){\makebox(0,0){400}}
\put(1183,113){\line(0,1){20}}
\put(1183,922){\line(0,-1){20}}
\put(1183,68){\makebox(0,0){500}}
\put(264,113){\line(1,0){1021}}
\put(1285,113){\line(0,1){809}}
\put(1285,922){\line(-1,0){1021}}
\put(264,922){\line(0,-1){809}}
\put(45,517){\makebox(0,0)[l]{\shortstack{$\sigma$(pb)}}}
\put(774,23){\makebox(0,0){ $m_{\tilde d_R}$(GeV)}}
\put(570,781){\makebox(0,0)[l]{ $\sqrt{s}$=1.28 TeV HERA+LC}}
\put(570,734){\makebox(0,0)[l]{ $\int {\cal L}_{\gamma p} dt$ = 20 $fb^{-1}$ / year}}
\put(570,652){\makebox(0,0)[l]{ $\lambda'_{31k}=0.11$}}
\sbox{\plotpoint}{\rule[-0.200pt]{0.400pt}{0.400pt}}%
\put(264,861){\usebox{\plotpoint}}
\put(264,859){\usebox{\plotpoint}}
\put(265,858){\usebox{\plotpoint}}
\put(266,857){\usebox{\plotpoint}}
\put(267,856){\usebox{\plotpoint}}
\put(268,855){\usebox{\plotpoint}}
\put(269,854){\usebox{\plotpoint}}
\put(270,853){\usebox{\plotpoint}}
\put(271,852){\usebox{\plotpoint}}
\put(272,851){\usebox{\plotpoint}}
\put(273,850){\usebox{\plotpoint}}
\put(274,849){\usebox{\plotpoint}}
\put(275,848){\usebox{\plotpoint}}
\put(276,847){\usebox{\plotpoint}}
\put(277,846){\usebox{\plotpoint}}
\put(278,845){\usebox{\plotpoint}}
\put(279,844){\usebox{\plotpoint}}
\put(280,842){\usebox{\plotpoint}}
\put(281,841){\usebox{\plotpoint}}
\put(282,840){\usebox{\plotpoint}}
\put(283,839){\usebox{\plotpoint}}
\put(284,838){\usebox{\plotpoint}}
\put(285,837){\usebox{\plotpoint}}
\put(286,836){\usebox{\plotpoint}}
\put(287,835){\usebox{\plotpoint}}
\put(288,834){\usebox{\plotpoint}}
\put(289,833){\usebox{\plotpoint}}
\put(290,832){\usebox{\plotpoint}}
\put(291,831){\usebox{\plotpoint}}
\put(292,830){\usebox{\plotpoint}}
\put(293,829){\usebox{\plotpoint}}
\put(294,828){\usebox{\plotpoint}}
\put(295,827){\usebox{\plotpoint}}
\put(296,826){\usebox{\plotpoint}}
\put(297,824){\usebox{\plotpoint}}
\put(298,823){\usebox{\plotpoint}}
\put(299,822){\usebox{\plotpoint}}
\put(300,821){\usebox{\plotpoint}}
\put(301,820){\usebox{\plotpoint}}
\put(302,819){\usebox{\plotpoint}}
\put(303,818){\usebox{\plotpoint}}
\put(304,817){\usebox{\plotpoint}}
\put(305,816){\usebox{\plotpoint}}
\put(306,815){\usebox{\plotpoint}}
\put(307,814){\usebox{\plotpoint}}
\put(308,813){\usebox{\plotpoint}}
\put(309,812){\usebox{\plotpoint}}
\put(310,811){\usebox{\plotpoint}}
\put(311,810){\usebox{\plotpoint}}
\put(312,809){\usebox{\plotpoint}}
\put(313,808){\usebox{\plotpoint}}
\put(314,807){\usebox{\plotpoint}}
\put(315,807){\usebox{\plotpoint}}
\put(316,806){\usebox{\plotpoint}}
\put(317,805){\usebox{\plotpoint}}
\put(318,804){\usebox{\plotpoint}}
\put(319,803){\usebox{\plotpoint}}
\put(320,802){\usebox{\plotpoint}}
\put(322,801){\usebox{\plotpoint}}
\put(323,800){\usebox{\plotpoint}}
\put(324,799){\usebox{\plotpoint}}
\put(325,798){\usebox{\plotpoint}}
\put(326,797){\usebox{\plotpoint}}
\put(328,796){\usebox{\plotpoint}}
\put(329,795){\usebox{\plotpoint}}
\put(330,794){\usebox{\plotpoint}}
\put(331,793){\usebox{\plotpoint}}
\put(332,792){\usebox{\plotpoint}}
\put(333,791){\usebox{\plotpoint}}
\put(335,790){\usebox{\plotpoint}}
\put(336,789){\usebox{\plotpoint}}
\put(337,788){\usebox{\plotpoint}}
\put(338,787){\usebox{\plotpoint}}
\put(339,786){\usebox{\plotpoint}}
\put(341,785){\usebox{\plotpoint}}
\put(342,784){\usebox{\plotpoint}}
\put(343,783){\usebox{\plotpoint}}
\put(344,782){\usebox{\plotpoint}}
\put(345,781){\usebox{\plotpoint}}
\put(347,780){\usebox{\plotpoint}}
\put(348,779){\usebox{\plotpoint}}
\put(349,778){\usebox{\plotpoint}}
\put(350,777){\usebox{\plotpoint}}
\put(351,776){\usebox{\plotpoint}}
\put(352,775){\usebox{\plotpoint}}
\put(354,774){\usebox{\plotpoint}}
\put(355,773){\usebox{\plotpoint}}
\put(356,772){\usebox{\plotpoint}}
\put(357,771){\usebox{\plotpoint}}
\put(358,770){\usebox{\plotpoint}}
\put(360,769){\usebox{\plotpoint}}
\put(361,768){\usebox{\plotpoint}}
\put(362,767){\usebox{\plotpoint}}
\put(363,766){\usebox{\plotpoint}}
\put(364,765){\usebox{\plotpoint}}
\put(365,764){\usebox{\plotpoint}}
\put(367,763){\usebox{\plotpoint}}
\put(368,762){\usebox{\plotpoint}}
\put(370,761){\usebox{\plotpoint}}
\put(371,760){\usebox{\plotpoint}}
\put(372,759){\usebox{\plotpoint}}
\put(374,758){\usebox{\plotpoint}}
\put(375,757){\usebox{\plotpoint}}
\put(376,756){\usebox{\plotpoint}}
\put(378,755){\usebox{\plotpoint}}
\put(379,754){\usebox{\plotpoint}}
\put(380,753){\usebox{\plotpoint}}
\put(382,752){\usebox{\plotpoint}}
\put(383,751){\usebox{\plotpoint}}
\put(384,750){\usebox{\plotpoint}}
\put(386,749){\usebox{\plotpoint}}
\put(387,748){\usebox{\plotpoint}}
\put(388,747){\usebox{\plotpoint}}
\put(390,746){\usebox{\plotpoint}}
\put(391,745){\usebox{\plotpoint}}
\put(392,744){\usebox{\plotpoint}}
\put(394,743){\usebox{\plotpoint}}
\put(395,742){\usebox{\plotpoint}}
\put(396,741){\usebox{\plotpoint}}
\put(398,740){\usebox{\plotpoint}}
\put(399,739){\usebox{\plotpoint}}
\put(400,738){\usebox{\plotpoint}}
\put(402,737){\usebox{\plotpoint}}
\put(403,736){\usebox{\plotpoint}}
\put(404,735){\usebox{\plotpoint}}
\put(406,734){\usebox{\plotpoint}}
\put(407,733){\usebox{\plotpoint}}
\put(408,732){\usebox{\plotpoint}}
\put(410,731){\usebox{\plotpoint}}
\put(411,730){\usebox{\plotpoint}}
\put(412,729){\usebox{\plotpoint}}
\put(414,728){\usebox{\plotpoint}}
\put(415,727){\usebox{\plotpoint}}
\put(416,726){\usebox{\plotpoint}}
\put(418,725){\usebox{\plotpoint}}
\put(420,724){\usebox{\plotpoint}}
\put(421,723){\usebox{\plotpoint}}
\put(423,722){\usebox{\plotpoint}}
\put(424,721){\usebox{\plotpoint}}
\put(426,720){\usebox{\plotpoint}}
\put(427,719){\usebox{\plotpoint}}
\put(429,718){\usebox{\plotpoint}}
\put(430,717){\usebox{\plotpoint}}
\put(432,716){\usebox{\plotpoint}}
\put(433,715){\usebox{\plotpoint}}
\put(435,714){\usebox{\plotpoint}}
\put(436,713){\usebox{\plotpoint}}
\put(438,712){\usebox{\plotpoint}}
\put(439,711){\usebox{\plotpoint}}
\put(441,710){\usebox{\plotpoint}}
\put(442,709){\usebox{\plotpoint}}
\put(444,708){\usebox{\plotpoint}}
\put(445,707){\usebox{\plotpoint}}
\put(447,706){\usebox{\plotpoint}}
\put(448,705){\usebox{\plotpoint}}
\put(450,704){\usebox{\plotpoint}}
\put(451,703){\usebox{\plotpoint}}
\put(453,702){\usebox{\plotpoint}}
\put(454,701){\usebox{\plotpoint}}
\put(456,700){\usebox{\plotpoint}}
\put(457,699){\usebox{\plotpoint}}
\put(459,698){\usebox{\plotpoint}}
\put(460,697){\usebox{\plotpoint}}
\put(462,696){\usebox{\plotpoint}}
\put(463,695){\usebox{\plotpoint}}
\put(465,694){\usebox{\plotpoint}}
\put(466,693){\usebox{\plotpoint}}
\put(468,692){\usebox{\plotpoint}}
\put(469,691){\usebox{\plotpoint}}
\put(471,690){\usebox{\plotpoint}}
\put(472,689){\usebox{\plotpoint}}
\put(474,688){\usebox{\plotpoint}}
\put(476,687){\usebox{\plotpoint}}
\put(477,686){\usebox{\plotpoint}}
\put(479,685){\usebox{\plotpoint}}
\put(481,684){\usebox{\plotpoint}}
\put(482,683){\usebox{\plotpoint}}
\put(484,682){\usebox{\plotpoint}}
\put(486,681){\usebox{\plotpoint}}
\put(487,680){\usebox{\plotpoint}}
\put(489,679){\usebox{\plotpoint}}
\put(491,678){\usebox{\plotpoint}}
\put(492,677){\usebox{\plotpoint}}
\put(494,676){\usebox{\plotpoint}}
\put(495,675){\usebox{\plotpoint}}
\put(497,674){\usebox{\plotpoint}}
\put(499,673){\usebox{\plotpoint}}
\put(500,672){\usebox{\plotpoint}}
\put(502,671){\usebox{\plotpoint}}
\put(504,670){\usebox{\plotpoint}}
\put(505,669){\usebox{\plotpoint}}
\put(507,668){\usebox{\plotpoint}}
\put(509,667){\usebox{\plotpoint}}
\put(510,666){\usebox{\plotpoint}}
\put(512,665){\usebox{\plotpoint}}
\put(514,664){\usebox{\plotpoint}}
\put(515,663){\usebox{\plotpoint}}
\put(517,662){\usebox{\plotpoint}}
\put(519,661){\rule[-0.200pt]{0.424pt}{0.400pt}}
\put(520,660){\rule[-0.200pt]{0.424pt}{0.400pt}}
\put(522,659){\rule[-0.200pt]{0.424pt}{0.400pt}}
\put(524,658){\rule[-0.200pt]{0.424pt}{0.400pt}}
\put(526,657){\rule[-0.200pt]{0.424pt}{0.400pt}}
\put(527,656){\rule[-0.200pt]{0.424pt}{0.400pt}}
\put(529,655){\rule[-0.200pt]{0.424pt}{0.400pt}}
\put(531,654){\rule[-0.200pt]{0.424pt}{0.400pt}}
\put(533,653){\rule[-0.200pt]{0.424pt}{0.400pt}}
\put(534,652){\rule[-0.200pt]{0.424pt}{0.400pt}}
\put(536,651){\rule[-0.200pt]{0.424pt}{0.400pt}}
\put(538,650){\rule[-0.200pt]{0.424pt}{0.400pt}}
\put(540,649){\rule[-0.200pt]{0.424pt}{0.400pt}}
\put(541,648){\rule[-0.200pt]{0.424pt}{0.400pt}}
\put(543,647){\rule[-0.200pt]{0.424pt}{0.400pt}}
\put(545,646){\rule[-0.200pt]{0.424pt}{0.400pt}}
\put(547,645){\rule[-0.200pt]{0.424pt}{0.400pt}}
\put(548,644){\rule[-0.200pt]{0.424pt}{0.400pt}}
\put(550,643){\rule[-0.200pt]{0.424pt}{0.400pt}}
\put(552,642){\rule[-0.200pt]{0.424pt}{0.400pt}}
\put(554,641){\rule[-0.200pt]{0.424pt}{0.400pt}}
\put(555,640){\rule[-0.200pt]{0.424pt}{0.400pt}}
\put(557,639){\rule[-0.200pt]{0.424pt}{0.400pt}}
\put(559,638){\rule[-0.200pt]{0.424pt}{0.400pt}}
\put(561,637){\rule[-0.200pt]{0.424pt}{0.400pt}}
\put(562,636){\rule[-0.200pt]{0.424pt}{0.400pt}}
\put(564,635){\rule[-0.200pt]{0.424pt}{0.400pt}}
\put(566,634){\rule[-0.200pt]{0.424pt}{0.400pt}}
\put(568,633){\rule[-0.200pt]{0.424pt}{0.400pt}}
\put(569,632){\rule[-0.200pt]{0.439pt}{0.400pt}}
\put(571,631){\rule[-0.200pt]{0.439pt}{0.400pt}}
\put(573,630){\rule[-0.200pt]{0.439pt}{0.400pt}}
\put(575,629){\rule[-0.200pt]{0.439pt}{0.400pt}}
\put(577,628){\rule[-0.200pt]{0.439pt}{0.400pt}}
\put(579,627){\rule[-0.200pt]{0.439pt}{0.400pt}}
\put(580,626){\rule[-0.200pt]{0.439pt}{0.400pt}}
\put(582,625){\rule[-0.200pt]{0.439pt}{0.400pt}}
\put(584,624){\rule[-0.200pt]{0.439pt}{0.400pt}}
\put(586,623){\rule[-0.200pt]{0.439pt}{0.400pt}}
\put(588,622){\rule[-0.200pt]{0.439pt}{0.400pt}}
\put(590,621){\rule[-0.200pt]{0.439pt}{0.400pt}}
\put(591,620){\rule[-0.200pt]{0.439pt}{0.400pt}}
\put(593,619){\rule[-0.200pt]{0.439pt}{0.400pt}}
\put(595,618){\rule[-0.200pt]{0.439pt}{0.400pt}}
\put(597,617){\rule[-0.200pt]{0.439pt}{0.400pt}}
\put(599,616){\rule[-0.200pt]{0.439pt}{0.400pt}}
\put(600,615){\rule[-0.200pt]{0.439pt}{0.400pt}}
\put(602,614){\rule[-0.200pt]{0.439pt}{0.400pt}}
\put(604,613){\rule[-0.200pt]{0.439pt}{0.400pt}}
\put(606,612){\rule[-0.200pt]{0.439pt}{0.400pt}}
\put(608,611){\rule[-0.200pt]{0.439pt}{0.400pt}}
\put(610,610){\rule[-0.200pt]{0.439pt}{0.400pt}}
\put(611,609){\rule[-0.200pt]{0.439pt}{0.400pt}}
\put(613,608){\rule[-0.200pt]{0.439pt}{0.400pt}}
\put(615,607){\rule[-0.200pt]{0.439pt}{0.400pt}}
\put(617,606){\rule[-0.200pt]{0.439pt}{0.400pt}}
\put(619,605){\rule[-0.200pt]{0.439pt}{0.400pt}}
\put(620,604){\rule[-0.200pt]{0.455pt}{0.400pt}}
\put(622,603){\rule[-0.200pt]{0.455pt}{0.400pt}}
\put(624,602){\rule[-0.200pt]{0.455pt}{0.400pt}}
\put(626,601){\rule[-0.200pt]{0.455pt}{0.400pt}}
\put(628,600){\rule[-0.200pt]{0.455pt}{0.400pt}}
\put(630,599){\rule[-0.200pt]{0.455pt}{0.400pt}}
\put(632,598){\rule[-0.200pt]{0.455pt}{0.400pt}}
\put(634,597){\rule[-0.200pt]{0.455pt}{0.400pt}}
\put(636,596){\rule[-0.200pt]{0.455pt}{0.400pt}}
\put(638,595){\rule[-0.200pt]{0.455pt}{0.400pt}}
\put(639,594){\rule[-0.200pt]{0.455pt}{0.400pt}}
\put(641,593){\rule[-0.200pt]{0.455pt}{0.400pt}}
\put(643,592){\rule[-0.200pt]{0.455pt}{0.400pt}}
\put(645,591){\rule[-0.200pt]{0.455pt}{0.400pt}}
\put(647,590){\rule[-0.200pt]{0.455pt}{0.400pt}}
\put(649,589){\rule[-0.200pt]{0.455pt}{0.400pt}}
\put(651,588){\rule[-0.200pt]{0.455pt}{0.400pt}}
\put(653,587){\rule[-0.200pt]{0.455pt}{0.400pt}}
\put(655,586){\rule[-0.200pt]{0.455pt}{0.400pt}}
\put(656,585){\rule[-0.200pt]{0.455pt}{0.400pt}}
\put(658,584){\rule[-0.200pt]{0.455pt}{0.400pt}}
\put(660,583){\rule[-0.200pt]{0.455pt}{0.400pt}}
\put(662,582){\rule[-0.200pt]{0.455pt}{0.400pt}}
\put(664,581){\rule[-0.200pt]{0.455pt}{0.400pt}}
\put(666,580){\rule[-0.200pt]{0.455pt}{0.400pt}}
\put(668,579){\rule[-0.200pt]{0.455pt}{0.400pt}}
\put(670,578){\rule[-0.200pt]{0.455pt}{0.400pt}}
\put(672,577){\rule[-0.200pt]{0.473pt}{0.400pt}}
\put(673,576){\rule[-0.200pt]{0.473pt}{0.400pt}}
\put(675,575){\rule[-0.200pt]{0.473pt}{0.400pt}}
\put(677,574){\rule[-0.200pt]{0.473pt}{0.400pt}}
\put(679,573){\rule[-0.200pt]{0.473pt}{0.400pt}}
\put(681,572){\rule[-0.200pt]{0.473pt}{0.400pt}}
\put(683,571){\rule[-0.200pt]{0.473pt}{0.400pt}}
\put(685,570){\rule[-0.200pt]{0.473pt}{0.400pt}}
\put(687,569){\rule[-0.200pt]{0.473pt}{0.400pt}}
\put(689,568){\rule[-0.200pt]{0.473pt}{0.400pt}}
\put(691,567){\rule[-0.200pt]{0.473pt}{0.400pt}}
\put(693,566){\rule[-0.200pt]{0.473pt}{0.400pt}}
\put(695,565){\rule[-0.200pt]{0.473pt}{0.400pt}}
\put(697,564){\rule[-0.200pt]{0.473pt}{0.400pt}}
\put(699,563){\rule[-0.200pt]{0.473pt}{0.400pt}}
\put(701,562){\rule[-0.200pt]{0.473pt}{0.400pt}}
\put(703,561){\rule[-0.200pt]{0.473pt}{0.400pt}}
\put(705,560){\rule[-0.200pt]{0.473pt}{0.400pt}}
\put(707,559){\rule[-0.200pt]{0.473pt}{0.400pt}}
\put(709,558){\rule[-0.200pt]{0.473pt}{0.400pt}}
\put(711,557){\rule[-0.200pt]{0.473pt}{0.400pt}}
\put(713,556){\rule[-0.200pt]{0.473pt}{0.400pt}}
\put(715,555){\rule[-0.200pt]{0.473pt}{0.400pt}}
\put(717,554){\rule[-0.200pt]{0.473pt}{0.400pt}}
\put(719,553){\rule[-0.200pt]{0.473pt}{0.400pt}}
\put(721,552){\rule[-0.200pt]{0.472pt}{0.400pt}}
\put(723,551){\rule[-0.200pt]{0.473pt}{0.400pt}}
\put(724,550){\rule[-0.200pt]{0.473pt}{0.400pt}}
\put(726,549){\rule[-0.200pt]{0.473pt}{0.400pt}}
\put(728,548){\rule[-0.200pt]{0.473pt}{0.400pt}}
\put(730,547){\rule[-0.200pt]{0.473pt}{0.400pt}}
\put(732,546){\rule[-0.200pt]{0.473pt}{0.400pt}}
\put(734,545){\rule[-0.200pt]{0.473pt}{0.400pt}}
\put(736,544){\rule[-0.200pt]{0.473pt}{0.400pt}}
\put(738,543){\rule[-0.200pt]{0.473pt}{0.400pt}}
\put(740,542){\rule[-0.200pt]{0.473pt}{0.400pt}}
\put(742,541){\rule[-0.200pt]{0.473pt}{0.400pt}}
\put(744,540){\rule[-0.200pt]{0.473pt}{0.400pt}}
\put(746,539){\rule[-0.200pt]{0.473pt}{0.400pt}}
\put(748,538){\rule[-0.200pt]{0.473pt}{0.400pt}}
\put(750,537){\rule[-0.200pt]{0.473pt}{0.400pt}}
\put(752,536){\rule[-0.200pt]{0.473pt}{0.400pt}}
\put(754,535){\rule[-0.200pt]{0.473pt}{0.400pt}}
\put(756,534){\rule[-0.200pt]{0.473pt}{0.400pt}}
\put(758,533){\rule[-0.200pt]{0.473pt}{0.400pt}}
\put(760,532){\rule[-0.200pt]{0.473pt}{0.400pt}}
\put(762,531){\rule[-0.200pt]{0.473pt}{0.400pt}}
\put(764,530){\rule[-0.200pt]{0.473pt}{0.400pt}}
\put(766,529){\rule[-0.200pt]{0.473pt}{0.400pt}}
\put(768,528){\rule[-0.200pt]{0.473pt}{0.400pt}}
\put(770,527){\rule[-0.200pt]{0.473pt}{0.400pt}}
\put(772,526){\rule[-0.200pt]{0.472pt}{0.400pt}}
\put(774,525){\rule[-0.200pt]{0.501pt}{0.400pt}}
\put(776,524){\rule[-0.200pt]{0.501pt}{0.400pt}}
\put(778,523){\rule[-0.200pt]{0.501pt}{0.400pt}}
\put(780,522){\rule[-0.200pt]{0.501pt}{0.400pt}}
\put(782,521){\rule[-0.200pt]{0.501pt}{0.400pt}}
\put(784,520){\rule[-0.200pt]{0.501pt}{0.400pt}}
\put(786,519){\rule[-0.200pt]{0.501pt}{0.400pt}}
\put(788,518){\rule[-0.200pt]{0.501pt}{0.400pt}}
\put(790,517){\rule[-0.200pt]{0.501pt}{0.400pt}}
\put(792,516){\rule[-0.200pt]{0.501pt}{0.400pt}}
\put(794,515){\rule[-0.200pt]{0.501pt}{0.400pt}}
\put(796,514){\rule[-0.200pt]{0.501pt}{0.400pt}}
\put(798,513){\rule[-0.200pt]{0.501pt}{0.400pt}}
\put(801,512){\rule[-0.200pt]{0.501pt}{0.400pt}}
\put(803,511){\rule[-0.200pt]{0.501pt}{0.400pt}}
\put(805,510){\rule[-0.200pt]{0.501pt}{0.400pt}}
\put(807,509){\rule[-0.200pt]{0.501pt}{0.400pt}}
\put(809,508){\rule[-0.200pt]{0.501pt}{0.400pt}}
\put(811,507){\rule[-0.200pt]{0.501pt}{0.400pt}}
\put(813,506){\rule[-0.200pt]{0.501pt}{0.400pt}}
\put(815,505){\rule[-0.200pt]{0.501pt}{0.400pt}}
\put(817,504){\rule[-0.200pt]{0.501pt}{0.400pt}}
\put(819,503){\rule[-0.200pt]{0.501pt}{0.400pt}}
\put(821,502){\rule[-0.200pt]{0.501pt}{0.400pt}}
\put(823,501){\rule[-0.200pt]{0.501pt}{0.400pt}}
\put(826,500){\rule[-0.200pt]{0.491pt}{0.400pt}}
\put(828,499){\rule[-0.200pt]{0.491pt}{0.400pt}}
\put(830,498){\rule[-0.200pt]{0.491pt}{0.400pt}}
\put(832,497){\rule[-0.200pt]{0.491pt}{0.400pt}}
\put(834,496){\rule[-0.200pt]{0.491pt}{0.400pt}}
\put(836,495){\rule[-0.200pt]{0.491pt}{0.400pt}}
\put(838,494){\rule[-0.200pt]{0.491pt}{0.400pt}}
\put(840,493){\rule[-0.200pt]{0.491pt}{0.400pt}}
\put(842,492){\rule[-0.200pt]{0.491pt}{0.400pt}}
\put(844,491){\rule[-0.200pt]{0.491pt}{0.400pt}}
\put(846,490){\rule[-0.200pt]{0.491pt}{0.400pt}}
\put(848,489){\rule[-0.200pt]{0.491pt}{0.400pt}}
\put(850,488){\rule[-0.200pt]{0.491pt}{0.400pt}}
\put(852,487){\rule[-0.200pt]{0.491pt}{0.400pt}}
\put(854,486){\rule[-0.200pt]{0.491pt}{0.400pt}}
\put(856,485){\rule[-0.200pt]{0.491pt}{0.400pt}}
\put(858,484){\rule[-0.200pt]{0.491pt}{0.400pt}}
\put(860,483){\rule[-0.200pt]{0.491pt}{0.400pt}}
\put(862,482){\rule[-0.200pt]{0.491pt}{0.400pt}}
\put(864,481){\rule[-0.200pt]{0.491pt}{0.400pt}}
\put(866,480){\rule[-0.200pt]{0.491pt}{0.400pt}}
\put(868,479){\rule[-0.200pt]{0.491pt}{0.400pt}}
\put(870,478){\rule[-0.200pt]{0.491pt}{0.400pt}}
\put(872,477){\rule[-0.200pt]{0.491pt}{0.400pt}}
\put(874,476){\rule[-0.200pt]{0.491pt}{0.400pt}}
\put(876,475){\rule[-0.200pt]{0.492pt}{0.400pt}}
\put(879,474){\rule[-0.200pt]{0.491pt}{0.400pt}}
\put(881,473){\rule[-0.200pt]{0.491pt}{0.400pt}}
\put(883,472){\rule[-0.200pt]{0.491pt}{0.400pt}}
\put(885,471){\rule[-0.200pt]{0.491pt}{0.400pt}}
\put(887,470){\rule[-0.200pt]{0.491pt}{0.400pt}}
\put(889,469){\rule[-0.200pt]{0.491pt}{0.400pt}}
\put(891,468){\rule[-0.200pt]{0.491pt}{0.400pt}}
\put(893,467){\rule[-0.200pt]{0.491pt}{0.400pt}}
\put(895,466){\rule[-0.200pt]{0.491pt}{0.400pt}}
\put(897,465){\rule[-0.200pt]{0.491pt}{0.400pt}}
\put(899,464){\rule[-0.200pt]{0.491pt}{0.400pt}}
\put(901,463){\rule[-0.200pt]{0.491pt}{0.400pt}}
\put(903,462){\rule[-0.200pt]{0.491pt}{0.400pt}}
\put(905,461){\rule[-0.200pt]{0.491pt}{0.400pt}}
\put(907,460){\rule[-0.200pt]{0.491pt}{0.400pt}}
\put(909,459){\rule[-0.200pt]{0.491pt}{0.400pt}}
\put(911,458){\rule[-0.200pt]{0.491pt}{0.400pt}}
\put(913,457){\rule[-0.200pt]{0.491pt}{0.400pt}}
\put(915,456){\rule[-0.200pt]{0.491pt}{0.400pt}}
\put(917,455){\rule[-0.200pt]{0.491pt}{0.400pt}}
\put(919,454){\rule[-0.200pt]{0.491pt}{0.400pt}}
\put(921,453){\rule[-0.200pt]{0.491pt}{0.400pt}}
\put(923,452){\rule[-0.200pt]{0.491pt}{0.400pt}}
\put(925,451){\rule[-0.200pt]{0.491pt}{0.400pt}}
\put(927,450){\rule[-0.200pt]{0.492pt}{0.400pt}}
\put(930,449){\rule[-0.200pt]{0.491pt}{0.400pt}}
\put(932,448){\rule[-0.200pt]{0.491pt}{0.400pt}}
\put(934,447){\rule[-0.200pt]{0.491pt}{0.400pt}}
\put(936,446){\rule[-0.200pt]{0.491pt}{0.400pt}}
\put(938,445){\rule[-0.200pt]{0.491pt}{0.400pt}}
\put(940,444){\rule[-0.200pt]{0.491pt}{0.400pt}}
\put(942,443){\rule[-0.200pt]{0.491pt}{0.400pt}}
\put(944,442){\rule[-0.200pt]{0.491pt}{0.400pt}}
\put(946,441){\rule[-0.200pt]{0.491pt}{0.400pt}}
\put(948,440){\rule[-0.200pt]{0.491pt}{0.400pt}}
\put(950,439){\rule[-0.200pt]{0.491pt}{0.400pt}}
\put(952,438){\rule[-0.200pt]{0.491pt}{0.400pt}}
\put(954,437){\rule[-0.200pt]{0.491pt}{0.400pt}}
\put(956,436){\rule[-0.200pt]{0.491pt}{0.400pt}}
\put(958,435){\rule[-0.200pt]{0.491pt}{0.400pt}}
\put(960,434){\rule[-0.200pt]{0.491pt}{0.400pt}}
\put(962,433){\rule[-0.200pt]{0.491pt}{0.400pt}}
\put(964,432){\rule[-0.200pt]{0.491pt}{0.400pt}}
\put(966,431){\rule[-0.200pt]{0.491pt}{0.400pt}}
\put(968,430){\rule[-0.200pt]{0.491pt}{0.400pt}}
\put(970,429){\rule[-0.200pt]{0.491pt}{0.400pt}}
\put(972,428){\rule[-0.200pt]{0.491pt}{0.400pt}}
\put(974,427){\rule[-0.200pt]{0.491pt}{0.400pt}}
\put(976,426){\rule[-0.200pt]{0.491pt}{0.400pt}}
\put(978,425){\rule[-0.200pt]{0.559pt}{0.400pt}}
\put(981,424){\rule[-0.200pt]{0.558pt}{0.400pt}}
\put(983,423){\rule[-0.200pt]{0.558pt}{0.400pt}}
\put(985,422){\rule[-0.200pt]{0.558pt}{0.400pt}}
\put(988,421){\rule[-0.200pt]{0.558pt}{0.400pt}}
\put(990,420){\rule[-0.200pt]{0.558pt}{0.400pt}}
\put(992,419){\rule[-0.200pt]{0.558pt}{0.400pt}}
\put(995,418){\rule[-0.200pt]{0.558pt}{0.400pt}}
\put(997,417){\rule[-0.200pt]{0.558pt}{0.400pt}}
\put(999,416){\rule[-0.200pt]{0.558pt}{0.400pt}}
\put(1002,415){\rule[-0.200pt]{0.558pt}{0.400pt}}
\put(1004,414){\rule[-0.200pt]{0.558pt}{0.400pt}}
\put(1006,413){\rule[-0.200pt]{0.558pt}{0.400pt}}
\put(1009,412){\rule[-0.200pt]{0.558pt}{0.400pt}}
\put(1011,411){\rule[-0.200pt]{0.558pt}{0.400pt}}
\put(1013,410){\rule[-0.200pt]{0.558pt}{0.400pt}}
\put(1016,409){\rule[-0.200pt]{0.558pt}{0.400pt}}
\put(1018,408){\rule[-0.200pt]{0.558pt}{0.400pt}}
\put(1020,407){\rule[-0.200pt]{0.558pt}{0.400pt}}
\put(1023,406){\rule[-0.200pt]{0.558pt}{0.400pt}}
\put(1025,405){\rule[-0.200pt]{0.558pt}{0.400pt}}
\put(1027,404){\rule[-0.200pt]{0.558pt}{0.400pt}}
\put(1030,403){\rule[-0.200pt]{0.439pt}{0.400pt}}
\put(1031,402){\rule[-0.200pt]{0.439pt}{0.400pt}}
\put(1033,401){\rule[-0.200pt]{0.439pt}{0.400pt}}
\put(1035,400){\rule[-0.200pt]{0.439pt}{0.400pt}}
\put(1037,399){\rule[-0.200pt]{0.439pt}{0.400pt}}
\put(1039,398){\rule[-0.200pt]{0.439pt}{0.400pt}}
\put(1040,397){\rule[-0.200pt]{0.439pt}{0.400pt}}
\put(1042,396){\rule[-0.200pt]{0.439pt}{0.400pt}}
\put(1044,395){\rule[-0.200pt]{0.439pt}{0.400pt}}
\put(1046,394){\rule[-0.200pt]{0.439pt}{0.400pt}}
\put(1048,393){\rule[-0.200pt]{0.439pt}{0.400pt}}
\put(1050,392){\rule[-0.200pt]{0.439pt}{0.400pt}}
\put(1051,391){\rule[-0.200pt]{0.439pt}{0.400pt}}
\put(1053,390){\rule[-0.200pt]{0.439pt}{0.400pt}}
\put(1055,389){\rule[-0.200pt]{0.439pt}{0.400pt}}
\put(1057,388){\rule[-0.200pt]{0.439pt}{0.400pt}}
\put(1059,387){\rule[-0.200pt]{0.439pt}{0.400pt}}
\put(1060,386){\rule[-0.200pt]{0.439pt}{0.400pt}}
\put(1062,385){\rule[-0.200pt]{0.439pt}{0.400pt}}
\put(1064,384){\rule[-0.200pt]{0.439pt}{0.400pt}}
\put(1066,383){\rule[-0.200pt]{0.439pt}{0.400pt}}
\put(1068,382){\rule[-0.200pt]{0.439pt}{0.400pt}}
\put(1070,381){\rule[-0.200pt]{0.439pt}{0.400pt}}
\put(1071,380){\rule[-0.200pt]{0.439pt}{0.400pt}}
\put(1073,379){\rule[-0.200pt]{0.439pt}{0.400pt}}
\put(1075,378){\rule[-0.200pt]{0.439pt}{0.400pt}}
\put(1077,377){\rule[-0.200pt]{0.439pt}{0.400pt}}
\put(1079,376){\rule[-0.200pt]{0.439pt}{0.400pt}}
\put(1080,375){\rule[-0.200pt]{0.473pt}{0.400pt}}
\put(1082,374){\rule[-0.200pt]{0.473pt}{0.400pt}}
\put(1084,373){\rule[-0.200pt]{0.473pt}{0.400pt}}
\put(1086,372){\rule[-0.200pt]{0.473pt}{0.400pt}}
\put(1088,371){\rule[-0.200pt]{0.473pt}{0.400pt}}
\put(1090,370){\rule[-0.200pt]{0.473pt}{0.400pt}}
\put(1092,369){\rule[-0.200pt]{0.473pt}{0.400pt}}
\put(1094,368){\rule[-0.200pt]{0.473pt}{0.400pt}}
\put(1096,367){\rule[-0.200pt]{0.473pt}{0.400pt}}
\put(1098,366){\rule[-0.200pt]{0.473pt}{0.400pt}}
\put(1100,365){\rule[-0.200pt]{0.473pt}{0.400pt}}
\put(1102,364){\rule[-0.200pt]{0.473pt}{0.400pt}}
\put(1104,363){\rule[-0.200pt]{0.473pt}{0.400pt}}
\put(1106,362){\rule[-0.200pt]{0.473pt}{0.400pt}}
\put(1108,361){\rule[-0.200pt]{0.473pt}{0.400pt}}
\put(1110,360){\rule[-0.200pt]{0.473pt}{0.400pt}}
\put(1112,359){\rule[-0.200pt]{0.473pt}{0.400pt}}
\put(1114,358){\rule[-0.200pt]{0.473pt}{0.400pt}}
\put(1116,357){\rule[-0.200pt]{0.473pt}{0.400pt}}
\put(1118,356){\rule[-0.200pt]{0.473pt}{0.400pt}}
\put(1120,355){\rule[-0.200pt]{0.473pt}{0.400pt}}
\put(1122,354){\rule[-0.200pt]{0.473pt}{0.400pt}}
\put(1124,353){\rule[-0.200pt]{0.473pt}{0.400pt}}
\put(1126,352){\rule[-0.200pt]{0.473pt}{0.400pt}}
\put(1128,351){\rule[-0.200pt]{0.473pt}{0.400pt}}
\put(1130,350){\rule[-0.200pt]{0.472pt}{0.400pt}}
\put(1132,349){\rule[-0.200pt]{0.473pt}{0.400pt}}
\put(1133,348){\rule[-0.200pt]{0.473pt}{0.400pt}}
\put(1135,347){\rule[-0.200pt]{0.473pt}{0.400pt}}
\put(1137,346){\rule[-0.200pt]{0.473pt}{0.400pt}}
\put(1139,345){\rule[-0.200pt]{0.473pt}{0.400pt}}
\put(1141,344){\rule[-0.200pt]{0.473pt}{0.400pt}}
\put(1143,343){\rule[-0.200pt]{0.473pt}{0.400pt}}
\put(1145,342){\rule[-0.200pt]{0.473pt}{0.400pt}}
\put(1147,341){\rule[-0.200pt]{0.473pt}{0.400pt}}
\put(1149,340){\rule[-0.200pt]{0.473pt}{0.400pt}}
\put(1151,339){\rule[-0.200pt]{0.473pt}{0.400pt}}
\put(1153,338){\rule[-0.200pt]{0.473pt}{0.400pt}}
\put(1155,337){\rule[-0.200pt]{0.473pt}{0.400pt}}
\put(1157,336){\rule[-0.200pt]{0.473pt}{0.400pt}}
\put(1159,335){\rule[-0.200pt]{0.473pt}{0.400pt}}
\put(1161,334){\rule[-0.200pt]{0.473pt}{0.400pt}}
\put(1163,333){\rule[-0.200pt]{0.473pt}{0.400pt}}
\put(1165,332){\rule[-0.200pt]{0.473pt}{0.400pt}}
\put(1167,331){\rule[-0.200pt]{0.473pt}{0.400pt}}
\put(1169,330){\rule[-0.200pt]{0.473pt}{0.400pt}}
\put(1171,329){\rule[-0.200pt]{0.473pt}{0.400pt}}
\put(1173,328){\rule[-0.200pt]{0.473pt}{0.400pt}}
\put(1175,327){\rule[-0.200pt]{0.473pt}{0.400pt}}
\put(1177,326){\rule[-0.200pt]{0.473pt}{0.400pt}}
\put(1179,325){\rule[-0.200pt]{0.473pt}{0.400pt}}
\put(1181,324){\rule[-0.200pt]{0.472pt}{0.400pt}}
\put(1183,323){\rule[-0.200pt]{0.473pt}{0.400pt}}
\put(1184,322){\rule[-0.200pt]{0.473pt}{0.400pt}}
\put(1186,321){\rule[-0.200pt]{0.473pt}{0.400pt}}
\put(1188,320){\rule[-0.200pt]{0.473pt}{0.400pt}}
\put(1190,319){\rule[-0.200pt]{0.473pt}{0.400pt}}
\put(1192,318){\rule[-0.200pt]{0.473pt}{0.400pt}}
\put(1194,317){\rule[-0.200pt]{0.473pt}{0.400pt}}
\put(1196,316){\rule[-0.200pt]{0.473pt}{0.400pt}}
\put(1198,315){\rule[-0.200pt]{0.473pt}{0.400pt}}
\put(1200,314){\rule[-0.200pt]{0.473pt}{0.400pt}}
\put(1202,313){\rule[-0.200pt]{0.473pt}{0.400pt}}
\put(1204,312){\rule[-0.200pt]{0.473pt}{0.400pt}}
\put(1206,311){\rule[-0.200pt]{0.473pt}{0.400pt}}
\put(1208,310){\rule[-0.200pt]{0.473pt}{0.400pt}}
\put(1210,309){\rule[-0.200pt]{0.473pt}{0.400pt}}
\put(1212,308){\rule[-0.200pt]{0.473pt}{0.400pt}}
\put(1214,307){\rule[-0.200pt]{0.473pt}{0.400pt}}
\put(1216,306){\rule[-0.200pt]{0.473pt}{0.400pt}}
\put(1218,305){\rule[-0.200pt]{0.473pt}{0.400pt}}
\put(1220,304){\rule[-0.200pt]{0.473pt}{0.400pt}}
\put(1222,303){\rule[-0.200pt]{0.473pt}{0.400pt}}
\put(1224,302){\rule[-0.200pt]{0.473pt}{0.400pt}}
\put(1226,301){\rule[-0.200pt]{0.473pt}{0.400pt}}
\put(1228,300){\rule[-0.200pt]{0.473pt}{0.400pt}}
\put(1230,299){\rule[-0.200pt]{0.473pt}{0.400pt}}
\put(1232,298){\rule[-0.200pt]{0.472pt}{0.400pt}}
\put(1234,297){\rule[-0.200pt]{0.455pt}{0.400pt}}
\put(1235,296){\rule[-0.200pt]{0.455pt}{0.400pt}}
\put(1237,295){\rule[-0.200pt]{0.455pt}{0.400pt}}
\put(1239,294){\rule[-0.200pt]{0.455pt}{0.400pt}}
\put(1241,293){\rule[-0.200pt]{0.455pt}{0.400pt}}
\put(1243,292){\rule[-0.200pt]{0.455pt}{0.400pt}}
\put(1245,291){\rule[-0.200pt]{0.455pt}{0.400pt}}
\put(1247,290){\rule[-0.200pt]{0.455pt}{0.400pt}}
\put(1249,289){\rule[-0.200pt]{0.455pt}{0.400pt}}
\put(1251,288){\rule[-0.200pt]{0.455pt}{0.400pt}}
\put(1252,287){\rule[-0.200pt]{0.455pt}{0.400pt}}
\put(1254,286){\rule[-0.200pt]{0.455pt}{0.400pt}}
\put(1256,285){\rule[-0.200pt]{0.455pt}{0.400pt}}
\put(1258,284){\rule[-0.200pt]{0.455pt}{0.400pt}}
\put(1260,283){\rule[-0.200pt]{0.455pt}{0.400pt}}
\put(1262,282){\rule[-0.200pt]{0.455pt}{0.400pt}}
\put(1264,281){\rule[-0.200pt]{0.455pt}{0.400pt}}
\put(1266,280){\rule[-0.200pt]{0.455pt}{0.400pt}}
\put(1268,279){\rule[-0.200pt]{0.455pt}{0.400pt}}
\put(1269,278){\rule[-0.200pt]{0.455pt}{0.400pt}}
\put(1271,277){\rule[-0.200pt]{0.455pt}{0.400pt}}
\put(1273,276){\rule[-0.200pt]{0.455pt}{0.400pt}}
\put(1275,275){\rule[-0.200pt]{0.455pt}{0.400pt}}
\put(1277,274){\rule[-0.200pt]{0.455pt}{0.400pt}}
\put(1279,273){\rule[-0.200pt]{0.455pt}{0.400pt}}
\put(1281,272){\rule[-0.200pt]{0.455pt}{0.400pt}}
\put(1283,271){\rule[-0.200pt]{0.455pt}{0.400pt}}
\end{picture}

%% file: fig2.tex
\setlength{\unitlength}{0.240900pt}
\begin{picture}(1349,990)(0,0)
\tenrm
\ifx\plotpoint\undefined\newsavebox{\plotpoint}\fi
\put(264,113){\line(1,0){20}}
\put(1285,113){\line(-1,0){20}}
\put(242,113){\makebox(0,0)[r]{0.1}}
\put(264,194){\line(1,0){10}}
\put(1285,194){\line(-1,0){10}}
\put(264,242){\line(1,0){10}}
\put(1285,242){\line(-1,0){10}}
\put(264,275){\line(1,0){10}}
\put(1285,275){\line(-1,0){10}}
\put(264,301){\line(1,0){10}}
\put(1285,301){\line(-1,0){10}}
\put(264,323){\line(1,0){10}}
\put(1285,323){\line(-1,0){10}}
\put(264,341){\line(1,0){10}}
\put(1285,341){\line(-1,0){10}}
\put(264,357){\line(1,0){10}}
\put(1285,357){\line(-1,0){10}}
\put(264,370){\line(1,0){10}}
\put(1285,370){\line(-1,0){10}}
\put(264,383){\line(1,0){20}}
\put(1285,383){\line(-1,0){20}}
\put(242,383){\makebox(0,0)[r]{1}}
\put(264,464){\line(1,0){10}}
\put(1285,464){\line(-1,0){10}}
\put(264,511){\line(1,0){10}}
\put(1285,511){\line(-1,0){10}}
\put(264,545){\line(1,0){10}}
\put(1285,545){\line(-1,0){10}}
\put(264,571){\line(1,0){10}}
\put(1285,571){\line(-1,0){10}}
\put(264,593){\line(1,0){10}}
\put(1285,593){\line(-1,0){10}}
\put(264,611){\line(1,0){10}}
\put(1285,611){\line(-1,0){10}}
\put(264,626){\line(1,0){10}}
\put(1285,626){\line(-1,0){10}}
\put(264,640){\line(1,0){10}}
\put(1285,640){\line(-1,0){10}}
\put(264,652){\line(1,0){20}}
\put(1285,652){\line(-1,0){20}}
\put(242,652){\makebox(0,0)[r]{10}}
\put(264,734){\line(1,0){10}}
\put(1285,734){\line(-1,0){10}}
\put(264,781){\line(1,0){10}}
\put(1285,781){\line(-1,0){10}}
\put(264,815){\line(1,0){10}}
\put(1285,815){\line(-1,0){10}}
\put(264,841){\line(1,0){10}}
\put(1285,841){\line(-1,0){10}}
\put(264,862){\line(1,0){10}}
\put(1285,862){\line(-1,0){10}}
\put(264,880){\line(1,0){10}}
\put(1285,880){\line(-1,0){10}}
\put(264,896){\line(1,0){10}}
\put(1285,896){\line(-1,0){10}}
\put(264,910){\line(1,0){10}}
\put(1285,910){\line(-1,0){10}}
\put(264,922){\line(1,0){20}}
\put(1285,922){\line(-1,0){20}}
\put(242,922){\makebox(0,0)[r]{100}}
\put(366,113){\line(0,1){20}}
\put(366,922){\line(0,-1){20}}
\put(366,68){\makebox(0,0){100}}
\put(570,113){\line(0,1){20}}
\put(570,922){\line(0,-1){20}}
\put(570,68){\makebox(0,0){200}}
\put(774,113){\line(0,1){20}}
\put(774,922){\line(0,-1){20}}
\put(774,68){\makebox(0,0){300}}
\put(979,113){\line(0,1){20}}
\put(979,922){\line(0,-1){20}}
\put(979,68){\makebox(0,0){400}}
\put(1183,113){\line(0,1){20}}
\put(1183,922){\line(0,-1){20}}
\put(1183,68){\makebox(0,0){500}}
\put(264,113){\line(1,0){1021}}
\put(1285,113){\line(0,1){809}}
\put(1285,922){\line(-1,0){1021}}
\put(264,922){\line(0,-1){809}}
\put(45,517){\makebox(0,0)[l]{\shortstack{$\sigma$(pb)}}}
\put(774,23){\makebox(0,0){ $m_{\tilde d_R}$(GeV)}}
\put(570,652){\makebox(0,0)[l]{ $\sqrt{s}$=1.28 TeV HERA+LC}}
\put(570,593){\makebox(0,0)[l]{ $\int {\cal L}_{\gamma p} dt$ = 20 $fb^{-1}$ / year}}
\put(570,511){\makebox(0,0)[l]{ $\lambda'_{31k}=0.11$}}
\sbox{\plotpoint}{\rule[-0.200pt]{0.400pt}{0.400pt}}%
\put(264,735){\usebox{\plotpoint}}
\put(264,735){\usebox{\plotpoint}}
\put(265,734){\usebox{\plotpoint}}
\put(266,733){\usebox{\plotpoint}}
\put(267,732){\usebox{\plotpoint}}
\put(268,731){\usebox{\plotpoint}}
\put(269,730){\usebox{\plotpoint}}
\put(270,729){\usebox{\plotpoint}}
\put(271,728){\usebox{\plotpoint}}
\put(272,727){\usebox{\plotpoint}}
\put(273,726){\usebox{\plotpoint}}
\put(274,725){\usebox{\plotpoint}}
\put(275,724){\usebox{\plotpoint}}
\put(276,723){\usebox{\plotpoint}}
\put(277,722){\usebox{\plotpoint}}
\put(278,721){\usebox{\plotpoint}}
\put(279,720){\usebox{\plotpoint}}
\put(280,719){\usebox{\plotpoint}}
\put(281,718){\usebox{\plotpoint}}
\put(282,717){\usebox{\plotpoint}}
\put(283,716){\usebox{\plotpoint}}
\put(284,715){\usebox{\plotpoint}}
\put(285,714){\usebox{\plotpoint}}
\put(286,713){\usebox{\plotpoint}}
\put(287,712){\usebox{\plotpoint}}
\put(288,711){\usebox{\plotpoint}}
\put(289,710){\usebox{\plotpoint}}
\put(290,709){\usebox{\plotpoint}}
\put(291,708){\usebox{\plotpoint}}
\put(292,707){\usebox{\plotpoint}}
\put(293,706){\usebox{\plotpoint}}
\put(294,705){\usebox{\plotpoint}}
\put(295,704){\usebox{\plotpoint}}
\put(296,703){\usebox{\plotpoint}}
\put(297,702){\usebox{\plotpoint}}
\put(298,701){\usebox{\plotpoint}}
\put(299,700){\usebox{\plotpoint}}
\put(300,699){\usebox{\plotpoint}}
\put(301,698){\usebox{\plotpoint}}
\put(302,697){\usebox{\plotpoint}}
\put(303,696){\usebox{\plotpoint}}
\put(304,695){\usebox{\plotpoint}}
\put(305,694){\usebox{\plotpoint}}
\put(306,693){\usebox{\plotpoint}}
\put(307,692){\usebox{\plotpoint}}
\put(308,691){\usebox{\plotpoint}}
\put(309,690){\usebox{\plotpoint}}
\put(310,689){\usebox{\plotpoint}}
\put(311,688){\usebox{\plotpoint}}
\put(312,687){\usebox{\plotpoint}}
\put(313,686){\usebox{\plotpoint}}
\put(314,685){\usebox{\plotpoint}}
\put(315,683){\usebox{\plotpoint}}
\put(316,682){\usebox{\plotpoint}}
\put(317,681){\usebox{\plotpoint}}
\put(318,680){\usebox{\plotpoint}}
\put(319,679){\usebox{\plotpoint}}
\put(320,678){\usebox{\plotpoint}}
\put(321,676){\usebox{\plotpoint}}
\put(322,675){\usebox{\plotpoint}}
\put(323,674){\usebox{\plotpoint}}
\put(324,673){\usebox{\plotpoint}}
\put(325,672){\usebox{\plotpoint}}
\put(326,671){\usebox{\plotpoint}}
\put(327,669){\usebox{\plotpoint}}
\put(328,668){\usebox{\plotpoint}}
\put(329,667){\usebox{\plotpoint}}
\put(330,666){\usebox{\plotpoint}}
\put(331,665){\usebox{\plotpoint}}
\put(332,664){\usebox{\plotpoint}}
\put(333,663){\usebox{\plotpoint}}
\put(334,661){\usebox{\plotpoint}}
\put(335,660){\usebox{\plotpoint}}
\put(336,659){\usebox{\plotpoint}}
\put(337,658){\usebox{\plotpoint}}
\put(338,657){\usebox{\plotpoint}}
\put(339,656){\usebox{\plotpoint}}
\put(340,654){\usebox{\plotpoint}}
\put(341,653){\usebox{\plotpoint}}
\put(342,652){\usebox{\plotpoint}}
\put(343,651){\usebox{\plotpoint}}
\put(344,650){\usebox{\plotpoint}}
\put(345,649){\usebox{\plotpoint}}
\put(346,647){\usebox{\plotpoint}}
\put(347,646){\usebox{\plotpoint}}
\put(348,645){\usebox{\plotpoint}}
\put(349,644){\usebox{\plotpoint}}
\put(350,643){\usebox{\plotpoint}}
\put(351,642){\usebox{\plotpoint}}
\put(352,641){\usebox{\plotpoint}}
\put(353,639){\usebox{\plotpoint}}
\put(354,638){\usebox{\plotpoint}}
\put(355,637){\usebox{\plotpoint}}
\put(356,636){\usebox{\plotpoint}}
\put(357,635){\usebox{\plotpoint}}
\put(358,634){\usebox{\plotpoint}}
\put(359,632){\usebox{\plotpoint}}
\put(360,631){\usebox{\plotpoint}}
\put(361,630){\usebox{\plotpoint}}
\put(362,629){\usebox{\plotpoint}}
\put(363,628){\usebox{\plotpoint}}
\put(364,627){\usebox{\plotpoint}}
\put(365,626){\usebox{\plotpoint}}
\put(366,626){\usebox{\plotpoint}}
\put(366,626){\usebox{\plotpoint}}
\put(367,625){\usebox{\plotpoint}}
\put(368,624){\usebox{\plotpoint}}
\put(369,623){\usebox{\plotpoint}}
\put(370,622){\usebox{\plotpoint}}
\put(371,621){\usebox{\plotpoint}}
\put(372,620){\usebox{\plotpoint}}
\put(373,619){\usebox{\plotpoint}}
\put(375,618){\usebox{\plotpoint}}
\put(376,617){\usebox{\plotpoint}}
\put(377,616){\usebox{\plotpoint}}
\put(378,615){\usebox{\plotpoint}}
\put(379,614){\usebox{\plotpoint}}
\put(380,613){\usebox{\plotpoint}}
\put(381,612){\usebox{\plotpoint}}
\put(382,611){\usebox{\plotpoint}}
\put(384,610){\usebox{\plotpoint}}
\put(385,609){\usebox{\plotpoint}}
\put(386,608){\usebox{\plotpoint}}
\put(387,607){\usebox{\plotpoint}}
\put(388,606){\usebox{\plotpoint}}
\put(389,605){\usebox{\plotpoint}}
\put(390,604){\usebox{\plotpoint}}
\put(392,603){\usebox{\plotpoint}}
\put(393,602){\usebox{\plotpoint}}
\put(394,601){\usebox{\plotpoint}}
\put(395,600){\usebox{\plotpoint}}
\put(396,599){\usebox{\plotpoint}}
\put(397,598){\usebox{\plotpoint}}
\put(398,597){\usebox{\plotpoint}}
\put(399,596){\usebox{\plotpoint}}
\put(401,595){\usebox{\plotpoint}}
\put(402,594){\usebox{\plotpoint}}
\put(403,593){\usebox{\plotpoint}}
\put(404,592){\usebox{\plotpoint}}
\put(405,591){\usebox{\plotpoint}}
\put(406,590){\usebox{\plotpoint}}
\put(407,589){\usebox{\plotpoint}}
\put(409,588){\usebox{\plotpoint}}
\put(410,587){\usebox{\plotpoint}}
\put(411,586){\usebox{\plotpoint}}
\put(412,585){\usebox{\plotpoint}}
\put(413,584){\usebox{\plotpoint}}
\put(414,583){\usebox{\plotpoint}}
\put(415,582){\usebox{\plotpoint}}
\put(416,581){\usebox{\plotpoint}}
\put(418,580){\usebox{\plotpoint}}
\put(419,579){\usebox{\plotpoint}}
\put(420,578){\usebox{\plotpoint}}
\put(422,577){\usebox{\plotpoint}}
\put(423,576){\usebox{\plotpoint}}
\put(424,575){\usebox{\plotpoint}}
\put(425,574){\usebox{\plotpoint}}
\put(427,573){\usebox{\plotpoint}}
\put(428,572){\usebox{\plotpoint}}
\put(429,571){\usebox{\plotpoint}}
\put(431,570){\usebox{\plotpoint}}
\put(432,569){\usebox{\plotpoint}}
\put(433,568){\usebox{\plotpoint}}
\put(434,567){\usebox{\plotpoint}}
\put(436,566){\usebox{\plotpoint}}
\put(437,565){\usebox{\plotpoint}}
\put(438,564){\usebox{\plotpoint}}
\put(439,563){\usebox{\plotpoint}}
\put(441,562){\usebox{\plotpoint}}
\put(442,561){\usebox{\plotpoint}}
\put(443,560){\usebox{\plotpoint}}
\put(445,559){\usebox{\plotpoint}}
\put(446,558){\usebox{\plotpoint}}
\put(447,557){\usebox{\plotpoint}}
\put(448,556){\usebox{\plotpoint}}
\put(450,555){\usebox{\plotpoint}}
\put(451,554){\usebox{\plotpoint}}
\put(452,553){\usebox{\plotpoint}}
\put(453,552){\usebox{\plotpoint}}
\put(455,551){\usebox{\plotpoint}}
\put(456,550){\usebox{\plotpoint}}
\put(457,549){\usebox{\plotpoint}}
\put(459,548){\usebox{\plotpoint}}
\put(460,547){\usebox{\plotpoint}}
\put(461,546){\usebox{\plotpoint}}
\put(462,545){\usebox{\plotpoint}}
\put(464,544){\usebox{\plotpoint}}
\put(465,543){\usebox{\plotpoint}}
\put(466,542){\usebox{\plotpoint}}
\put(467,541){\usebox{\plotpoint}}
\put(469,540){\usebox{\plotpoint}}
\put(470,539){\usebox{\plotpoint}}
\put(472,538){\usebox{\plotpoint}}
\put(473,537){\usebox{\plotpoint}}
\put(474,536){\usebox{\plotpoint}}
\put(476,535){\usebox{\plotpoint}}
\put(477,534){\usebox{\plotpoint}}
\put(479,533){\usebox{\plotpoint}}
\put(480,532){\usebox{\plotpoint}}
\put(481,531){\usebox{\plotpoint}}
\put(483,530){\usebox{\plotpoint}}
\put(484,529){\usebox{\plotpoint}}
\put(485,528){\usebox{\plotpoint}}
\put(487,527){\usebox{\plotpoint}}
\put(488,526){\usebox{\plotpoint}}
\put(490,525){\usebox{\plotpoint}}
\put(491,524){\usebox{\plotpoint}}
\put(492,523){\usebox{\plotpoint}}
\put(494,522){\usebox{\plotpoint}}
\put(495,521){\usebox{\plotpoint}}
\put(496,520){\usebox{\plotpoint}}
\put(498,519){\usebox{\plotpoint}}
\put(499,518){\usebox{\plotpoint}}
\put(501,517){\usebox{\plotpoint}}
\put(502,516){\usebox{\plotpoint}}
\put(503,515){\usebox{\plotpoint}}
\put(505,514){\usebox{\plotpoint}}
\put(506,513){\usebox{\plotpoint}}
\put(507,512){\usebox{\plotpoint}}
\put(509,511){\usebox{\plotpoint}}
\put(510,510){\usebox{\plotpoint}}
\put(512,509){\usebox{\plotpoint}}
\put(513,508){\usebox{\plotpoint}}
\put(514,507){\usebox{\plotpoint}}
\put(516,506){\usebox{\plotpoint}}
\put(517,505){\usebox{\plotpoint}}
\put(519,504){\usebox{\plotpoint}}
\put(520,503){\usebox{\plotpoint}}
\put(522,502){\usebox{\plotpoint}}
\put(523,501){\usebox{\plotpoint}}
\put(525,500){\usebox{\plotpoint}}
\put(526,499){\usebox{\plotpoint}}
\put(528,498){\usebox{\plotpoint}}
\put(529,497){\usebox{\plotpoint}}
\put(531,496){\usebox{\plotpoint}}
\put(532,495){\usebox{\plotpoint}}
\put(534,494){\usebox{\plotpoint}}
\put(535,493){\usebox{\plotpoint}}
\put(537,492){\usebox{\plotpoint}}
\put(538,491){\usebox{\plotpoint}}
\put(540,490){\usebox{\plotpoint}}
\put(541,489){\usebox{\plotpoint}}
\put(543,488){\usebox{\plotpoint}}
\put(544,487){\usebox{\plotpoint}}
\put(546,486){\usebox{\plotpoint}}
\put(547,485){\usebox{\plotpoint}}
\put(549,484){\usebox{\plotpoint}}
\put(550,483){\usebox{\plotpoint}}
\put(552,482){\usebox{\plotpoint}}
\put(553,481){\usebox{\plotpoint}}
\put(555,480){\usebox{\plotpoint}}
\put(556,479){\usebox{\plotpoint}}
\put(558,478){\usebox{\plotpoint}}
\put(559,477){\usebox{\plotpoint}}
\put(561,476){\usebox{\plotpoint}}
\put(562,475){\usebox{\plotpoint}}
\put(564,474){\usebox{\plotpoint}}
\put(565,473){\usebox{\plotpoint}}
\put(567,472){\usebox{\plotpoint}}
\put(568,471){\usebox{\plotpoint}}
\put(570,470){\usebox{\plotpoint}}
\put(571,469){\usebox{\plotpoint}}
\put(573,468){\usebox{\plotpoint}}
\put(574,467){\usebox{\plotpoint}}
\put(576,466){\usebox{\plotpoint}}
\put(577,465){\usebox{\plotpoint}}
\put(579,464){\usebox{\plotpoint}}
\put(581,463){\usebox{\plotpoint}}
\put(582,462){\usebox{\plotpoint}}
\put(584,461){\usebox{\plotpoint}}
\put(585,460){\usebox{\plotpoint}}
\put(587,459){\usebox{\plotpoint}}
\put(589,458){\usebox{\plotpoint}}
\put(590,457){\usebox{\plotpoint}}
\put(592,456){\usebox{\plotpoint}}
\put(593,455){\usebox{\plotpoint}}
\put(595,454){\usebox{\plotpoint}}
\put(597,453){\usebox{\plotpoint}}
\put(598,452){\usebox{\plotpoint}}
\put(600,451){\usebox{\plotpoint}}
\put(601,450){\usebox{\plotpoint}}
\put(603,449){\usebox{\plotpoint}}
\put(605,448){\usebox{\plotpoint}}
\put(606,447){\usebox{\plotpoint}}
\put(608,446){\usebox{\plotpoint}}
\put(609,445){\usebox{\plotpoint}}
\put(611,444){\usebox{\plotpoint}}
\put(613,443){\usebox{\plotpoint}}
\put(614,442){\usebox{\plotpoint}}
\put(616,441){\usebox{\plotpoint}}
\put(617,440){\usebox{\plotpoint}}
\put(619,439){\usebox{\plotpoint}}
\put(621,438){\rule[-0.200pt]{0.410pt}{0.400pt}}
\put(622,437){\rule[-0.200pt]{0.410pt}{0.400pt}}
\put(624,436){\rule[-0.200pt]{0.410pt}{0.400pt}}
\put(626,435){\rule[-0.200pt]{0.410pt}{0.400pt}}
\put(627,434){\rule[-0.200pt]{0.410pt}{0.400pt}}
\put(629,433){\rule[-0.200pt]{0.410pt}{0.400pt}}
\put(631,432){\rule[-0.200pt]{0.410pt}{0.400pt}}
\put(632,431){\rule[-0.200pt]{0.410pt}{0.400pt}}
\put(634,430){\rule[-0.200pt]{0.410pt}{0.400pt}}
\put(636,429){\rule[-0.200pt]{0.410pt}{0.400pt}}
\put(638,428){\rule[-0.200pt]{0.410pt}{0.400pt}}
\put(639,427){\rule[-0.200pt]{0.410pt}{0.400pt}}
\put(641,426){\rule[-0.200pt]{0.410pt}{0.400pt}}
\put(643,425){\rule[-0.200pt]{0.410pt}{0.400pt}}
\put(644,424){\rule[-0.200pt]{0.410pt}{0.400pt}}
\put(646,423){\rule[-0.200pt]{0.410pt}{0.400pt}}
\put(648,422){\rule[-0.200pt]{0.410pt}{0.400pt}}
\put(649,421){\rule[-0.200pt]{0.410pt}{0.400pt}}
\put(651,420){\rule[-0.200pt]{0.410pt}{0.400pt}}
\put(653,419){\rule[-0.200pt]{0.410pt}{0.400pt}}
\put(655,418){\rule[-0.200pt]{0.410pt}{0.400pt}}
\put(656,417){\rule[-0.200pt]{0.410pt}{0.400pt}}
\put(658,416){\rule[-0.200pt]{0.410pt}{0.400pt}}
\put(660,415){\rule[-0.200pt]{0.410pt}{0.400pt}}
\put(661,414){\rule[-0.200pt]{0.410pt}{0.400pt}}
\put(663,413){\rule[-0.200pt]{0.410pt}{0.400pt}}
\put(665,412){\rule[-0.200pt]{0.410pt}{0.400pt}}
\put(666,411){\rule[-0.200pt]{0.410pt}{0.400pt}}
\put(668,410){\rule[-0.200pt]{0.410pt}{0.400pt}}
\put(670,409){\rule[-0.200pt]{0.409pt}{0.400pt}}
\put(672,408){\rule[-0.200pt]{0.424pt}{0.400pt}}
\put(673,407){\rule[-0.200pt]{0.424pt}{0.400pt}}
\put(675,406){\rule[-0.200pt]{0.424pt}{0.400pt}}
\put(677,405){\rule[-0.200pt]{0.424pt}{0.400pt}}
\put(679,404){\rule[-0.200pt]{0.424pt}{0.400pt}}
\put(680,403){\rule[-0.200pt]{0.424pt}{0.400pt}}
\put(682,402){\rule[-0.200pt]{0.424pt}{0.400pt}}
\put(684,401){\rule[-0.200pt]{0.424pt}{0.400pt}}
\put(686,400){\rule[-0.200pt]{0.424pt}{0.400pt}}
\put(687,399){\rule[-0.200pt]{0.424pt}{0.400pt}}
\put(689,398){\rule[-0.200pt]{0.424pt}{0.400pt}}
\put(691,397){\rule[-0.200pt]{0.424pt}{0.400pt}}
\put(693,396){\rule[-0.200pt]{0.424pt}{0.400pt}}
\put(694,395){\rule[-0.200pt]{0.424pt}{0.400pt}}
\put(696,394){\rule[-0.200pt]{0.424pt}{0.400pt}}
\put(698,393){\rule[-0.200pt]{0.424pt}{0.400pt}}
\put(700,392){\rule[-0.200pt]{0.424pt}{0.400pt}}
\put(701,391){\rule[-0.200pt]{0.424pt}{0.400pt}}
\put(703,390){\rule[-0.200pt]{0.424pt}{0.400pt}}
\put(705,389){\rule[-0.200pt]{0.424pt}{0.400pt}}
\put(707,388){\rule[-0.200pt]{0.424pt}{0.400pt}}
\put(708,387){\rule[-0.200pt]{0.424pt}{0.400pt}}
\put(710,386){\rule[-0.200pt]{0.424pt}{0.400pt}}
\put(712,385){\rule[-0.200pt]{0.424pt}{0.400pt}}
\put(714,384){\rule[-0.200pt]{0.424pt}{0.400pt}}
\put(715,383){\rule[-0.200pt]{0.424pt}{0.400pt}}
\put(717,382){\rule[-0.200pt]{0.424pt}{0.400pt}}
\put(719,381){\rule[-0.200pt]{0.424pt}{0.400pt}}
\put(721,380){\rule[-0.200pt]{0.424pt}{0.400pt}}
\put(722,379){\rule[-0.200pt]{0.424pt}{0.400pt}}
\put(724,378){\rule[-0.200pt]{0.424pt}{0.400pt}}
\put(726,377){\rule[-0.200pt]{0.424pt}{0.400pt}}
\put(728,376){\rule[-0.200pt]{0.424pt}{0.400pt}}
\put(730,375){\rule[-0.200pt]{0.424pt}{0.400pt}}
\put(731,374){\rule[-0.200pt]{0.424pt}{0.400pt}}
\put(733,373){\rule[-0.200pt]{0.424pt}{0.400pt}}
\put(735,372){\rule[-0.200pt]{0.424pt}{0.400pt}}
\put(737,371){\rule[-0.200pt]{0.424pt}{0.400pt}}
\put(738,370){\rule[-0.200pt]{0.424pt}{0.400pt}}
\put(740,369){\rule[-0.200pt]{0.424pt}{0.400pt}}
\put(742,368){\rule[-0.200pt]{0.424pt}{0.400pt}}
\put(744,367){\rule[-0.200pt]{0.424pt}{0.400pt}}
\put(745,366){\rule[-0.200pt]{0.424pt}{0.400pt}}
\put(747,365){\rule[-0.200pt]{0.424pt}{0.400pt}}
\put(749,364){\rule[-0.200pt]{0.424pt}{0.400pt}}
\put(751,363){\rule[-0.200pt]{0.424pt}{0.400pt}}
\put(752,362){\rule[-0.200pt]{0.424pt}{0.400pt}}
\put(754,361){\rule[-0.200pt]{0.424pt}{0.400pt}}
\put(756,360){\rule[-0.200pt]{0.424pt}{0.400pt}}
\put(758,359){\rule[-0.200pt]{0.424pt}{0.400pt}}
\put(759,358){\rule[-0.200pt]{0.424pt}{0.400pt}}
\put(761,357){\rule[-0.200pt]{0.424pt}{0.400pt}}
\put(763,356){\rule[-0.200pt]{0.424pt}{0.400pt}}
\put(765,355){\rule[-0.200pt]{0.424pt}{0.400pt}}
\put(766,354){\rule[-0.200pt]{0.424pt}{0.400pt}}
\put(768,353){\rule[-0.200pt]{0.424pt}{0.400pt}}
\put(770,352){\rule[-0.200pt]{0.424pt}{0.400pt}}
\put(772,351){\rule[-0.200pt]{0.424pt}{0.400pt}}
\put(773,350){\rule[-0.200pt]{0.447pt}{0.400pt}}
\put(775,349){\rule[-0.200pt]{0.447pt}{0.400pt}}
\put(777,348){\rule[-0.200pt]{0.447pt}{0.400pt}}
\put(779,347){\rule[-0.200pt]{0.447pt}{0.400pt}}
\put(781,346){\rule[-0.200pt]{0.447pt}{0.400pt}}
\put(783,345){\rule[-0.200pt]{0.447pt}{0.400pt}}
\put(785,344){\rule[-0.200pt]{0.447pt}{0.400pt}}
\put(786,343){\rule[-0.200pt]{0.447pt}{0.400pt}}
\put(788,342){\rule[-0.200pt]{0.447pt}{0.400pt}}
\put(790,341){\rule[-0.200pt]{0.447pt}{0.400pt}}
\put(792,340){\rule[-0.200pt]{0.447pt}{0.400pt}}
\put(794,339){\rule[-0.200pt]{0.447pt}{0.400pt}}
\put(796,338){\rule[-0.200pt]{0.447pt}{0.400pt}}
\put(798,337){\rule[-0.200pt]{0.447pt}{0.400pt}}
\put(799,336){\rule[-0.200pt]{0.447pt}{0.400pt}}
\put(801,335){\rule[-0.200pt]{0.447pt}{0.400pt}}
\put(803,334){\rule[-0.200pt]{0.447pt}{0.400pt}}
\put(805,333){\rule[-0.200pt]{0.447pt}{0.400pt}}
\put(807,332){\rule[-0.200pt]{0.447pt}{0.400pt}}
\put(809,331){\rule[-0.200pt]{0.447pt}{0.400pt}}
\put(811,330){\rule[-0.200pt]{0.447pt}{0.400pt}}
\put(812,329){\rule[-0.200pt]{0.447pt}{0.400pt}}
\put(814,328){\rule[-0.200pt]{0.447pt}{0.400pt}}
\put(816,327){\rule[-0.200pt]{0.447pt}{0.400pt}}
\put(818,326){\rule[-0.200pt]{0.447pt}{0.400pt}}
\put(820,325){\rule[-0.200pt]{0.447pt}{0.400pt}}
\put(822,324){\rule[-0.200pt]{0.447pt}{0.400pt}}
\put(824,323){\rule[-0.200pt]{0.447pt}{0.400pt}}
\put(825,322){\rule[-0.200pt]{0.455pt}{0.400pt}}
\put(827,321){\rule[-0.200pt]{0.455pt}{0.400pt}}
\put(829,320){\rule[-0.200pt]{0.455pt}{0.400pt}}
\put(831,319){\rule[-0.200pt]{0.455pt}{0.400pt}}
\put(833,318){\rule[-0.200pt]{0.455pt}{0.400pt}}
\put(835,317){\rule[-0.200pt]{0.455pt}{0.400pt}}
\put(837,316){\rule[-0.200pt]{0.455pt}{0.400pt}}
\put(839,315){\rule[-0.200pt]{0.455pt}{0.400pt}}
\put(841,314){\rule[-0.200pt]{0.455pt}{0.400pt}}
\put(843,313){\rule[-0.200pt]{0.455pt}{0.400pt}}
\put(844,312){\rule[-0.200pt]{0.455pt}{0.400pt}}
\put(846,311){\rule[-0.200pt]{0.455pt}{0.400pt}}
\put(848,310){\rule[-0.200pt]{0.455pt}{0.400pt}}
\put(850,309){\rule[-0.200pt]{0.455pt}{0.400pt}}
\put(852,308){\rule[-0.200pt]{0.455pt}{0.400pt}}
\put(854,307){\rule[-0.200pt]{0.455pt}{0.400pt}}
\put(856,306){\rule[-0.200pt]{0.455pt}{0.400pt}}
\put(858,305){\rule[-0.200pt]{0.455pt}{0.400pt}}
\put(860,304){\rule[-0.200pt]{0.455pt}{0.400pt}}
\put(861,303){\rule[-0.200pt]{0.455pt}{0.400pt}}
\put(863,302){\rule[-0.200pt]{0.455pt}{0.400pt}}
\put(865,301){\rule[-0.200pt]{0.455pt}{0.400pt}}
\put(867,300){\rule[-0.200pt]{0.455pt}{0.400pt}}
\put(869,299){\rule[-0.200pt]{0.455pt}{0.400pt}}
\put(871,298){\rule[-0.200pt]{0.455pt}{0.400pt}}
\put(873,297){\rule[-0.200pt]{0.455pt}{0.400pt}}
\put(875,296){\rule[-0.200pt]{0.455pt}{0.400pt}}
\put(877,295){\rule[-0.200pt]{0.455pt}{0.400pt}}
\put(878,294){\rule[-0.200pt]{0.455pt}{0.400pt}}
\put(880,293){\rule[-0.200pt]{0.455pt}{0.400pt}}
\put(882,292){\rule[-0.200pt]{0.455pt}{0.400pt}}
\put(884,291){\rule[-0.200pt]{0.455pt}{0.400pt}}
\put(886,290){\rule[-0.200pt]{0.455pt}{0.400pt}}
\put(888,289){\rule[-0.200pt]{0.455pt}{0.400pt}}
\put(890,288){\rule[-0.200pt]{0.455pt}{0.400pt}}
\put(892,287){\rule[-0.200pt]{0.455pt}{0.400pt}}
\put(894,286){\rule[-0.200pt]{0.455pt}{0.400pt}}
\put(895,285){\rule[-0.200pt]{0.455pt}{0.400pt}}
\put(897,284){\rule[-0.200pt]{0.455pt}{0.400pt}}
\put(899,283){\rule[-0.200pt]{0.455pt}{0.400pt}}
\put(901,282){\rule[-0.200pt]{0.455pt}{0.400pt}}
\put(903,281){\rule[-0.200pt]{0.455pt}{0.400pt}}
\put(905,280){\rule[-0.200pt]{0.455pt}{0.400pt}}
\put(907,279){\rule[-0.200pt]{0.455pt}{0.400pt}}
\put(909,278){\rule[-0.200pt]{0.455pt}{0.400pt}}
\put(911,277){\rule[-0.200pt]{0.455pt}{0.400pt}}
\put(912,276){\rule[-0.200pt]{0.455pt}{0.400pt}}
\put(914,275){\rule[-0.200pt]{0.455pt}{0.400pt}}
\put(916,274){\rule[-0.200pt]{0.455pt}{0.400pt}}
\put(918,273){\rule[-0.200pt]{0.455pt}{0.400pt}}
\put(920,272){\rule[-0.200pt]{0.455pt}{0.400pt}}
\put(922,271){\rule[-0.200pt]{0.455pt}{0.400pt}}
\put(924,270){\rule[-0.200pt]{0.455pt}{0.400pt}}
\put(926,269){\rule[-0.200pt]{0.455pt}{0.400pt}}
\put(928,268){\rule[-0.200pt]{0.455pt}{0.400pt}}
\put(929,267){\rule[-0.200pt]{0.455pt}{0.400pt}}
\put(931,266){\rule[-0.200pt]{0.455pt}{0.400pt}}
\put(933,265){\rule[-0.200pt]{0.455pt}{0.400pt}}
\put(935,264){\rule[-0.200pt]{0.455pt}{0.400pt}}
\put(937,263){\rule[-0.200pt]{0.455pt}{0.400pt}}
\put(939,262){\rule[-0.200pt]{0.455pt}{0.400pt}}
\put(941,261){\rule[-0.200pt]{0.455pt}{0.400pt}}
\put(943,260){\rule[-0.200pt]{0.455pt}{0.400pt}}
\put(945,259){\rule[-0.200pt]{0.455pt}{0.400pt}}
\put(946,258){\rule[-0.200pt]{0.455pt}{0.400pt}}
\put(948,257){\rule[-0.200pt]{0.455pt}{0.400pt}}
\put(950,256){\rule[-0.200pt]{0.455pt}{0.400pt}}
\put(952,255){\rule[-0.200pt]{0.455pt}{0.400pt}}
\put(954,254){\rule[-0.200pt]{0.455pt}{0.400pt}}
\put(956,253){\rule[-0.200pt]{0.455pt}{0.400pt}}
\put(958,252){\rule[-0.200pt]{0.455pt}{0.400pt}}
\put(960,251){\rule[-0.200pt]{0.455pt}{0.400pt}}
\put(962,250){\rule[-0.200pt]{0.455pt}{0.400pt}}
\put(963,249){\rule[-0.200pt]{0.455pt}{0.400pt}}
\put(965,248){\rule[-0.200pt]{0.455pt}{0.400pt}}
\put(967,247){\rule[-0.200pt]{0.455pt}{0.400pt}}
\put(969,246){\rule[-0.200pt]{0.455pt}{0.400pt}}
\put(971,245){\rule[-0.200pt]{0.455pt}{0.400pt}}
\put(973,244){\rule[-0.200pt]{0.455pt}{0.400pt}}
\put(975,243){\rule[-0.200pt]{0.455pt}{0.400pt}}
\put(977,242){\rule[-0.200pt]{0.455pt}{0.400pt}}
\put(979,241){\rule[-0.200pt]{0.455pt}{0.400pt}}
\put(980,240){\rule[-0.200pt]{0.455pt}{0.400pt}}
\put(982,239){\rule[-0.200pt]{0.455pt}{0.400pt}}
\put(984,238){\rule[-0.200pt]{0.455pt}{0.400pt}}
\put(986,237){\rule[-0.200pt]{0.455pt}{0.400pt}}
\put(988,236){\rule[-0.200pt]{0.455pt}{0.400pt}}
\put(990,235){\rule[-0.200pt]{0.455pt}{0.400pt}}
\put(992,234){\rule[-0.200pt]{0.455pt}{0.400pt}}
\put(994,233){\rule[-0.200pt]{0.455pt}{0.400pt}}
\put(996,232){\rule[-0.200pt]{0.455pt}{0.400pt}}
\put(997,231){\rule[-0.200pt]{0.455pt}{0.400pt}}
\put(999,230){\rule[-0.200pt]{0.455pt}{0.400pt}}
\put(1001,229){\rule[-0.200pt]{0.455pt}{0.400pt}}
\put(1003,228){\rule[-0.200pt]{0.455pt}{0.400pt}}
\put(1005,227){\rule[-0.200pt]{0.455pt}{0.400pt}}
\put(1007,226){\rule[-0.200pt]{0.455pt}{0.400pt}}
\put(1009,225){\rule[-0.200pt]{0.455pt}{0.400pt}}
\put(1011,224){\rule[-0.200pt]{0.455pt}{0.400pt}}
\put(1013,223){\rule[-0.200pt]{0.455pt}{0.400pt}}
\put(1014,222){\rule[-0.200pt]{0.455pt}{0.400pt}}
\put(1016,221){\rule[-0.200pt]{0.455pt}{0.400pt}}
\put(1018,220){\rule[-0.200pt]{0.455pt}{0.400pt}}
\put(1020,219){\rule[-0.200pt]{0.455pt}{0.400pt}}
\put(1022,218){\rule[-0.200pt]{0.455pt}{0.400pt}}
\put(1024,217){\rule[-0.200pt]{0.455pt}{0.400pt}}
\put(1026,216){\rule[-0.200pt]{0.455pt}{0.400pt}}
\put(1028,215){\rule[-0.200pt]{0.455pt}{0.400pt}}
\put(1030,214){\rule[-0.200pt]{0.439pt}{0.400pt}}
\put(1031,213){\rule[-0.200pt]{0.439pt}{0.400pt}}
\put(1033,212){\rule[-0.200pt]{0.439pt}{0.400pt}}
\put(1035,211){\rule[-0.200pt]{0.439pt}{0.400pt}}
\put(1037,210){\rule[-0.200pt]{0.439pt}{0.400pt}}
\put(1039,209){\rule[-0.200pt]{0.439pt}{0.400pt}}
\put(1040,208){\rule[-0.200pt]{0.439pt}{0.400pt}}
\put(1042,207){\rule[-0.200pt]{0.439pt}{0.400pt}}
\put(1044,206){\rule[-0.200pt]{0.439pt}{0.400pt}}
\put(1046,205){\rule[-0.200pt]{0.439pt}{0.400pt}}
\put(1048,204){\rule[-0.200pt]{0.439pt}{0.400pt}}
\put(1050,203){\rule[-0.200pt]{0.439pt}{0.400pt}}
\put(1051,202){\rule[-0.200pt]{0.439pt}{0.400pt}}
\put(1053,201){\rule[-0.200pt]{0.439pt}{0.400pt}}
\put(1055,200){\rule[-0.200pt]{0.439pt}{0.400pt}}
\put(1057,199){\rule[-0.200pt]{0.439pt}{0.400pt}}
\put(1059,198){\rule[-0.200pt]{0.439pt}{0.400pt}}
\put(1060,197){\rule[-0.200pt]{0.439pt}{0.400pt}}
\put(1062,196){\rule[-0.200pt]{0.439pt}{0.400pt}}
\put(1064,195){\rule[-0.200pt]{0.439pt}{0.400pt}}
\put(1066,194){\rule[-0.200pt]{0.439pt}{0.400pt}}
\put(1068,193){\rule[-0.200pt]{0.439pt}{0.400pt}}
\put(1070,192){\rule[-0.200pt]{0.439pt}{0.400pt}}
\put(1071,191){\rule[-0.200pt]{0.439pt}{0.400pt}}
\put(1073,190){\rule[-0.200pt]{0.439pt}{0.400pt}}
\put(1075,189){\rule[-0.200pt]{0.439pt}{0.400pt}}
\put(1077,188){\rule[-0.200pt]{0.439pt}{0.400pt}}
\put(1079,187){\rule[-0.200pt]{0.439pt}{0.400pt}}
\put(1080,186){\rule[-0.200pt]{0.455pt}{0.400pt}}
\put(1082,185){\rule[-0.200pt]{0.455pt}{0.400pt}}
\put(1084,184){\rule[-0.200pt]{0.455pt}{0.400pt}}
\put(1086,183){\rule[-0.200pt]{0.455pt}{0.400pt}}
\put(1088,182){\rule[-0.200pt]{0.455pt}{0.400pt}}
\put(1090,181){\rule[-0.200pt]{0.455pt}{0.400pt}}
\put(1092,180){\rule[-0.200pt]{0.455pt}{0.400pt}}
\put(1094,179){\rule[-0.200pt]{0.455pt}{0.400pt}}
\put(1096,178){\rule[-0.200pt]{0.455pt}{0.400pt}}
\put(1098,177){\rule[-0.200pt]{0.455pt}{0.400pt}}
\put(1099,176){\rule[-0.200pt]{0.455pt}{0.400pt}}
\put(1101,175){\rule[-0.200pt]{0.455pt}{0.400pt}}
\put(1103,174){\rule[-0.200pt]{0.455pt}{0.400pt}}
\put(1105,173){\rule[-0.200pt]{0.455pt}{0.400pt}}
\put(1107,172){\rule[-0.200pt]{0.455pt}{0.400pt}}
\put(1109,171){\rule[-0.200pt]{0.455pt}{0.400pt}}
\put(1111,170){\rule[-0.200pt]{0.455pt}{0.400pt}}
\put(1113,169){\rule[-0.200pt]{0.455pt}{0.400pt}}
\put(1115,168){\rule[-0.200pt]{0.455pt}{0.400pt}}
\put(1116,167){\rule[-0.200pt]{0.455pt}{0.400pt}}
\put(1118,166){\rule[-0.200pt]{0.455pt}{0.400pt}}
\put(1120,165){\rule[-0.200pt]{0.455pt}{0.400pt}}
\put(1122,164){\rule[-0.200pt]{0.455pt}{0.400pt}}
\put(1124,163){\rule[-0.200pt]{0.455pt}{0.400pt}}
\put(1126,162){\rule[-0.200pt]{0.455pt}{0.400pt}}
\put(1128,161){\rule[-0.200pt]{0.455pt}{0.400pt}}
\put(1130,160){\rule[-0.200pt]{0.455pt}{0.400pt}}
\put(1132,159){\rule[-0.200pt]{0.439pt}{0.400pt}}
\put(1133,158){\rule[-0.200pt]{0.439pt}{0.400pt}}
\put(1135,157){\rule[-0.200pt]{0.439pt}{0.400pt}}
\put(1137,156){\rule[-0.200pt]{0.439pt}{0.400pt}}
\put(1139,155){\rule[-0.200pt]{0.439pt}{0.400pt}}
\put(1141,154){\rule[-0.200pt]{0.439pt}{0.400pt}}
\put(1142,153){\rule[-0.200pt]{0.439pt}{0.400pt}}
\put(1144,152){\rule[-0.200pt]{0.439pt}{0.400pt}}
\put(1146,151){\rule[-0.200pt]{0.439pt}{0.400pt}}
\put(1148,150){\rule[-0.200pt]{0.439pt}{0.400pt}}
\put(1150,149){\rule[-0.200pt]{0.439pt}{0.400pt}}
\put(1152,148){\rule[-0.200pt]{0.439pt}{0.400pt}}
\put(1153,147){\rule[-0.200pt]{0.439pt}{0.400pt}}
\put(1155,146){\rule[-0.200pt]{0.439pt}{0.400pt}}
\put(1157,145){\rule[-0.200pt]{0.439pt}{0.400pt}}
\put(1159,144){\rule[-0.200pt]{0.439pt}{0.400pt}}
\put(1161,143){\rule[-0.200pt]{0.439pt}{0.400pt}}
\put(1162,142){\rule[-0.200pt]{0.439pt}{0.400pt}}
\put(1164,141){\rule[-0.200pt]{0.439pt}{0.400pt}}
\put(1166,140){\rule[-0.200pt]{0.439pt}{0.400pt}}
\put(1168,139){\rule[-0.200pt]{0.439pt}{0.400pt}}
\put(1170,138){\rule[-0.200pt]{0.439pt}{0.400pt}}
\put(1172,137){\rule[-0.200pt]{0.439pt}{0.400pt}}
\put(1173,136){\rule[-0.200pt]{0.439pt}{0.400pt}}
\put(1175,135){\rule[-0.200pt]{0.439pt}{0.400pt}}
\put(1177,134){\rule[-0.200pt]{0.439pt}{0.400pt}}
\put(1179,133){\rule[-0.200pt]{0.439pt}{0.400pt}}
\put(1181,132){\rule[-0.200pt]{0.439pt}{0.400pt}}
\put(1182,131){\rule[-0.200pt]{0.428pt}{0.400pt}}
\put(1184,130){\rule[-0.200pt]{0.428pt}{0.400pt}}
\put(1186,129){\rule[-0.200pt]{0.428pt}{0.400pt}}
\put(1188,128){\rule[-0.200pt]{0.428pt}{0.400pt}}
\put(1190,127){\rule[-0.200pt]{0.428pt}{0.400pt}}
\put(1191,126){\rule[-0.200pt]{0.428pt}{0.400pt}}
\put(1193,125){\rule[-0.200pt]{0.428pt}{0.400pt}}
\put(1195,124){\rule[-0.200pt]{0.428pt}{0.400pt}}
\put(1197,123){\rule[-0.200pt]{0.428pt}{0.400pt}}
\put(1199,122){\rule[-0.200pt]{0.428pt}{0.400pt}}
\put(1200,121){\rule[-0.200pt]{0.428pt}{0.400pt}}
\put(1202,120){\rule[-0.200pt]{0.428pt}{0.400pt}}
\put(1204,119){\rule[-0.200pt]{0.428pt}{0.400pt}}
\put(1206,118){\rule[-0.200pt]{0.428pt}{0.400pt}}
\put(1207,117){\rule[-0.200pt]{0.428pt}{0.400pt}}
\put(1209,116){\rule[-0.200pt]{0.428pt}{0.400pt}}
\put(1211,115){\rule[-0.200pt]{0.428pt}{0.400pt}}
\put(1213,114){\rule[-0.200pt]{0.428pt}{0.400pt}}
\end{picture}

%% file: fig3.tex
\setlength{\unitlength}{0.240900pt}
\begin{picture}(1349,990)(0,0)
\tenrm
\ifx\plotpoint\undefined\newsavebox{\plotpoint}\fi
\put(264,113){\line(1,0){20}}
\put(1285,113){\line(-1,0){20}}
\put(242,113){\makebox(0,0)[r]{0.1}}
\put(264,194){\line(1,0){10}}
\put(1285,194){\line(-1,0){10}}
\put(264,242){\line(1,0){10}}
\put(1285,242){\line(-1,0){10}}
\put(264,275){\line(1,0){10}}
\put(1285,275){\line(-1,0){10}}
\put(264,301){\line(1,0){10}}
\put(1285,301){\line(-1,0){10}}
\put(264,323){\line(1,0){10}}
\put(1285,323){\line(-1,0){10}}
\put(264,341){\line(1,0){10}}
\put(1285,341){\line(-1,0){10}}
\put(264,357){\line(1,0){10}}
\put(1285,357){\line(-1,0){10}}
\put(264,370){\line(1,0){10}}
\put(1285,370){\line(-1,0){10}}
\put(264,383){\line(1,0){20}}
\put(1285,383){\line(-1,0){20}}
\put(242,383){\makebox(0,0)[r]{1}}
\put(264,464){\line(1,0){10}}
\put(1285,464){\line(-1,0){10}}
\put(264,511){\line(1,0){10}}
\put(1285,511){\line(-1,0){10}}
\put(264,545){\line(1,0){10}}
\put(1285,545){\line(-1,0){10}}
\put(264,571){\line(1,0){10}}
\put(1285,571){\line(-1,0){10}}
\put(264,593){\line(1,0){10}}
\put(1285,593){\line(-1,0){10}}
\put(264,611){\line(1,0){10}}
\put(1285,611){\line(-1,0){10}}
\put(264,626){\line(1,0){10}}
\put(1285,626){\line(-1,0){10}}
\put(264,640){\line(1,0){10}}
\put(1285,640){\line(-1,0){10}}
\put(264,652){\line(1,0){20}}
\put(1285,652){\line(-1,0){20}}
\put(242,652){\makebox(0,0)[r]{10}}
\put(264,734){\line(1,0){10}}
\put(1285,734){\line(-1,0){10}}
\put(264,781){\line(1,0){10}}
\put(1285,781){\line(-1,0){10}}
\put(264,815){\line(1,0){10}}
\put(1285,815){\line(-1,0){10}}
\put(264,841){\line(1,0){10}}
\put(1285,841){\line(-1,0){10}}
\put(264,862){\line(1,0){10}}
\put(1285,862){\line(-1,0){10}}
\put(264,880){\line(1,0){10}}
\put(1285,880){\line(-1,0){10}}
\put(264,896){\line(1,0){10}}
\put(1285,896){\line(-1,0){10}}
\put(264,910){\line(1,0){10}}
\put(1285,910){\line(-1,0){10}}
\put(264,922){\line(1,0){20}}
\put(1285,922){\line(-1,0){20}}
\put(242,922){\makebox(0,0)[r]{100}}
\put(264,113){\line(0,1){20}}
\put(264,922){\line(0,-1){20}}
\put(264,68){\makebox(0,0){50}}
\put(468,113){\line(0,1){20}}
\put(468,922){\line(0,-1){20}}
\put(468,68){\makebox(0,0){100}}
\put(672,113){\line(0,1){20}}
\put(672,922){\line(0,-1){20}}
\put(672,68){\makebox(0,0){150}}
\put(877,113){\line(0,1){20}}
\put(877,922){\line(0,-1){20}}
\put(877,68){\makebox(0,0){200}}
\put(1081,113){\line(0,1){20}}
\put(1081,922){\line(0,-1){20}}
\put(1081,68){\makebox(0,0){250}}
\put(1285,113){\line(0,1){20}}
\put(1285,922){\line(0,-1){20}}
\put(1285,68){\makebox(0,0){300}}
\put(264,113){\line(1,0){1021}}
\put(1285,113){\line(0,1){809}}
\put(1285,922){\line(-1,0){1021}}
\put(264,922){\line(0,-1){809}}
\put(45,517){\makebox(0,0)[l]{\shortstack{$\sigma$(pb)}}}
\put(774,23){\makebox(0,0){ $m_{\tilde t_R}$(GeV)}}
\put(672,781){\makebox(0,0)[l]{ $\sqrt{s}$=500 GeV NLC}}
\put(672,734){\makebox(0,0)[l]{ $\int {\cal L}_{\gamma e} dt$ = 200 $fb^{-1}$ / year}}
\put(754,652){\makebox(0,0)[l]{ $\lambda'_{132}=0.33$}}
\sbox{\plotpoint}{\rule[-0.200pt]{0.400pt}{0.400pt}}%
\put(264,817){\usebox{\plotpoint}}
\put(264,817){\rule[-0.200pt]{0.702pt}{0.400pt}}
\put(266,816){\rule[-0.200pt]{0.702pt}{0.400pt}}
\put(269,815){\rule[-0.200pt]{0.702pt}{0.400pt}}
\put(272,814){\rule[-0.200pt]{0.702pt}{0.400pt}}
\put(275,813){\rule[-0.200pt]{0.702pt}{0.400pt}}
\put(278,812){\rule[-0.200pt]{0.702pt}{0.400pt}}
\put(281,811){\rule[-0.200pt]{0.702pt}{0.400pt}}
\put(284,810){\rule[-0.200pt]{0.702pt}{0.400pt}}
\put(287,809){\rule[-0.200pt]{0.702pt}{0.400pt}}
\put(290,808){\rule[-0.200pt]{0.702pt}{0.400pt}}
\put(293,807){\rule[-0.200pt]{0.702pt}{0.400pt}}
\put(296,806){\rule[-0.200pt]{0.702pt}{0.400pt}}
\put(298,805){\rule[-0.200pt]{0.702pt}{0.400pt}}
\put(301,804){\rule[-0.200pt]{0.702pt}{0.400pt}}
\put(304,803){\rule[-0.200pt]{0.702pt}{0.400pt}}
\put(307,802){\rule[-0.200pt]{0.702pt}{0.400pt}}
\put(310,801){\rule[-0.200pt]{0.702pt}{0.400pt}}
\put(313,800){\rule[-0.200pt]{0.702pt}{0.400pt}}
\put(316,799){\rule[-0.200pt]{0.702pt}{0.400pt}}
\put(319,798){\rule[-0.200pt]{0.702pt}{0.400pt}}
\put(322,797){\rule[-0.200pt]{0.702pt}{0.400pt}}
\put(325,796){\rule[-0.200pt]{0.702pt}{0.400pt}}
\put(328,795){\rule[-0.200pt]{0.702pt}{0.400pt}}
\put(331,794){\rule[-0.200pt]{0.702pt}{0.400pt}}
\put(333,793){\rule[-0.200pt]{0.702pt}{0.400pt}}
\put(336,792){\rule[-0.200pt]{0.702pt}{0.400pt}}
\put(339,791){\rule[-0.200pt]{0.702pt}{0.400pt}}
\put(342,790){\rule[-0.200pt]{0.702pt}{0.400pt}}
\put(345,789){\rule[-0.200pt]{0.702pt}{0.400pt}}
\put(348,788){\rule[-0.200pt]{0.702pt}{0.400pt}}
\put(351,787){\rule[-0.200pt]{0.702pt}{0.400pt}}
\put(354,786){\rule[-0.200pt]{0.702pt}{0.400pt}}
\put(357,785){\rule[-0.200pt]{0.702pt}{0.400pt}}
\put(360,784){\rule[-0.200pt]{0.702pt}{0.400pt}}
\put(363,783){\rule[-0.200pt]{0.702pt}{0.400pt}}
\put(365,782){\rule[-0.200pt]{0.523pt}{0.400pt}}
\put(368,781){\rule[-0.200pt]{0.523pt}{0.400pt}}
\put(370,780){\rule[-0.200pt]{0.523pt}{0.400pt}}
\put(372,779){\rule[-0.200pt]{0.523pt}{0.400pt}}
\put(374,778){\rule[-0.200pt]{0.523pt}{0.400pt}}
\put(376,777){\rule[-0.200pt]{0.523pt}{0.400pt}}
\put(379,776){\rule[-0.200pt]{0.523pt}{0.400pt}}
\put(381,775){\rule[-0.200pt]{0.523pt}{0.400pt}}
\put(383,774){\rule[-0.200pt]{0.523pt}{0.400pt}}
\put(385,773){\rule[-0.200pt]{0.523pt}{0.400pt}}
\put(387,772){\rule[-0.200pt]{0.523pt}{0.400pt}}
\put(389,771){\rule[-0.200pt]{0.523pt}{0.400pt}}
\put(392,770){\rule[-0.200pt]{0.523pt}{0.400pt}}
\put(394,769){\rule[-0.200pt]{0.523pt}{0.400pt}}
\put(396,768){\rule[-0.200pt]{0.523pt}{0.400pt}}
\put(398,767){\rule[-0.200pt]{0.523pt}{0.400pt}}
\put(400,766){\rule[-0.200pt]{0.523pt}{0.400pt}}
\put(402,765){\rule[-0.200pt]{0.523pt}{0.400pt}}
\put(405,764){\rule[-0.200pt]{0.523pt}{0.400pt}}
\put(407,763){\rule[-0.200pt]{0.523pt}{0.400pt}}
\put(409,762){\rule[-0.200pt]{0.523pt}{0.400pt}}
\put(411,761){\rule[-0.200pt]{0.523pt}{0.400pt}}
\put(413,760){\rule[-0.200pt]{0.523pt}{0.400pt}}
\put(415,759){\rule[-0.200pt]{0.523pt}{0.400pt}}
\put(418,758){\rule[-0.200pt]{0.523pt}{0.400pt}}
\put(420,757){\rule[-0.200pt]{0.523pt}{0.400pt}}
\put(422,756){\rule[-0.200pt]{0.523pt}{0.400pt}}
\put(424,755){\rule[-0.200pt]{0.523pt}{0.400pt}}
\put(426,754){\rule[-0.200pt]{0.523pt}{0.400pt}}
\put(428,753){\rule[-0.200pt]{0.523pt}{0.400pt}}
\put(431,752){\rule[-0.200pt]{0.523pt}{0.400pt}}
\put(433,751){\rule[-0.200pt]{0.523pt}{0.400pt}}
\put(435,750){\rule[-0.200pt]{0.523pt}{0.400pt}}
\put(437,749){\rule[-0.200pt]{0.523pt}{0.400pt}}
\put(439,748){\rule[-0.200pt]{0.523pt}{0.400pt}}
\put(441,747){\rule[-0.200pt]{0.523pt}{0.400pt}}
\put(444,746){\rule[-0.200pt]{0.523pt}{0.400pt}}
\put(446,745){\rule[-0.200pt]{0.523pt}{0.400pt}}
\put(448,744){\rule[-0.200pt]{0.523pt}{0.400pt}}
\put(450,743){\rule[-0.200pt]{0.523pt}{0.400pt}}
\put(452,742){\rule[-0.200pt]{0.523pt}{0.400pt}}
\put(454,741){\rule[-0.200pt]{0.523pt}{0.400pt}}
\put(457,740){\rule[-0.200pt]{0.523pt}{0.400pt}}
\put(459,739){\rule[-0.200pt]{0.523pt}{0.400pt}}
\put(461,738){\rule[-0.200pt]{0.523pt}{0.400pt}}
\put(463,737){\rule[-0.200pt]{0.523pt}{0.400pt}}
\put(465,736){\rule[-0.200pt]{0.523pt}{0.400pt}}
\put(468,735){\rule[-0.200pt]{0.439pt}{0.400pt}}
\put(469,734){\rule[-0.200pt]{0.439pt}{0.400pt}}
\put(471,733){\rule[-0.200pt]{0.439pt}{0.400pt}}
\put(473,732){\rule[-0.200pt]{0.439pt}{0.400pt}}
\put(475,731){\rule[-0.200pt]{0.439pt}{0.400pt}}
\put(477,730){\rule[-0.200pt]{0.439pt}{0.400pt}}
\put(478,729){\rule[-0.200pt]{0.439pt}{0.400pt}}
\put(480,728){\rule[-0.200pt]{0.439pt}{0.400pt}}
\put(482,727){\rule[-0.200pt]{0.439pt}{0.400pt}}
\put(484,726){\rule[-0.200pt]{0.439pt}{0.400pt}}
\put(486,725){\rule[-0.200pt]{0.439pt}{0.400pt}}
\put(488,724){\rule[-0.200pt]{0.439pt}{0.400pt}}
\put(489,723){\rule[-0.200pt]{0.439pt}{0.400pt}}
\put(491,722){\rule[-0.200pt]{0.439pt}{0.400pt}}
\put(493,721){\rule[-0.200pt]{0.439pt}{0.400pt}}
\put(495,720){\rule[-0.200pt]{0.439pt}{0.400pt}}
\put(497,719){\rule[-0.200pt]{0.439pt}{0.400pt}}
\put(498,718){\rule[-0.200pt]{0.439pt}{0.400pt}}
\put(500,717){\rule[-0.200pt]{0.439pt}{0.400pt}}
\put(502,716){\rule[-0.200pt]{0.439pt}{0.400pt}}
\put(504,715){\rule[-0.200pt]{0.439pt}{0.400pt}}
\put(506,714){\rule[-0.200pt]{0.439pt}{0.400pt}}
\put(508,713){\rule[-0.200pt]{0.439pt}{0.400pt}}
\put(509,712){\rule[-0.200pt]{0.439pt}{0.400pt}}
\put(511,711){\rule[-0.200pt]{0.439pt}{0.400pt}}
\put(513,710){\rule[-0.200pt]{0.439pt}{0.400pt}}
\put(515,709){\rule[-0.200pt]{0.439pt}{0.400pt}}
\put(517,708){\rule[-0.200pt]{0.439pt}{0.400pt}}
\put(519,707){\rule[-0.200pt]{0.439pt}{0.400pt}}
\put(520,706){\rule[-0.200pt]{0.439pt}{0.400pt}}
\put(522,705){\rule[-0.200pt]{0.439pt}{0.400pt}}
\put(524,704){\rule[-0.200pt]{0.439pt}{0.400pt}}
\put(526,703){\rule[-0.200pt]{0.439pt}{0.400pt}}
\put(528,702){\rule[-0.200pt]{0.439pt}{0.400pt}}
\put(529,701){\rule[-0.200pt]{0.439pt}{0.400pt}}
\put(531,700){\rule[-0.200pt]{0.439pt}{0.400pt}}
\put(533,699){\rule[-0.200pt]{0.439pt}{0.400pt}}
\put(535,698){\rule[-0.200pt]{0.439pt}{0.400pt}}
\put(537,697){\rule[-0.200pt]{0.439pt}{0.400pt}}
\put(539,696){\rule[-0.200pt]{0.439pt}{0.400pt}}
\put(540,695){\rule[-0.200pt]{0.439pt}{0.400pt}}
\put(542,694){\rule[-0.200pt]{0.439pt}{0.400pt}}
\put(544,693){\rule[-0.200pt]{0.439pt}{0.400pt}}
\put(546,692){\rule[-0.200pt]{0.439pt}{0.400pt}}
\put(548,691){\rule[-0.200pt]{0.439pt}{0.400pt}}
\put(549,690){\rule[-0.200pt]{0.439pt}{0.400pt}}
\put(551,689){\rule[-0.200pt]{0.439pt}{0.400pt}}
\put(553,688){\rule[-0.200pt]{0.439pt}{0.400pt}}
\put(555,687){\rule[-0.200pt]{0.439pt}{0.400pt}}
\put(557,686){\rule[-0.200pt]{0.439pt}{0.400pt}}
\put(559,685){\rule[-0.200pt]{0.439pt}{0.400pt}}
\put(560,684){\rule[-0.200pt]{0.439pt}{0.400pt}}
\put(562,683){\rule[-0.200pt]{0.439pt}{0.400pt}}
\put(564,682){\rule[-0.200pt]{0.439pt}{0.400pt}}
\put(566,681){\rule[-0.200pt]{0.439pt}{0.400pt}}
\put(568,680){\rule[-0.200pt]{0.439pt}{0.400pt}}
\put(569,679){\rule[-0.200pt]{0.473pt}{0.400pt}}
\put(571,678){\rule[-0.200pt]{0.473pt}{0.400pt}}
\put(573,677){\rule[-0.200pt]{0.473pt}{0.400pt}}
\put(575,676){\rule[-0.200pt]{0.473pt}{0.400pt}}
\put(577,675){\rule[-0.200pt]{0.473pt}{0.400pt}}
\put(579,674){\rule[-0.200pt]{0.473pt}{0.400pt}}
\put(581,673){\rule[-0.200pt]{0.473pt}{0.400pt}}
\put(583,672){\rule[-0.200pt]{0.473pt}{0.400pt}}
\put(585,671){\rule[-0.200pt]{0.473pt}{0.400pt}}
\put(587,670){\rule[-0.200pt]{0.473pt}{0.400pt}}
\put(589,669){\rule[-0.200pt]{0.473pt}{0.400pt}}
\put(591,668){\rule[-0.200pt]{0.473pt}{0.400pt}}
\put(593,667){\rule[-0.200pt]{0.473pt}{0.400pt}}
\put(595,666){\rule[-0.200pt]{0.473pt}{0.400pt}}
\put(597,665){\rule[-0.200pt]{0.473pt}{0.400pt}}
\put(599,664){\rule[-0.200pt]{0.473pt}{0.400pt}}
\put(601,663){\rule[-0.200pt]{0.473pt}{0.400pt}}
\put(603,662){\rule[-0.200pt]{0.473pt}{0.400pt}}
\put(605,661){\rule[-0.200pt]{0.473pt}{0.400pt}}
\put(607,660){\rule[-0.200pt]{0.473pt}{0.400pt}}
\put(609,659){\rule[-0.200pt]{0.473pt}{0.400pt}}
\put(611,658){\rule[-0.200pt]{0.473pt}{0.400pt}}
\put(613,657){\rule[-0.200pt]{0.473pt}{0.400pt}}
\put(615,656){\rule[-0.200pt]{0.473pt}{0.400pt}}
\put(617,655){\rule[-0.200pt]{0.473pt}{0.400pt}}
\put(619,654){\rule[-0.200pt]{0.473pt}{0.400pt}}
\put(621,653){\rule[-0.200pt]{0.473pt}{0.400pt}}
\put(622,652){\rule[-0.200pt]{0.473pt}{0.400pt}}
\put(624,651){\rule[-0.200pt]{0.473pt}{0.400pt}}
\put(626,650){\rule[-0.200pt]{0.473pt}{0.400pt}}
\put(628,649){\rule[-0.200pt]{0.473pt}{0.400pt}}
\put(630,648){\rule[-0.200pt]{0.473pt}{0.400pt}}
\put(632,647){\rule[-0.200pt]{0.473pt}{0.400pt}}
\put(634,646){\rule[-0.200pt]{0.473pt}{0.400pt}}
\put(636,645){\rule[-0.200pt]{0.473pt}{0.400pt}}
\put(638,644){\rule[-0.200pt]{0.473pt}{0.400pt}}
\put(640,643){\rule[-0.200pt]{0.473pt}{0.400pt}}
\put(642,642){\rule[-0.200pt]{0.473pt}{0.400pt}}
\put(644,641){\rule[-0.200pt]{0.473pt}{0.400pt}}
\put(646,640){\rule[-0.200pt]{0.473pt}{0.400pt}}
\put(648,639){\rule[-0.200pt]{0.473pt}{0.400pt}}
\put(650,638){\rule[-0.200pt]{0.473pt}{0.400pt}}
\put(652,637){\rule[-0.200pt]{0.473pt}{0.400pt}}
\put(654,636){\rule[-0.200pt]{0.473pt}{0.400pt}}
\put(656,635){\rule[-0.200pt]{0.473pt}{0.400pt}}
\put(658,634){\rule[-0.200pt]{0.473pt}{0.400pt}}
\put(660,633){\rule[-0.200pt]{0.473pt}{0.400pt}}
\put(662,632){\rule[-0.200pt]{0.473pt}{0.400pt}}
\put(664,631){\rule[-0.200pt]{0.473pt}{0.400pt}}
\put(666,630){\rule[-0.200pt]{0.473pt}{0.400pt}}
\put(668,629){\rule[-0.200pt]{0.473pt}{0.400pt}}
\put(670,628){\rule[-0.200pt]{0.472pt}{0.400pt}}
\put(672,627){\rule[-0.200pt]{0.819pt}{0.400pt}}
\put(675,626){\rule[-0.200pt]{0.819pt}{0.400pt}}
\put(678,625){\rule[-0.200pt]{0.819pt}{0.400pt}}
\put(682,624){\rule[-0.200pt]{0.819pt}{0.400pt}}
\put(685,623){\rule[-0.200pt]{0.819pt}{0.400pt}}
\put(689,622){\rule[-0.200pt]{0.819pt}{0.400pt}}
\put(692,621){\rule[-0.200pt]{0.819pt}{0.400pt}}
\put(695,620){\rule[-0.200pt]{0.819pt}{0.400pt}}
\put(699,619){\rule[-0.200pt]{0.819pt}{0.400pt}}
\put(702,618){\rule[-0.200pt]{0.819pt}{0.400pt}}
\put(706,617){\rule[-0.200pt]{0.819pt}{0.400pt}}
\put(709,616){\rule[-0.200pt]{0.819pt}{0.400pt}}
\put(712,615){\rule[-0.200pt]{0.819pt}{0.400pt}}
\put(716,614){\rule[-0.200pt]{0.819pt}{0.400pt}}
\put(719,613){\rule[-0.200pt]{0.819pt}{0.400pt}}
\put(723,612){\rule[-0.200pt]{0.819pt}{0.400pt}}
\put(726,611){\rule[-0.200pt]{0.819pt}{0.400pt}}
\put(729,610){\rule[-0.200pt]{0.819pt}{0.400pt}}
\put(733,609){\rule[-0.200pt]{0.819pt}{0.400pt}}
\put(736,608){\rule[-0.200pt]{0.819pt}{0.400pt}}
\put(740,607){\rule[-0.200pt]{0.819pt}{0.400pt}}
\put(743,606){\rule[-0.200pt]{0.819pt}{0.400pt}}
\put(746,605){\rule[-0.200pt]{0.819pt}{0.400pt}}
\put(750,604){\rule[-0.200pt]{0.819pt}{0.400pt}}
\put(753,603){\rule[-0.200pt]{0.819pt}{0.400pt}}
\put(757,602){\rule[-0.200pt]{0.819pt}{0.400pt}}
\put(760,601){\rule[-0.200pt]{0.819pt}{0.400pt}}
\put(763,600){\rule[-0.200pt]{0.819pt}{0.400pt}}
\put(767,599){\rule[-0.200pt]{0.819pt}{0.400pt}}
\put(770,598){\rule[-0.200pt]{0.819pt}{0.400pt}}
\put(774,597){\rule[-0.200pt]{0.451pt}{0.400pt}}
\put(775,596){\rule[-0.200pt]{0.451pt}{0.400pt}}
\put(777,595){\rule[-0.200pt]{0.451pt}{0.400pt}}
\put(779,594){\rule[-0.200pt]{0.451pt}{0.400pt}}
\put(781,593){\rule[-0.200pt]{0.451pt}{0.400pt}}
\put(783,592){\rule[-0.200pt]{0.451pt}{0.400pt}}
\put(785,591){\rule[-0.200pt]{0.451pt}{0.400pt}}
\put(787,590){\rule[-0.200pt]{0.451pt}{0.400pt}}
\put(788,589){\rule[-0.200pt]{0.451pt}{0.400pt}}
\put(790,588){\rule[-0.200pt]{0.451pt}{0.400pt}}
\put(792,587){\rule[-0.200pt]{0.451pt}{0.400pt}}
\put(794,586){\rule[-0.200pt]{0.451pt}{0.400pt}}
\put(796,585){\rule[-0.200pt]{0.451pt}{0.400pt}}
\put(798,584){\rule[-0.200pt]{0.451pt}{0.400pt}}
\put(800,583){\rule[-0.200pt]{0.451pt}{0.400pt}}
\put(802,582){\rule[-0.200pt]{0.451pt}{0.400pt}}
\put(803,581){\rule[-0.200pt]{0.451pt}{0.400pt}}
\put(805,580){\rule[-0.200pt]{0.451pt}{0.400pt}}
\put(807,579){\rule[-0.200pt]{0.451pt}{0.400pt}}
\put(809,578){\rule[-0.200pt]{0.451pt}{0.400pt}}
\put(811,577){\rule[-0.200pt]{0.451pt}{0.400pt}}
\put(813,576){\rule[-0.200pt]{0.451pt}{0.400pt}}
\put(815,575){\rule[-0.200pt]{0.451pt}{0.400pt}}
\put(817,574){\rule[-0.200pt]{0.451pt}{0.400pt}}
\put(818,573){\rule[-0.200pt]{0.451pt}{0.400pt}}
\put(820,572){\rule[-0.200pt]{0.451pt}{0.400pt}}
\put(822,571){\rule[-0.200pt]{0.451pt}{0.400pt}}
\put(824,570){\rule[-0.200pt]{0.451pt}{0.400pt}}
\put(826,569){\rule[-0.200pt]{0.451pt}{0.400pt}}
\put(828,568){\rule[-0.200pt]{0.451pt}{0.400pt}}
\put(830,567){\rule[-0.200pt]{0.451pt}{0.400pt}}
\put(832,566){\rule[-0.200pt]{0.451pt}{0.400pt}}
\put(833,565){\rule[-0.200pt]{0.451pt}{0.400pt}}
\put(835,564){\rule[-0.200pt]{0.451pt}{0.400pt}}
\put(837,563){\rule[-0.200pt]{0.451pt}{0.400pt}}
\put(839,562){\rule[-0.200pt]{0.451pt}{0.400pt}}
\put(841,561){\rule[-0.200pt]{0.451pt}{0.400pt}}
\put(843,560){\rule[-0.200pt]{0.451pt}{0.400pt}}
\put(845,559){\rule[-0.200pt]{0.451pt}{0.400pt}}
\put(847,558){\rule[-0.200pt]{0.451pt}{0.400pt}}
\put(848,557){\rule[-0.200pt]{0.451pt}{0.400pt}}
\put(850,556){\rule[-0.200pt]{0.451pt}{0.400pt}}
\put(852,555){\rule[-0.200pt]{0.451pt}{0.400pt}}
\put(854,554){\rule[-0.200pt]{0.451pt}{0.400pt}}
\put(856,553){\rule[-0.200pt]{0.451pt}{0.400pt}}
\put(858,552){\rule[-0.200pt]{0.451pt}{0.400pt}}
\put(860,551){\rule[-0.200pt]{0.451pt}{0.400pt}}
\put(862,550){\rule[-0.200pt]{0.451pt}{0.400pt}}
\put(863,549){\rule[-0.200pt]{0.451pt}{0.400pt}}
\put(865,548){\rule[-0.200pt]{0.451pt}{0.400pt}}
\put(867,547){\rule[-0.200pt]{0.451pt}{0.400pt}}
\put(869,546){\rule[-0.200pt]{0.451pt}{0.400pt}}
\put(871,545){\rule[-0.200pt]{0.451pt}{0.400pt}}
\put(873,544){\rule[-0.200pt]{0.451pt}{0.400pt}}
\put(875,543){\rule[-0.200pt]{0.451pt}{0.400pt}}
\put(877,542){\usebox{\plotpoint}}
\put(878,541){\usebox{\plotpoint}}
\put(880,540){\usebox{\plotpoint}}
\put(881,539){\usebox{\plotpoint}}
\put(883,538){\usebox{\plotpoint}}
\put(885,537){\usebox{\plotpoint}}
\put(886,536){\usebox{\plotpoint}}
\put(888,535){\usebox{\plotpoint}}
\put(890,534){\usebox{\plotpoint}}
\put(891,533){\usebox{\plotpoint}}
\put(893,532){\usebox{\plotpoint}}
\put(895,531){\usebox{\plotpoint}}
\put(896,530){\usebox{\plotpoint}}
\put(898,529){\usebox{\plotpoint}}
\put(900,528){\usebox{\plotpoint}}
\put(901,527){\usebox{\plotpoint}}
\put(903,526){\usebox{\plotpoint}}
\put(904,525){\usebox{\plotpoint}}
\put(906,524){\usebox{\plotpoint}}
\put(908,523){\usebox{\plotpoint}}
\put(909,522){\usebox{\plotpoint}}
\put(911,521){\usebox{\plotpoint}}
\put(913,520){\usebox{\plotpoint}}
\put(914,519){\usebox{\plotpoint}}
\put(916,518){\usebox{\plotpoint}}
\put(918,517){\usebox{\plotpoint}}
\put(919,516){\usebox{\plotpoint}}
\put(921,515){\usebox{\plotpoint}}
\put(923,514){\usebox{\plotpoint}}
\put(924,513){\usebox{\plotpoint}}
\put(926,512){\usebox{\plotpoint}}
\put(927,511){\usebox{\plotpoint}}
\put(929,510){\usebox{\plotpoint}}
\put(931,509){\usebox{\plotpoint}}
\put(932,508){\usebox{\plotpoint}}
\put(934,507){\usebox{\plotpoint}}
\put(936,506){\usebox{\plotpoint}}
\put(937,505){\usebox{\plotpoint}}
\put(939,504){\usebox{\plotpoint}}
\put(941,503){\usebox{\plotpoint}}
\put(942,502){\usebox{\plotpoint}}
\put(944,501){\usebox{\plotpoint}}
\put(946,500){\usebox{\plotpoint}}
\put(947,499){\usebox{\plotpoint}}
\put(949,498){\usebox{\plotpoint}}
\put(951,497){\usebox{\plotpoint}}
\put(952,496){\usebox{\plotpoint}}
\put(954,495){\usebox{\plotpoint}}
\put(955,494){\usebox{\plotpoint}}
\put(957,493){\usebox{\plotpoint}}
\put(959,492){\usebox{\plotpoint}}
\put(960,491){\usebox{\plotpoint}}
\put(962,490){\usebox{\plotpoint}}
\put(964,489){\usebox{\plotpoint}}
\put(965,488){\usebox{\plotpoint}}
\put(967,487){\usebox{\plotpoint}}
\put(969,486){\usebox{\plotpoint}}
\put(970,485){\usebox{\plotpoint}}
\put(972,484){\usebox{\plotpoint}}
\put(974,483){\usebox{\plotpoint}}
\put(975,482){\usebox{\plotpoint}}
\put(977,481){\usebox{\plotpoint}}
\put(978,480){\rule[-0.200pt]{0.502pt}{0.400pt}}
\put(981,479){\rule[-0.200pt]{0.501pt}{0.400pt}}
\put(983,478){\rule[-0.200pt]{0.501pt}{0.400pt}}
\put(985,477){\rule[-0.200pt]{0.501pt}{0.400pt}}
\put(987,476){\rule[-0.200pt]{0.501pt}{0.400pt}}
\put(989,475){\rule[-0.200pt]{0.501pt}{0.400pt}}
\put(991,474){\rule[-0.200pt]{0.501pt}{0.400pt}}
\put(993,473){\rule[-0.200pt]{0.501pt}{0.400pt}}
\put(995,472){\rule[-0.200pt]{0.501pt}{0.400pt}}
\put(997,471){\rule[-0.200pt]{0.501pt}{0.400pt}}
\put(999,470){\rule[-0.200pt]{0.501pt}{0.400pt}}
\put(1001,469){\rule[-0.200pt]{0.501pt}{0.400pt}}
\put(1003,468){\rule[-0.200pt]{0.501pt}{0.400pt}}
\put(1006,467){\rule[-0.200pt]{0.501pt}{0.400pt}}
\put(1008,466){\rule[-0.200pt]{0.501pt}{0.400pt}}
\put(1010,465){\rule[-0.200pt]{0.501pt}{0.400pt}}
\put(1012,464){\rule[-0.200pt]{0.501pt}{0.400pt}}
\put(1014,463){\rule[-0.200pt]{0.501pt}{0.400pt}}
\put(1016,462){\rule[-0.200pt]{0.501pt}{0.400pt}}
\put(1018,461){\rule[-0.200pt]{0.501pt}{0.400pt}}
\put(1020,460){\rule[-0.200pt]{0.501pt}{0.400pt}}
\put(1022,459){\rule[-0.200pt]{0.501pt}{0.400pt}}
\put(1024,458){\rule[-0.200pt]{0.501pt}{0.400pt}}
\put(1026,457){\rule[-0.200pt]{0.501pt}{0.400pt}}
\put(1028,456){\rule[-0.200pt]{0.501pt}{0.400pt}}
\put(1031,455){\rule[-0.200pt]{0.501pt}{0.400pt}}
\put(1033,454){\rule[-0.200pt]{0.501pt}{0.400pt}}
\put(1035,453){\rule[-0.200pt]{0.501pt}{0.400pt}}
\put(1037,452){\rule[-0.200pt]{0.501pt}{0.400pt}}
\put(1039,451){\rule[-0.200pt]{0.501pt}{0.400pt}}
\put(1041,450){\rule[-0.200pt]{0.501pt}{0.400pt}}
\put(1043,449){\rule[-0.200pt]{0.501pt}{0.400pt}}
\put(1045,448){\rule[-0.200pt]{0.501pt}{0.400pt}}
\put(1047,447){\rule[-0.200pt]{0.501pt}{0.400pt}}
\put(1049,446){\rule[-0.200pt]{0.501pt}{0.400pt}}
\put(1051,445){\rule[-0.200pt]{0.501pt}{0.400pt}}
\put(1053,444){\rule[-0.200pt]{0.501pt}{0.400pt}}
\put(1056,443){\rule[-0.200pt]{0.501pt}{0.400pt}}
\put(1058,442){\rule[-0.200pt]{0.501pt}{0.400pt}}
\put(1060,441){\rule[-0.200pt]{0.501pt}{0.400pt}}
\put(1062,440){\rule[-0.200pt]{0.501pt}{0.400pt}}
\put(1064,439){\rule[-0.200pt]{0.501pt}{0.400pt}}
\put(1066,438){\rule[-0.200pt]{0.501pt}{0.400pt}}
\put(1068,437){\rule[-0.200pt]{0.501pt}{0.400pt}}
\put(1070,436){\rule[-0.200pt]{0.501pt}{0.400pt}}
\put(1072,435){\rule[-0.200pt]{0.501pt}{0.400pt}}
\put(1074,434){\rule[-0.200pt]{0.501pt}{0.400pt}}
\put(1076,433){\rule[-0.200pt]{0.501pt}{0.400pt}}
\put(1078,432){\rule[-0.200pt]{0.501pt}{0.400pt}}
\put(1081,431){\rule[-0.200pt]{0.512pt}{0.400pt}}
\put(1083,430){\rule[-0.200pt]{0.512pt}{0.400pt}}
\put(1085,429){\rule[-0.200pt]{0.512pt}{0.400pt}}
\put(1087,428){\rule[-0.200pt]{0.512pt}{0.400pt}}
\put(1089,427){\rule[-0.200pt]{0.512pt}{0.400pt}}
\put(1091,426){\rule[-0.200pt]{0.512pt}{0.400pt}}
\put(1093,425){\rule[-0.200pt]{0.512pt}{0.400pt}}
\put(1095,424){\rule[-0.200pt]{0.512pt}{0.400pt}}
\put(1098,423){\rule[-0.200pt]{0.512pt}{0.400pt}}
\put(1100,422){\rule[-0.200pt]{0.512pt}{0.400pt}}
\put(1102,421){\rule[-0.200pt]{0.512pt}{0.400pt}}
\put(1104,420){\rule[-0.200pt]{0.512pt}{0.400pt}}
\put(1106,419){\rule[-0.200pt]{0.512pt}{0.400pt}}
\put(1108,418){\rule[-0.200pt]{0.512pt}{0.400pt}}
\put(1110,417){\rule[-0.200pt]{0.512pt}{0.400pt}}
\put(1112,416){\rule[-0.200pt]{0.512pt}{0.400pt}}
\put(1115,415){\rule[-0.200pt]{0.512pt}{0.400pt}}
\put(1117,414){\rule[-0.200pt]{0.512pt}{0.400pt}}
\put(1119,413){\rule[-0.200pt]{0.512pt}{0.400pt}}
\put(1121,412){\rule[-0.200pt]{0.512pt}{0.400pt}}
\put(1123,411){\rule[-0.200pt]{0.512pt}{0.400pt}}
\put(1125,410){\rule[-0.200pt]{0.512pt}{0.400pt}}
\put(1127,409){\rule[-0.200pt]{0.512pt}{0.400pt}}
\put(1129,408){\rule[-0.200pt]{0.512pt}{0.400pt}}
\put(1132,407){\rule[-0.200pt]{0.512pt}{0.400pt}}
\put(1134,406){\rule[-0.200pt]{0.512pt}{0.400pt}}
\put(1136,405){\rule[-0.200pt]{0.512pt}{0.400pt}}
\put(1138,404){\rule[-0.200pt]{0.512pt}{0.400pt}}
\put(1140,403){\rule[-0.200pt]{0.512pt}{0.400pt}}
\put(1142,402){\rule[-0.200pt]{0.512pt}{0.400pt}}
\put(1144,401){\rule[-0.200pt]{0.512pt}{0.400pt}}
\put(1146,400){\rule[-0.200pt]{0.512pt}{0.400pt}}
\put(1149,399){\rule[-0.200pt]{0.512pt}{0.400pt}}
\put(1151,398){\rule[-0.200pt]{0.512pt}{0.400pt}}
\put(1153,397){\rule[-0.200pt]{0.512pt}{0.400pt}}
\put(1155,396){\rule[-0.200pt]{0.512pt}{0.400pt}}
\put(1157,395){\rule[-0.200pt]{0.512pt}{0.400pt}}
\put(1159,394){\rule[-0.200pt]{0.512pt}{0.400pt}}
\put(1161,393){\rule[-0.200pt]{0.512pt}{0.400pt}}
\put(1163,392){\rule[-0.200pt]{0.512pt}{0.400pt}}
\put(1166,391){\rule[-0.200pt]{0.512pt}{0.400pt}}
\put(1168,390){\rule[-0.200pt]{0.512pt}{0.400pt}}
\put(1170,389){\rule[-0.200pt]{0.512pt}{0.400pt}}
\put(1172,388){\rule[-0.200pt]{0.512pt}{0.400pt}}
\put(1174,387){\rule[-0.200pt]{0.512pt}{0.400pt}}
\put(1176,386){\rule[-0.200pt]{0.512pt}{0.400pt}}
\put(1178,385){\rule[-0.200pt]{0.512pt}{0.400pt}}
\put(1180,384){\rule[-0.200pt]{0.512pt}{0.400pt}}
\put(1183,383){\rule[-0.200pt]{0.464pt}{0.400pt}}
\put(1184,382){\rule[-0.200pt]{0.464pt}{0.400pt}}
\put(1186,381){\rule[-0.200pt]{0.464pt}{0.400pt}}
\put(1188,380){\rule[-0.200pt]{0.464pt}{0.400pt}}
\put(1190,379){\rule[-0.200pt]{0.464pt}{0.400pt}}
\put(1192,378){\rule[-0.200pt]{0.464pt}{0.400pt}}
\put(1194,377){\rule[-0.200pt]{0.464pt}{0.400pt}}
\put(1196,376){\rule[-0.200pt]{0.464pt}{0.400pt}}
\put(1198,375){\rule[-0.200pt]{0.464pt}{0.400pt}}
\put(1200,374){\rule[-0.200pt]{0.464pt}{0.400pt}}
\put(1202,373){\rule[-0.200pt]{0.464pt}{0.400pt}}
\put(1204,372){\rule[-0.200pt]{0.464pt}{0.400pt}}
\put(1206,371){\rule[-0.200pt]{0.464pt}{0.400pt}}
\put(1208,370){\rule[-0.200pt]{0.464pt}{0.400pt}}
\put(1209,369){\rule[-0.200pt]{0.464pt}{0.400pt}}
\put(1211,368){\rule[-0.200pt]{0.464pt}{0.400pt}}
\put(1213,367){\rule[-0.200pt]{0.464pt}{0.400pt}}
\put(1215,366){\rule[-0.200pt]{0.464pt}{0.400pt}}
\put(1217,365){\rule[-0.200pt]{0.464pt}{0.400pt}}
\put(1219,364){\rule[-0.200pt]{0.464pt}{0.400pt}}
\put(1221,363){\rule[-0.200pt]{0.464pt}{0.400pt}}
\put(1223,362){\rule[-0.200pt]{0.464pt}{0.400pt}}
\put(1225,361){\rule[-0.200pt]{0.464pt}{0.400pt}}
\put(1227,360){\rule[-0.200pt]{0.464pt}{0.400pt}}
\put(1229,359){\rule[-0.200pt]{0.464pt}{0.400pt}}
\put(1231,358){\rule[-0.200pt]{0.464pt}{0.400pt}}
\put(1233,357){\rule[-0.200pt]{0.464pt}{0.400pt}}
\put(1234,356){\rule[-0.200pt]{0.464pt}{0.400pt}}
\put(1236,355){\rule[-0.200pt]{0.464pt}{0.400pt}}
\put(1238,354){\rule[-0.200pt]{0.464pt}{0.400pt}}
\put(1240,353){\rule[-0.200pt]{0.464pt}{0.400pt}}
\put(1242,352){\rule[-0.200pt]{0.464pt}{0.400pt}}
\put(1244,351){\rule[-0.200pt]{0.464pt}{0.400pt}}
\put(1246,350){\rule[-0.200pt]{0.464pt}{0.400pt}}
\put(1248,349){\rule[-0.200pt]{0.464pt}{0.400pt}}
\put(1250,348){\rule[-0.200pt]{0.464pt}{0.400pt}}
\put(1252,347){\rule[-0.200pt]{0.464pt}{0.400pt}}
\put(1254,346){\rule[-0.200pt]{0.464pt}{0.400pt}}
\put(1256,345){\rule[-0.200pt]{0.464pt}{0.400pt}}
\put(1258,344){\rule[-0.200pt]{0.464pt}{0.400pt}}
\put(1259,343){\rule[-0.200pt]{0.464pt}{0.400pt}}
\put(1261,342){\rule[-0.200pt]{0.464pt}{0.400pt}}
\put(1263,341){\rule[-0.200pt]{0.464pt}{0.400pt}}
\put(1265,340){\rule[-0.200pt]{0.464pt}{0.400pt}}
\put(1267,339){\rule[-0.200pt]{0.464pt}{0.400pt}}
\put(1269,338){\rule[-0.200pt]{0.464pt}{0.400pt}}
\put(1271,337){\rule[-0.200pt]{0.464pt}{0.400pt}}
\put(1273,336){\rule[-0.200pt]{0.464pt}{0.400pt}}
\put(1275,335){\rule[-0.200pt]{0.464pt}{0.400pt}}
\put(1277,334){\rule[-0.200pt]{0.464pt}{0.400pt}}
\put(1279,333){\rule[-0.200pt]{0.464pt}{0.400pt}}
\put(1281,332){\rule[-0.200pt]{0.464pt}{0.400pt}}
\put(1283,331){\rule[-0.200pt]{0.463pt}{0.400pt}}
\end{picture}

%% file: fig4.tex
\setlength{\unitlength}{0.240900pt}
\begin{picture}(1349,990)(0,0)
\tenrm
\ifx\plotpoint\undefined\newsavebox{\plotpoint}\fi
\put(264,113){\line(1,0){20}}
\put(1285,113){\line(-1,0){20}}
\put(242,113){\makebox(0,0)[r]{0.01}}
\put(264,194){\line(1,0){10}}
\put(1285,194){\line(-1,0){10}}
\put(264,242){\line(1,0){10}}
\put(1285,242){\line(-1,0){10}}
\put(264,275){\line(1,0){10}}
\put(1285,275){\line(-1,0){10}}
\put(264,301){\line(1,0){10}}
\put(1285,301){\line(-1,0){10}}
\put(264,323){\line(1,0){10}}
\put(1285,323){\line(-1,0){10}}
\put(264,341){\line(1,0){10}}
\put(1285,341){\line(-1,0){10}}
\put(264,357){\line(1,0){10}}
\put(1285,357){\line(-1,0){10}}
\put(264,370){\line(1,0){10}}
\put(1285,370){\line(-1,0){10}}
\put(264,383){\line(1,0){20}}
\put(1285,383){\line(-1,0){20}}
\put(242,383){\makebox(0,0)[r]{0.1}}
\put(264,464){\line(1,0){10}}
\put(1285,464){\line(-1,0){10}}
\put(264,511){\line(1,0){10}}
\put(1285,511){\line(-1,0){10}}
\put(264,545){\line(1,0){10}}
\put(1285,545){\line(-1,0){10}}
\put(264,571){\line(1,0){10}}
\put(1285,571){\line(-1,0){10}}
\put(264,593){\line(1,0){10}}
\put(1285,593){\line(-1,0){10}}
\put(264,611){\line(1,0){10}}
\put(1285,611){\line(-1,0){10}}
\put(264,626){\line(1,0){10}}
\put(1285,626){\line(-1,0){10}}
\put(264,640){\line(1,0){10}}
\put(1285,640){\line(-1,0){10}}
\put(264,652){\line(1,0){20}}
\put(1285,652){\line(-1,0){20}}
\put(242,652){\makebox(0,0)[r]{1}}
\put(264,734){\line(1,0){10}}
\put(1285,734){\line(-1,0){10}}
\put(264,781){\line(1,0){10}}
\put(1285,781){\line(-1,0){10}}
\put(264,815){\line(1,0){10}}
\put(1285,815){\line(-1,0){10}}
\put(264,841){\line(1,0){10}}
\put(1285,841){\line(-1,0){10}}
\put(264,862){\line(1,0){10}}
\put(1285,862){\line(-1,0){10}}
\put(264,880){\line(1,0){10}}
\put(1285,880){\line(-1,0){10}}
\put(264,896){\line(1,0){10}}
\put(1285,896){\line(-1,0){10}}
\put(264,910){\line(1,0){10}}
\put(1285,910){\line(-1,0){10}}
\put(264,922){\line(1,0){20}}
\put(1285,922){\line(-1,0){20}}
\put(242,922){\makebox(0,0)[r]{10}}
\put(264,113){\line(0,1){20}}
\put(264,922){\line(0,-1){20}}
\put(264,68){\makebox(0,0){50}}
\put(468,113){\line(0,1){20}}
\put(468,922){\line(0,-1){20}}
\put(468,68){\makebox(0,0){100}}
\put(672,113){\line(0,1){20}}
\put(672,922){\line(0,-1){20}}
\put(672,68){\makebox(0,0){150}}
\put(877,113){\line(0,1){20}}
\put(877,922){\line(0,-1){20}}
\put(877,68){\makebox(0,0){200}}
\put(1081,113){\line(0,1){20}}
\put(1081,922){\line(0,-1){20}}
\put(1081,68){\makebox(0,0){250}}
\put(1285,113){\line(0,1){20}}
\put(1285,922){\line(0,-1){20}}
\put(1285,68){\makebox(0,0){300}}
\put(264,113){\line(1,0){1021}}
\put(1285,113){\line(0,1){809}}
\put(1285,922){\line(-1,0){1021}}
\put(264,922){\line(0,-1){809}}
\put(45,517){\makebox(0,0)[l]{\shortstack{$\sigma$(pb)}}}
\put(774,23){\makebox(0,0){ $m_{\tilde t_R}$(GeV)}}
\put(570,652){\makebox(0,0)[l]{ $\sqrt{s}$=500 GeV NLC}}
\put(570,593){\makebox(0,0)[l]{ $\gamma^* e$ mode with W-W photons}}
\put(570,511){\makebox(0,0)[l]{ $\lambda'_{132}=0.33$}}
\sbox{\plotpoint}{\rule[-0.200pt]{0.400pt}{0.400pt}}%
\put(264,674){\usebox{\plotpoint}}
\put(264,674){\usebox{\plotpoint}}
\put(265,673){\usebox{\plotpoint}}
\put(266,672){\usebox{\plotpoint}}
\put(267,671){\usebox{\plotpoint}}
\put(268,670){\usebox{\plotpoint}}
\put(269,669){\usebox{\plotpoint}}
\put(270,668){\usebox{\plotpoint}}
\put(271,667){\usebox{\plotpoint}}
\put(272,666){\usebox{\plotpoint}}
\put(273,665){\usebox{\plotpoint}}
\put(274,664){\usebox{\plotpoint}}
\put(276,663){\usebox{\plotpoint}}
\put(277,662){\usebox{\plotpoint}}
\put(278,661){\usebox{\plotpoint}}
\put(279,660){\usebox{\plotpoint}}
\put(280,659){\usebox{\plotpoint}}
\put(281,658){\usebox{\plotpoint}}
\put(282,657){\usebox{\plotpoint}}
\put(283,656){\usebox{\plotpoint}}
\put(284,655){\usebox{\plotpoint}}
\put(285,654){\usebox{\plotpoint}}
\put(287,653){\usebox{\plotpoint}}
\put(288,652){\usebox{\plotpoint}}
\put(289,651){\usebox{\plotpoint}}
\put(290,650){\usebox{\plotpoint}}
\put(291,649){\usebox{\plotpoint}}
\put(292,648){\usebox{\plotpoint}}
\put(293,647){\usebox{\plotpoint}}
\put(294,646){\usebox{\plotpoint}}
\put(295,645){\usebox{\plotpoint}}
\put(296,644){\usebox{\plotpoint}}
\put(297,643){\usebox{\plotpoint}}
\put(299,642){\usebox{\plotpoint}}
\put(300,641){\usebox{\plotpoint}}
\put(301,640){\usebox{\plotpoint}}
\put(302,639){\usebox{\plotpoint}}
\put(303,638){\usebox{\plotpoint}}
\put(304,637){\usebox{\plotpoint}}
\put(305,636){\usebox{\plotpoint}}
\put(306,635){\usebox{\plotpoint}}
\put(307,634){\usebox{\plotpoint}}
\put(308,633){\usebox{\plotpoint}}
\put(310,632){\usebox{\plotpoint}}
\put(311,631){\usebox{\plotpoint}}
\put(312,630){\usebox{\plotpoint}}
\put(313,629){\usebox{\plotpoint}}
\put(314,628){\usebox{\plotpoint}}
\put(315,627){\usebox{\plotpoint}}
\put(316,626){\usebox{\plotpoint}}
\put(317,625){\usebox{\plotpoint}}
\put(318,624){\usebox{\plotpoint}}
\put(319,623){\usebox{\plotpoint}}
\put(321,622){\usebox{\plotpoint}}
\put(322,621){\usebox{\plotpoint}}
\put(323,620){\usebox{\plotpoint}}
\put(324,619){\usebox{\plotpoint}}
\put(325,618){\usebox{\plotpoint}}
\put(326,617){\usebox{\plotpoint}}
\put(327,616){\usebox{\plotpoint}}
\put(328,615){\usebox{\plotpoint}}
\put(329,614){\usebox{\plotpoint}}
\put(330,613){\usebox{\plotpoint}}
\put(331,612){\usebox{\plotpoint}}
\put(333,611){\usebox{\plotpoint}}
\put(334,610){\usebox{\plotpoint}}
\put(335,609){\usebox{\plotpoint}}
\put(336,608){\usebox{\plotpoint}}
\put(337,607){\usebox{\plotpoint}}
\put(338,606){\usebox{\plotpoint}}
\put(339,605){\usebox{\plotpoint}}
\put(340,604){\usebox{\plotpoint}}
\put(341,603){\usebox{\plotpoint}}
\put(342,602){\usebox{\plotpoint}}
\put(344,601){\usebox{\plotpoint}}
\put(345,600){\usebox{\plotpoint}}
\put(346,599){\usebox{\plotpoint}}
\put(347,598){\usebox{\plotpoint}}
\put(348,597){\usebox{\plotpoint}}
\put(349,596){\usebox{\plotpoint}}
\put(350,595){\usebox{\plotpoint}}
\put(351,594){\usebox{\plotpoint}}
\put(352,593){\usebox{\plotpoint}}
\put(353,592){\usebox{\plotpoint}}
\put(355,591){\usebox{\plotpoint}}
\put(356,590){\usebox{\plotpoint}}
\put(357,589){\usebox{\plotpoint}}
\put(358,588){\usebox{\plotpoint}}
\put(359,587){\usebox{\plotpoint}}
\put(360,586){\usebox{\plotpoint}}
\put(361,585){\usebox{\plotpoint}}
\put(362,584){\usebox{\plotpoint}}
\put(363,583){\usebox{\plotpoint}}
\put(364,582){\usebox{\plotpoint}}
\put(365,581){\usebox{\plotpoint}}
\put(367,580){\usebox{\plotpoint}}
\put(368,579){\usebox{\plotpoint}}
\put(370,578){\usebox{\plotpoint}}
\put(371,577){\usebox{\plotpoint}}
\put(373,576){\usebox{\plotpoint}}
\put(374,575){\usebox{\plotpoint}}
\put(376,574){\usebox{\plotpoint}}
\put(377,573){\usebox{\plotpoint}}
\put(379,572){\usebox{\plotpoint}}
\put(380,571){\usebox{\plotpoint}}
\put(382,570){\usebox{\plotpoint}}
\put(383,569){\usebox{\plotpoint}}
\put(384,568){\usebox{\plotpoint}}
\put(386,567){\usebox{\plotpoint}}
\put(387,566){\usebox{\plotpoint}}
\put(389,565){\usebox{\plotpoint}}
\put(390,564){\usebox{\plotpoint}}
\put(392,563){\usebox{\plotpoint}}
\put(393,562){\usebox{\plotpoint}}
\put(395,561){\usebox{\plotpoint}}
\put(396,560){\usebox{\plotpoint}}
\put(398,559){\usebox{\plotpoint}}
\put(399,558){\usebox{\plotpoint}}
\put(400,557){\usebox{\plotpoint}}
\put(402,556){\usebox{\plotpoint}}
\put(403,555){\usebox{\plotpoint}}
\put(405,554){\usebox{\plotpoint}}
\put(406,553){\usebox{\plotpoint}}
\put(408,552){\usebox{\plotpoint}}
\put(409,551){\usebox{\plotpoint}}
\put(411,550){\usebox{\plotpoint}}
\put(412,549){\usebox{\plotpoint}}
\put(414,548){\usebox{\plotpoint}}
\put(415,547){\usebox{\plotpoint}}
\put(417,546){\usebox{\plotpoint}}
\put(418,545){\usebox{\plotpoint}}
\put(419,544){\usebox{\plotpoint}}
\put(421,543){\usebox{\plotpoint}}
\put(422,542){\usebox{\plotpoint}}
\put(424,541){\usebox{\plotpoint}}
\put(425,540){\usebox{\plotpoint}}
\put(427,539){\usebox{\plotpoint}}
\put(428,538){\usebox{\plotpoint}}
\put(430,537){\usebox{\plotpoint}}
\put(431,536){\usebox{\plotpoint}}
\put(433,535){\usebox{\plotpoint}}
\put(434,534){\usebox{\plotpoint}}
\put(435,533){\usebox{\plotpoint}}
\put(437,532){\usebox{\plotpoint}}
\put(438,531){\usebox{\plotpoint}}
\put(440,530){\usebox{\plotpoint}}
\put(441,529){\usebox{\plotpoint}}
\put(443,528){\usebox{\plotpoint}}
\put(444,527){\usebox{\plotpoint}}
\put(446,526){\usebox{\plotpoint}}
\put(447,525){\usebox{\plotpoint}}
\put(449,524){\usebox{\plotpoint}}
\put(450,523){\usebox{\plotpoint}}
\put(451,522){\usebox{\plotpoint}}
\put(453,521){\usebox{\plotpoint}}
\put(454,520){\usebox{\plotpoint}}
\put(456,519){\usebox{\plotpoint}}
\put(457,518){\usebox{\plotpoint}}
\put(459,517){\usebox{\plotpoint}}
\put(460,516){\usebox{\plotpoint}}
\put(462,515){\usebox{\plotpoint}}
\put(463,514){\usebox{\plotpoint}}
\put(465,513){\usebox{\plotpoint}}
\put(466,512){\usebox{\plotpoint}}
\put(468,511){\rule[-0.200pt]{0.431pt}{0.400pt}}
\put(469,510){\rule[-0.200pt]{0.431pt}{0.400pt}}
\put(471,509){\rule[-0.200pt]{0.431pt}{0.400pt}}
\put(473,508){\rule[-0.200pt]{0.431pt}{0.400pt}}
\put(475,507){\rule[-0.200pt]{0.431pt}{0.400pt}}
\put(476,506){\rule[-0.200pt]{0.431pt}{0.400pt}}
\put(478,505){\rule[-0.200pt]{0.431pt}{0.400pt}}
\put(480,504){\rule[-0.200pt]{0.431pt}{0.400pt}}
\put(482,503){\rule[-0.200pt]{0.431pt}{0.400pt}}
\put(484,502){\rule[-0.200pt]{0.431pt}{0.400pt}}
\put(485,501){\rule[-0.200pt]{0.431pt}{0.400pt}}
\put(487,500){\rule[-0.200pt]{0.431pt}{0.400pt}}
\put(489,499){\rule[-0.200pt]{0.431pt}{0.400pt}}
\put(491,498){\rule[-0.200pt]{0.431pt}{0.400pt}}
\put(493,497){\rule[-0.200pt]{0.431pt}{0.400pt}}
\put(494,496){\rule[-0.200pt]{0.431pt}{0.400pt}}
\put(496,495){\rule[-0.200pt]{0.431pt}{0.400pt}}
\put(498,494){\rule[-0.200pt]{0.431pt}{0.400pt}}
\put(500,493){\rule[-0.200pt]{0.431pt}{0.400pt}}
\put(501,492){\rule[-0.200pt]{0.431pt}{0.400pt}}
\put(503,491){\rule[-0.200pt]{0.431pt}{0.400pt}}
\put(505,490){\rule[-0.200pt]{0.431pt}{0.400pt}}
\put(507,489){\rule[-0.200pt]{0.431pt}{0.400pt}}
\put(509,488){\rule[-0.200pt]{0.431pt}{0.400pt}}
\put(510,487){\rule[-0.200pt]{0.431pt}{0.400pt}}
\put(512,486){\rule[-0.200pt]{0.431pt}{0.400pt}}
\put(514,485){\rule[-0.200pt]{0.431pt}{0.400pt}}
\put(516,484){\rule[-0.200pt]{0.431pt}{0.400pt}}
\put(518,483){\rule[-0.200pt]{0.431pt}{0.400pt}}
\put(519,482){\rule[-0.200pt]{0.431pt}{0.400pt}}
\put(521,481){\rule[-0.200pt]{0.431pt}{0.400pt}}
\put(523,480){\rule[-0.200pt]{0.431pt}{0.400pt}}
\put(525,479){\rule[-0.200pt]{0.431pt}{0.400pt}}
\put(527,478){\rule[-0.200pt]{0.431pt}{0.400pt}}
\put(528,477){\rule[-0.200pt]{0.431pt}{0.400pt}}
\put(530,476){\rule[-0.200pt]{0.431pt}{0.400pt}}
\put(532,475){\rule[-0.200pt]{0.431pt}{0.400pt}}
\put(534,474){\rule[-0.200pt]{0.431pt}{0.400pt}}
\put(535,473){\rule[-0.200pt]{0.431pt}{0.400pt}}
\put(537,472){\rule[-0.200pt]{0.431pt}{0.400pt}}
\put(539,471){\rule[-0.200pt]{0.431pt}{0.400pt}}
\put(541,470){\rule[-0.200pt]{0.431pt}{0.400pt}}
\put(543,469){\rule[-0.200pt]{0.431pt}{0.400pt}}
\put(544,468){\rule[-0.200pt]{0.431pt}{0.400pt}}
\put(546,467){\rule[-0.200pt]{0.431pt}{0.400pt}}
\put(548,466){\rule[-0.200pt]{0.431pt}{0.400pt}}
\put(550,465){\rule[-0.200pt]{0.431pt}{0.400pt}}
\put(552,464){\rule[-0.200pt]{0.431pt}{0.400pt}}
\put(553,463){\rule[-0.200pt]{0.431pt}{0.400pt}}
\put(555,462){\rule[-0.200pt]{0.431pt}{0.400pt}}
\put(557,461){\rule[-0.200pt]{0.431pt}{0.400pt}}
\put(559,460){\rule[-0.200pt]{0.431pt}{0.400pt}}
\put(561,459){\rule[-0.200pt]{0.431pt}{0.400pt}}
\put(562,458){\rule[-0.200pt]{0.431pt}{0.400pt}}
\put(564,457){\rule[-0.200pt]{0.431pt}{0.400pt}}
\put(566,456){\rule[-0.200pt]{0.431pt}{0.400pt}}
\put(568,455){\rule[-0.200pt]{0.431pt}{0.400pt}}
\put(570,454){\rule[-0.200pt]{0.491pt}{0.400pt}}
\put(572,453){\rule[-0.200pt]{0.491pt}{0.400pt}}
\put(574,452){\rule[-0.200pt]{0.491pt}{0.400pt}}
\put(576,451){\rule[-0.200pt]{0.491pt}{0.400pt}}
\put(578,450){\rule[-0.200pt]{0.491pt}{0.400pt}}
\put(580,449){\rule[-0.200pt]{0.491pt}{0.400pt}}
\put(582,448){\rule[-0.200pt]{0.491pt}{0.400pt}}
\put(584,447){\rule[-0.200pt]{0.491pt}{0.400pt}}
\put(586,446){\rule[-0.200pt]{0.491pt}{0.400pt}}
\put(588,445){\rule[-0.200pt]{0.491pt}{0.400pt}}
\put(590,444){\rule[-0.200pt]{0.491pt}{0.400pt}}
\put(592,443){\rule[-0.200pt]{0.491pt}{0.400pt}}
\put(594,442){\rule[-0.200pt]{0.491pt}{0.400pt}}
\put(596,441){\rule[-0.200pt]{0.491pt}{0.400pt}}
\put(598,440){\rule[-0.200pt]{0.491pt}{0.400pt}}
\put(600,439){\rule[-0.200pt]{0.491pt}{0.400pt}}
\put(602,438){\rule[-0.200pt]{0.491pt}{0.400pt}}
\put(604,437){\rule[-0.200pt]{0.491pt}{0.400pt}}
\put(606,436){\rule[-0.200pt]{0.491pt}{0.400pt}}
\put(608,435){\rule[-0.200pt]{0.491pt}{0.400pt}}
\put(610,434){\rule[-0.200pt]{0.491pt}{0.400pt}}
\put(612,433){\rule[-0.200pt]{0.491pt}{0.400pt}}
\put(614,432){\rule[-0.200pt]{0.491pt}{0.400pt}}
\put(616,431){\rule[-0.200pt]{0.491pt}{0.400pt}}
\put(618,430){\rule[-0.200pt]{0.491pt}{0.400pt}}
\put(620,429){\rule[-0.200pt]{0.491pt}{0.400pt}}
\put(623,428){\rule[-0.200pt]{0.491pt}{0.400pt}}
\put(625,427){\rule[-0.200pt]{0.491pt}{0.400pt}}
\put(627,426){\rule[-0.200pt]{0.491pt}{0.400pt}}
\put(629,425){\rule[-0.200pt]{0.491pt}{0.400pt}}
\put(631,424){\rule[-0.200pt]{0.491pt}{0.400pt}}
\put(633,423){\rule[-0.200pt]{0.491pt}{0.400pt}}
\put(635,422){\rule[-0.200pt]{0.491pt}{0.400pt}}
\put(637,421){\rule[-0.200pt]{0.491pt}{0.400pt}}
\put(639,420){\rule[-0.200pt]{0.491pt}{0.400pt}}
\put(641,419){\rule[-0.200pt]{0.491pt}{0.400pt}}
\put(643,418){\rule[-0.200pt]{0.491pt}{0.400pt}}
\put(645,417){\rule[-0.200pt]{0.491pt}{0.400pt}}
\put(647,416){\rule[-0.200pt]{0.491pt}{0.400pt}}
\put(649,415){\rule[-0.200pt]{0.491pt}{0.400pt}}
\put(651,414){\rule[-0.200pt]{0.491pt}{0.400pt}}
\put(653,413){\rule[-0.200pt]{0.491pt}{0.400pt}}
\put(655,412){\rule[-0.200pt]{0.491pt}{0.400pt}}
\put(657,411){\rule[-0.200pt]{0.491pt}{0.400pt}}
\put(659,410){\rule[-0.200pt]{0.491pt}{0.400pt}}
\put(661,409){\rule[-0.200pt]{0.491pt}{0.400pt}}
\put(663,408){\rule[-0.200pt]{0.491pt}{0.400pt}}
\put(665,407){\rule[-0.200pt]{0.491pt}{0.400pt}}
\put(667,406){\rule[-0.200pt]{0.491pt}{0.400pt}}
\put(669,405){\rule[-0.200pt]{0.491pt}{0.400pt}}
\put(671,404){\rule[-0.200pt]{0.559pt}{0.400pt}}
\put(674,403){\rule[-0.200pt]{0.558pt}{0.400pt}}
\put(676,402){\rule[-0.200pt]{0.558pt}{0.400pt}}
\put(678,401){\rule[-0.200pt]{0.558pt}{0.400pt}}
\put(681,400){\rule[-0.200pt]{0.558pt}{0.400pt}}
\put(683,399){\rule[-0.200pt]{0.558pt}{0.400pt}}
\put(685,398){\rule[-0.200pt]{0.558pt}{0.400pt}}
\put(688,397){\rule[-0.200pt]{0.558pt}{0.400pt}}
\put(690,396){\rule[-0.200pt]{0.558pt}{0.400pt}}
\put(692,395){\rule[-0.200pt]{0.558pt}{0.400pt}}
\put(695,394){\rule[-0.200pt]{0.558pt}{0.400pt}}
\put(697,393){\rule[-0.200pt]{0.558pt}{0.400pt}}
\put(699,392){\rule[-0.200pt]{0.558pt}{0.400pt}}
\put(702,391){\rule[-0.200pt]{0.558pt}{0.400pt}}
\put(704,390){\rule[-0.200pt]{0.558pt}{0.400pt}}
\put(706,389){\rule[-0.200pt]{0.558pt}{0.400pt}}
\put(709,388){\rule[-0.200pt]{0.558pt}{0.400pt}}
\put(711,387){\rule[-0.200pt]{0.558pt}{0.400pt}}
\put(713,386){\rule[-0.200pt]{0.558pt}{0.400pt}}
\put(716,385){\rule[-0.200pt]{0.558pt}{0.400pt}}
\put(718,384){\rule[-0.200pt]{0.558pt}{0.400pt}}
\put(720,383){\rule[-0.200pt]{0.558pt}{0.400pt}}
\put(722,382){\rule[-0.200pt]{0.558pt}{0.400pt}}
\put(725,381){\rule[-0.200pt]{0.558pt}{0.400pt}}
\put(727,380){\rule[-0.200pt]{0.558pt}{0.400pt}}
\put(729,379){\rule[-0.200pt]{0.558pt}{0.400pt}}
\put(732,378){\rule[-0.200pt]{0.558pt}{0.400pt}}
\put(734,377){\rule[-0.200pt]{0.558pt}{0.400pt}}
\put(736,376){\rule[-0.200pt]{0.558pt}{0.400pt}}
\put(739,375){\rule[-0.200pt]{0.558pt}{0.400pt}}
\put(741,374){\rule[-0.200pt]{0.558pt}{0.400pt}}
\put(743,373){\rule[-0.200pt]{0.558pt}{0.400pt}}
\put(746,372){\rule[-0.200pt]{0.558pt}{0.400pt}}
\put(748,371){\rule[-0.200pt]{0.558pt}{0.400pt}}
\put(750,370){\rule[-0.200pt]{0.558pt}{0.400pt}}
\put(753,369){\rule[-0.200pt]{0.558pt}{0.400pt}}
\put(755,368){\rule[-0.200pt]{0.558pt}{0.400pt}}
\put(757,367){\rule[-0.200pt]{0.558pt}{0.400pt}}
\put(760,366){\rule[-0.200pt]{0.558pt}{0.400pt}}
\put(762,365){\rule[-0.200pt]{0.558pt}{0.400pt}}
\put(764,364){\rule[-0.200pt]{0.558pt}{0.400pt}}
\put(767,363){\rule[-0.200pt]{0.558pt}{0.400pt}}
\put(769,362){\rule[-0.200pt]{0.558pt}{0.400pt}}
\put(771,361){\rule[-0.200pt]{0.558pt}{0.400pt}}
\put(773,360){\rule[-0.200pt]{0.591pt}{0.400pt}}
\put(776,359){\rule[-0.200pt]{0.591pt}{0.400pt}}
\put(778,358){\rule[-0.200pt]{0.591pt}{0.400pt}}
\put(781,357){\rule[-0.200pt]{0.591pt}{0.400pt}}
\put(783,356){\rule[-0.200pt]{0.591pt}{0.400pt}}
\put(786,355){\rule[-0.200pt]{0.591pt}{0.400pt}}
\put(788,354){\rule[-0.200pt]{0.591pt}{0.400pt}}
\put(791,353){\rule[-0.200pt]{0.591pt}{0.400pt}}
\put(793,352){\rule[-0.200pt]{0.591pt}{0.400pt}}
\put(796,351){\rule[-0.200pt]{0.591pt}{0.400pt}}
\put(798,350){\rule[-0.200pt]{0.591pt}{0.400pt}}
\put(800,349){\rule[-0.200pt]{0.591pt}{0.400pt}}
\put(803,348){\rule[-0.200pt]{0.591pt}{0.400pt}}
\put(805,347){\rule[-0.200pt]{0.591pt}{0.400pt}}
\put(808,346){\rule[-0.200pt]{0.591pt}{0.400pt}}
\put(810,345){\rule[-0.200pt]{0.591pt}{0.400pt}}
\put(813,344){\rule[-0.200pt]{0.591pt}{0.400pt}}
\put(815,343){\rule[-0.200pt]{0.591pt}{0.400pt}}
\put(818,342){\rule[-0.200pt]{0.591pt}{0.400pt}}
\put(820,341){\rule[-0.200pt]{0.591pt}{0.400pt}}
\put(823,340){\rule[-0.200pt]{0.591pt}{0.400pt}}
\put(825,339){\rule[-0.200pt]{0.591pt}{0.400pt}}
\put(827,338){\rule[-0.200pt]{0.591pt}{0.400pt}}
\put(830,337){\rule[-0.200pt]{0.591pt}{0.400pt}}
\put(832,336){\rule[-0.200pt]{0.591pt}{0.400pt}}
\put(835,335){\rule[-0.200pt]{0.591pt}{0.400pt}}
\put(837,334){\rule[-0.200pt]{0.591pt}{0.400pt}}
\put(840,333){\rule[-0.200pt]{0.591pt}{0.400pt}}
\put(842,332){\rule[-0.200pt]{0.591pt}{0.400pt}}
\put(845,331){\rule[-0.200pt]{0.591pt}{0.400pt}}
\put(847,330){\rule[-0.200pt]{0.591pt}{0.400pt}}
\put(850,329){\rule[-0.200pt]{0.591pt}{0.400pt}}
\put(852,328){\rule[-0.200pt]{0.591pt}{0.400pt}}
\put(854,327){\rule[-0.200pt]{0.591pt}{0.400pt}}
\put(857,326){\rule[-0.200pt]{0.591pt}{0.400pt}}
\put(859,325){\rule[-0.200pt]{0.591pt}{0.400pt}}
\put(862,324){\rule[-0.200pt]{0.591pt}{0.400pt}}
\put(864,323){\rule[-0.200pt]{0.591pt}{0.400pt}}
\put(867,322){\rule[-0.200pt]{0.591pt}{0.400pt}}
\put(869,321){\rule[-0.200pt]{0.591pt}{0.400pt}}
\put(872,320){\rule[-0.200pt]{0.591pt}{0.400pt}}
\put(874,319){\rule[-0.200pt]{0.591pt}{0.400pt}}
\put(877,318){\rule[-0.200pt]{0.614pt}{0.400pt}}
\put(879,317){\rule[-0.200pt]{0.614pt}{0.400pt}}
\put(882,316){\rule[-0.200pt]{0.614pt}{0.400pt}}
\put(884,315){\rule[-0.200pt]{0.614pt}{0.400pt}}
\put(887,314){\rule[-0.200pt]{0.614pt}{0.400pt}}
\put(889,313){\rule[-0.200pt]{0.614pt}{0.400pt}}
\put(892,312){\rule[-0.200pt]{0.614pt}{0.400pt}}
\put(894,311){\rule[-0.200pt]{0.614pt}{0.400pt}}
\put(897,310){\rule[-0.200pt]{0.614pt}{0.400pt}}
\put(899,309){\rule[-0.200pt]{0.614pt}{0.400pt}}
\put(902,308){\rule[-0.200pt]{0.614pt}{0.400pt}}
\put(905,307){\rule[-0.200pt]{0.614pt}{0.400pt}}
\put(907,306){\rule[-0.200pt]{0.614pt}{0.400pt}}
\put(910,305){\rule[-0.200pt]{0.614pt}{0.400pt}}
\put(912,304){\rule[-0.200pt]{0.614pt}{0.400pt}}
\put(915,303){\rule[-0.200pt]{0.614pt}{0.400pt}}
\put(917,302){\rule[-0.200pt]{0.614pt}{0.400pt}}
\put(920,301){\rule[-0.200pt]{0.614pt}{0.400pt}}
\put(922,300){\rule[-0.200pt]{0.614pt}{0.400pt}}
\put(925,299){\rule[-0.200pt]{0.614pt}{0.400pt}}
\put(927,298){\rule[-0.200pt]{0.614pt}{0.400pt}}
\put(930,297){\rule[-0.200pt]{0.614pt}{0.400pt}}
\put(933,296){\rule[-0.200pt]{0.614pt}{0.400pt}}
\put(935,295){\rule[-0.200pt]{0.614pt}{0.400pt}}
\put(938,294){\rule[-0.200pt]{0.614pt}{0.400pt}}
\put(940,293){\rule[-0.200pt]{0.614pt}{0.400pt}}
\put(943,292){\rule[-0.200pt]{0.614pt}{0.400pt}}
\put(945,291){\rule[-0.200pt]{0.614pt}{0.400pt}}
\put(948,290){\rule[-0.200pt]{0.614pt}{0.400pt}}
\put(950,289){\rule[-0.200pt]{0.614pt}{0.400pt}}
\put(953,288){\rule[-0.200pt]{0.614pt}{0.400pt}}
\put(956,287){\rule[-0.200pt]{0.614pt}{0.400pt}}
\put(958,286){\rule[-0.200pt]{0.614pt}{0.400pt}}
\put(961,285){\rule[-0.200pt]{0.614pt}{0.400pt}}
\put(963,284){\rule[-0.200pt]{0.614pt}{0.400pt}}
\put(966,283){\rule[-0.200pt]{0.614pt}{0.400pt}}
\put(968,282){\rule[-0.200pt]{0.614pt}{0.400pt}}
\put(971,281){\rule[-0.200pt]{0.614pt}{0.400pt}}
\put(973,280){\rule[-0.200pt]{0.614pt}{0.400pt}}
\put(976,279){\rule[-0.200pt]{0.614pt}{0.400pt}}
\put(978,278){\rule[-0.200pt]{0.630pt}{0.400pt}}
\put(981,277){\rule[-0.200pt]{0.630pt}{0.400pt}}
\put(984,276){\rule[-0.200pt]{0.630pt}{0.400pt}}
\put(986,275){\rule[-0.200pt]{0.630pt}{0.400pt}}
\put(989,274){\rule[-0.200pt]{0.630pt}{0.400pt}}
\put(992,273){\rule[-0.200pt]{0.630pt}{0.400pt}}
\put(994,272){\rule[-0.200pt]{0.630pt}{0.400pt}}
\put(997,271){\rule[-0.200pt]{0.630pt}{0.400pt}}
\put(999,270){\rule[-0.200pt]{0.630pt}{0.400pt}}
\put(1002,269){\rule[-0.200pt]{0.630pt}{0.400pt}}
\put(1005,268){\rule[-0.200pt]{0.630pt}{0.400pt}}
\put(1007,267){\rule[-0.200pt]{0.630pt}{0.400pt}}
\put(1010,266){\rule[-0.200pt]{0.630pt}{0.400pt}}
\put(1012,265){\rule[-0.200pt]{0.630pt}{0.400pt}}
\put(1015,264){\rule[-0.200pt]{0.630pt}{0.400pt}}
\put(1018,263){\rule[-0.200pt]{0.630pt}{0.400pt}}
\put(1020,262){\rule[-0.200pt]{0.630pt}{0.400pt}}
\put(1023,261){\rule[-0.200pt]{0.630pt}{0.400pt}}
\put(1026,260){\rule[-0.200pt]{0.630pt}{0.400pt}}
\put(1028,259){\rule[-0.200pt]{0.630pt}{0.400pt}}
\put(1031,258){\rule[-0.200pt]{0.630pt}{0.400pt}}
\put(1033,257){\rule[-0.200pt]{0.630pt}{0.400pt}}
\put(1036,256){\rule[-0.200pt]{0.630pt}{0.400pt}}
\put(1039,255){\rule[-0.200pt]{0.630pt}{0.400pt}}
\put(1041,254){\rule[-0.200pt]{0.630pt}{0.400pt}}
\put(1044,253){\rule[-0.200pt]{0.630pt}{0.400pt}}
\put(1046,252){\rule[-0.200pt]{0.630pt}{0.400pt}}
\put(1049,251){\rule[-0.200pt]{0.630pt}{0.400pt}}
\put(1052,250){\rule[-0.200pt]{0.630pt}{0.400pt}}
\put(1054,249){\rule[-0.200pt]{0.630pt}{0.400pt}}
\put(1057,248){\rule[-0.200pt]{0.630pt}{0.400pt}}
\put(1060,247){\rule[-0.200pt]{0.630pt}{0.400pt}}
\put(1062,246){\rule[-0.200pt]{0.630pt}{0.400pt}}
\put(1065,245){\rule[-0.200pt]{0.630pt}{0.400pt}}
\put(1067,244){\rule[-0.200pt]{0.630pt}{0.400pt}}
\put(1070,243){\rule[-0.200pt]{0.630pt}{0.400pt}}
\put(1073,242){\rule[-0.200pt]{0.630pt}{0.400pt}}
\put(1075,241){\rule[-0.200pt]{0.630pt}{0.400pt}}
\put(1078,240){\rule[-0.200pt]{0.630pt}{0.400pt}}
\put(1080,239){\rule[-0.200pt]{0.630pt}{0.400pt}}
\put(1083,238){\rule[-0.200pt]{0.630pt}{0.400pt}}
\put(1086,237){\rule[-0.200pt]{0.630pt}{0.400pt}}
\put(1088,236){\rule[-0.200pt]{0.630pt}{0.400pt}}
\put(1091,235){\rule[-0.200pt]{0.630pt}{0.400pt}}
\put(1094,234){\rule[-0.200pt]{0.630pt}{0.400pt}}
\put(1096,233){\rule[-0.200pt]{0.630pt}{0.400pt}}
\put(1099,232){\rule[-0.200pt]{0.630pt}{0.400pt}}
\put(1101,231){\rule[-0.200pt]{0.630pt}{0.400pt}}
\put(1104,230){\rule[-0.200pt]{0.630pt}{0.400pt}}
\put(1107,229){\rule[-0.200pt]{0.630pt}{0.400pt}}
\put(1109,228){\rule[-0.200pt]{0.630pt}{0.400pt}}
\put(1112,227){\rule[-0.200pt]{0.630pt}{0.400pt}}
\put(1114,226){\rule[-0.200pt]{0.630pt}{0.400pt}}
\put(1117,225){\rule[-0.200pt]{0.630pt}{0.400pt}}
\put(1120,224){\rule[-0.200pt]{0.630pt}{0.400pt}}
\put(1122,223){\rule[-0.200pt]{0.630pt}{0.400pt}}
\put(1125,222){\rule[-0.200pt]{0.630pt}{0.400pt}}
\put(1128,221){\rule[-0.200pt]{0.630pt}{0.400pt}}
\put(1130,220){\rule[-0.200pt]{0.630pt}{0.400pt}}
\put(1133,219){\rule[-0.200pt]{0.630pt}{0.400pt}}
\put(1135,218){\rule[-0.200pt]{0.630pt}{0.400pt}}
\put(1138,217){\rule[-0.200pt]{0.630pt}{0.400pt}}
\put(1141,216){\rule[-0.200pt]{0.630pt}{0.400pt}}
\put(1143,215){\rule[-0.200pt]{0.630pt}{0.400pt}}
\put(1146,214){\rule[-0.200pt]{0.630pt}{0.400pt}}
\put(1148,213){\rule[-0.200pt]{0.630pt}{0.400pt}}
\put(1151,212){\rule[-0.200pt]{0.630pt}{0.400pt}}
\put(1154,211){\rule[-0.200pt]{0.630pt}{0.400pt}}
\put(1156,210){\rule[-0.200pt]{0.630pt}{0.400pt}}
\put(1159,209){\rule[-0.200pt]{0.630pt}{0.400pt}}
\put(1162,208){\rule[-0.200pt]{0.630pt}{0.400pt}}
\put(1164,207){\rule[-0.200pt]{0.630pt}{0.400pt}}
\put(1167,206){\rule[-0.200pt]{0.630pt}{0.400pt}}
\put(1169,205){\rule[-0.200pt]{0.630pt}{0.400pt}}
\put(1172,204){\rule[-0.200pt]{0.630pt}{0.400pt}}
\put(1175,203){\rule[-0.200pt]{0.630pt}{0.400pt}}
\put(1177,202){\rule[-0.200pt]{0.630pt}{0.400pt}}
\put(1180,201){\rule[-0.200pt]{0.630pt}{0.400pt}}
\put(1182,200){\rule[-0.200pt]{0.600pt}{0.400pt}}
\put(1185,199){\rule[-0.200pt]{0.599pt}{0.400pt}}
\put(1187,198){\rule[-0.200pt]{0.599pt}{0.400pt}}
\put(1190,197){\rule[-0.200pt]{0.599pt}{0.400pt}}
\put(1192,196){\rule[-0.200pt]{0.599pt}{0.400pt}}
\put(1195,195){\rule[-0.200pt]{0.599pt}{0.400pt}}
\put(1197,194){\rule[-0.200pt]{0.599pt}{0.400pt}}
\put(1200,193){\rule[-0.200pt]{0.599pt}{0.400pt}}
\put(1202,192){\rule[-0.200pt]{0.599pt}{0.400pt}}
\put(1205,191){\rule[-0.200pt]{0.599pt}{0.400pt}}
\put(1207,190){\rule[-0.200pt]{0.599pt}{0.400pt}}
\put(1210,189){\rule[-0.200pt]{0.599pt}{0.400pt}}
\put(1212,188){\rule[-0.200pt]{0.599pt}{0.400pt}}
\put(1215,187){\rule[-0.200pt]{0.599pt}{0.400pt}}
\put(1217,186){\rule[-0.200pt]{0.599pt}{0.400pt}}
\put(1220,185){\rule[-0.200pt]{0.599pt}{0.400pt}}
\put(1222,184){\rule[-0.200pt]{0.599pt}{0.400pt}}
\put(1225,183){\rule[-0.200pt]{0.599pt}{0.400pt}}
\put(1227,182){\rule[-0.200pt]{0.599pt}{0.400pt}}
\put(1230,181){\rule[-0.200pt]{0.599pt}{0.400pt}}
\put(1232,180){\rule[-0.200pt]{0.599pt}{0.400pt}}
\put(1235,179){\rule[-0.200pt]{0.599pt}{0.400pt}}
\put(1237,178){\rule[-0.200pt]{0.599pt}{0.400pt}}
\put(1240,177){\rule[-0.200pt]{0.599pt}{0.400pt}}
\put(1242,176){\rule[-0.200pt]{0.599pt}{0.400pt}}
\put(1245,175){\rule[-0.200pt]{0.599pt}{0.400pt}}
\put(1247,174){\rule[-0.200pt]{0.599pt}{0.400pt}}
\put(1250,173){\rule[-0.200pt]{0.599pt}{0.400pt}}
\put(1252,172){\rule[-0.200pt]{0.599pt}{0.400pt}}
\put(1255,171){\rule[-0.200pt]{0.599pt}{0.400pt}}
\put(1257,170){\rule[-0.200pt]{0.599pt}{0.400pt}}
\put(1260,169){\rule[-0.200pt]{0.599pt}{0.400pt}}
\put(1262,168){\rule[-0.200pt]{0.599pt}{0.400pt}}
\put(1265,167){\rule[-0.200pt]{0.599pt}{0.400pt}}
\put(1267,166){\rule[-0.200pt]{0.599pt}{0.400pt}}
\put(1270,165){\rule[-0.200pt]{0.599pt}{0.400pt}}
\put(1272,164){\rule[-0.200pt]{0.599pt}{0.400pt}}
\put(1275,163){\rule[-0.200pt]{0.599pt}{0.400pt}}
\put(1277,162){\rule[-0.200pt]{0.599pt}{0.400pt}}
\put(1280,161){\rule[-0.200pt]{0.599pt}{0.400pt}}
\put(1282,160){\rule[-0.200pt]{0.599pt}{0.400pt}}
\put(1284,159){\usebox{\plotpoint}}
\end{picture}